%% file: paper327_final.tex
\documentstyle[10pt,epsf,epsfig,hangcaption,xspace,amssymb,amsfonts,amsmath,amsthm,cite,
dp_delphititle,lineno]{dp_delphi}
\makeindex
\def\DpPaperGroup{EP}
\def\DpPaperRef{2004-007}
\def\DpDate{06 January 2004}
\def\DpAuthors{DELPHI Collaboration}
\def\DpSubmit{(Accepted by Euro. Phys. J. C)}
\def\DpTitle{{  The measurement of \boldmath\as\ from event shapes with the
DELPHI detector at the highest LEP energies}}
\def\DpComment{  }
\def\DpEMail{   }

\newcommand{\fig}{Figure~\ref}
\newcommand{\tab}{Table~\ref}
\newcommand{\as}{$\alpha_s$}
\newcommand{\oas}{$\cal O$($\alpha_s^2$)}

\newcommand{\gev}{\mbox{\,Ge\kern-0.2exV}}
\newcommand{\mev}{\mbox{\,Me\kern-0.2exV}}
\newcommand{\bmin}{$B_{\mathrm min}$}

\newcommand{\qcd}
{$e^+e^- \rightarrow {\rm Z}/\gamma \rightarrow q\bar{q}$\hspace{0.1cm}}
\newcommand{\DW}{Dokshitzer and Webber}
\newcommand{\asb}{${\alpha}_0$}
\newcommand{\beq}{\begin{equation}}
\newcommand{\eeq}{\end{equation}}
\def\jetset{{\sc Jetset}}
\def\pythia{{\sc Pythia}}
\def\herwig{{\sc Herwig}}
\def\ariadne{{\sc Ariadne}}
\def\sprime{{\sc Sprime}}
\def\delsim{{\sc Delsim}}

\begin{document}
\makeatletter
\newcount\@tempcntc
\def\@citex[#1]#2{\if@filesw\immediate\write\@auxout{\string\citation{#2}}\fi
  \@tempcnta\z@\@tempcntb\m@ne\def\@citea{}\@cite{\@for\@citeb:=#2\do
    {\@ifundefined
       {b@\@citeb}{\@citeo\@tempcntb\m@ne\@citea\def\@citea{,}{\bf ?}\@warning
       {Citation `\@citeb' on page \thepage \space undefined}}%
    {\setbox\z@\hbox{\global\@tempcntc0\csname b@\@citeb\endcsname\relax}%
     \ifnum\@tempcntc=\z@ \@citeo\@tempcntb\m@ne
       \@citea\def\@citea{,}\hbox{\csname b@\@citeb\endcsname}%
     \else
      \advance\@tempcntb\@ne
      \ifnum\@tempcntb=\@tempcntc
      \else\advance\@tempcntb\m@ne\@citeo
      \@tempcnta\@tempcntc\@tempcntb\@tempcntc\fi\fi}}\@citeo}{#1}}
\def\@citeo{\ifnum\@tempcnta>\@tempcntb\else\@citea\def\@citea{,}%
  \ifnum\@tempcnta=\@tempcntb\the\@tempcnta\else
   {\advance\@tempcnta\@ne\ifnum\@tempcnta=\@tempcntb \else \def\@citea{--}\fi
    \advance\@tempcnta\m@ne\the\@tempcnta\@citea\the\@tempcntb}\fi\fi}
 
\makeatother
\begin{titlepage}
\pagenumbering{roman}
\CERNpreprint{\DpPaperGroup}{\DpPaperRef} 
\date{{\small\DpDate}} 
\title{\DpTitle} 
\address{\DpAuthors} 
\begin{shortabs} 
\noindent
\noindent
Hadronic event shape distributions are determined from data in $e^+e^-$ 
collisions between 183 and 207 \gev.
From these the strong coupling $\alpha_s$ is extracted
in ${\cal{O}}(\alpha_s^2)$, NLLA and matched ${\cal{O}}(\alpha_s^2)$+NLLA
theory. Hadronisation corrections evaluated with fragmentation model
generators as well as an analytical  power ansatz are applied.
Comparing these measurements to those obtained at and around $M_Z$ allows
a combined measurement of $\alpha_s$ from all DELPHI data and a test of the 
energy dependence of the strong coupling.
\end{shortabs}
\vfill
\begin{center}
\DpSubmit \ \\ 
\DpComment \ \\
\DpEMail \ \\
\end{center}
\vfill
\clearpage
\headsep 10.0pt
\addtolength{\textheight}{10mm}
\addtolength{\footskip}{-5mm}
\begingroup
%
\newcommand{\DpName}[2]{\hbox{#1$^{\ref{#2}}$},\hfill}
\newcommand{\DpNameTwo}[3]{\hbox{#1$^{\ref{#2},\ref{#3}}$},\hfill}
\newcommand{\DpNameThree}[4]{\hbox{#1$^{\ref{#2},\ref{#3},\ref{#4}}$},\hfill}
\newskip\Bigfill \Bigfill = 0pt plus 1000fill
\newcommand{\DpNameLast}[2]{\hbox{#1$^{\ref{#2}}$}\hspace{\Bigfill}}
%
\footnotesize
\noindent
\DpName{J.Abdallah}{LPNHE}
\DpName{P.Abreu}{LIP}
\DpName{W.Adam}{VIENNA}
\DpName{P.Adzic}{DEMOKRITOS}
\DpName{T.Albrecht}{KARLSRUHE}
\DpName{T.Alderweireld}{AIM}
\DpName{R.Alemany-Fernandez}{CERN}
\DpName{T.Allmendinger}{KARLSRUHE}
\DpName{P.P.Allport}{LIVERPOOL}
\DpName{U.Amaldi}{MILANO2}
% ...
\normalsize
\endgroup
\titlefoot{Department of Physics and Astronomy, Iowa State
     University, Ames IA 50011-3160, USA
    \label{AMES}}
\titlefoot{Physics Department, Universiteit Antwerpen,
     Universiteitsplein 1, B-2610 Antwerpen, Belgium \\
     \indent~~and IIHE, ULB-VUB,
     Pleinlaan 2, B-1050 Brussels, Belgium \\
     \indent~~and Facult\'e des Sciences,
     Univ. de l'Etat Mons, Av. Maistriau 19, B-7000 Mons, Belgium
    \label{AIM}}
\titlefoot{Physics Laboratory, University of Athens, Solonos Str.
     104, GR-10680 Athens, Greece
    \label{ATHENS}}
\titlefoot{Department of Physics, University of Bergen,
     All\'egaten 55, NO-5007 Bergen, Norway
    \label{BERGEN}}
\titlefoot{Dipartimento di Fisica, Universit\`a di Bologna and INFN,
     Via Irnerio 46, IT-40126 Bologna, Italy
    \label{BOLOGNA}}
\titlefoot{Centro Brasileiro de Pesquisas F\'{\i}sicas, rua Xavier Sigaud 150,
     BR-22290 Rio de Janeiro, Brazil \\
     \indent~~and Depto. de F\'{\i}sica, Pont. Univ. Cat\'olica,
     C.P. 38071 BR-22453 Rio de Janeiro, Brazil \\
     \indent~~and Inst. de F\'{\i}sica, Univ. Estadual do Rio de Janeiro,
     rua S\~{a}o Francisco Xavier 524, Rio de Janeiro, Brazil
    \label{BRASIL}}
\titlefoot{Coll\`ege de France, Lab. de Physique Corpusculaire, IN2P3-CNRS,
     FR-75231 Paris Cedex 05, France
    \label{CDF}}
\titlefoot{CERN, CH-1211 Geneva 23, Switzerland
    \label{CERN}}
\titlefoot{Institut de Recherches Subatomiques, IN2P3 - CNRS/ULP - BP20,
     FR-67037 Strasbourg Cedex, France
    \label{CRN}}
\titlefoot{Now at DESY-Zeuthen, Platanenallee 6, D-15735 Zeuthen, Germany
    \label{DESY}}
\titlefoot{Institute of Nuclear Physics, N.C.S.R. Demokritos,
     P.O. Box 60228, GR-15310 Athens, Greece
    \label{DEMOKRITOS}}
\titlefoot{FZU, Inst. of Phys. of the C.A.S. High Energy Physics Division,
     Na Slovance 2, CZ-180 40, Praha 8, Czech Republic
    \label{FZU}}
\titlefoot{Dipartimento di Fisica, Universit\`a di Genova and INFN,
     Via Dodecaneso 33, IT-16146 Genova, Italy
    \label{GENOVA}}
\titlefoot{Institut des Sciences Nucl\'eaires, IN2P3-CNRS, Universit\'e
     de Grenoble 1, FR-38026 Grenoble Cedex, France
    \label{GRENOBLE}}
\titlefoot{Helsinki Institute of Physics, P.O. Box 64,
     FIN-00014 University of Helsinki, Finland
    \label{HELSINKI}}
\titlefoot{Joint Institute for Nuclear Research, Dubna, Head Post
     Office, P.O. Box 79, RU-101 000 Moscow, Russian Federation
    \label{JINR}}
\titlefoot{Institut f\"ur Experimentelle Kernphysik,
     Universit\"at Karlsruhe, Postfach 6980, DE-76128 Karlsruhe,
     Germany
    \label{KARLSRUHE}}
\titlefoot{Institute of Nuclear Physics PAN,Ul. Radzikowskiego 152,
     PL-31142 Krakow, Poland
    \label{KRAKOW1}}
\titlefoot{Faculty of Physics and Nuclear Techniques, University of Mining
     and Metallurgy, PL-30055 Krakow, Poland
    \label{KRAKOW2}}
\titlefoot{Universit\'e de Paris-Sud, Lab. de l'Acc\'el\'erateur
     Lin\'eaire, IN2P3-CNRS, B\^{a}t. 200, FR-91405 Orsay Cedex, France
    \label{LAL}}
\titlefoot{School of Physics and Chemistry, University of Lancaster,
     Lancaster LA1 4YB, UK
    \label{LANCASTER}}
\titlefoot{LIP, IST, FCUL - Av. Elias Garcia, 14-$1^{o}$,
     PT-1000 Lisboa Codex, Portugal
    \label{LIP}}
\titlefoot{Department of Physics, University of Liverpool, P.O.
     Box 147, Liverpool L69 3BX, UK
    \label{LIVERPOOL}}
\titlefoot{Dept. of Physics and Astronomy, Kelvin Building,
     University of Glasgow, Glasgow G12 8QQ
    \label{GLASGOW}}
\titlefoot{LPNHE, IN2P3-CNRS, Univ.~Paris VI et VII, Tour 33 (RdC),
     4 place Jussieu, FR-75252 Paris Cedex 05, France
    \label{LPNHE}}
\titlefoot{Department of Physics, University of Lund,
     S\"olvegatan 14, SE-223 63 Lund, Sweden
    \label{LUND}}
\titlefoot{Universit\'e Claude Bernard de Lyon, IPNL, IN2P3-CNRS,
     FR-69622 Villeurbanne Cedex, France
    \label{LYON}}
\titlefoot{Dipartimento di Fisica, Universit\`a di Milano and INFN-MILANO,
     Via Celoria 16, IT-20133 Milan, Italy
    \label{MILANO}}
\titlefoot{Dipartimento di Fisica, Univ. di Milano-Bicocca and
     INFN-MILANO, Piazza della Scienza 2, IT-20126 Milan, Italy
    \label{MILANO2}}
\titlefoot{IPNP of MFF, Charles Univ., Areal MFF,
     V Holesovickach 2, CZ-180 00, Praha 8, Czech Republic
    \label{NC}}
\titlefoot{NIKHEF, Postbus 41882, NL-1009 DB
     Amsterdam, The Netherlands
    \label{NIKHEF}}
\titlefoot{National Technical University, Physics Department,
     Zografou Campus, GR-15773 Athens, Greece
    \label{NTU-ATHENS}}
\titlefoot{Physics Department, University of Oslo, Blindern,
     NO-0316 Oslo, Norway
    \label{OSLO}}
\titlefoot{Dpto. Fisica, Univ. Oviedo, Avda. Calvo Sotelo
     s/n, ES-33007 Oviedo, Spain
    \label{OVIEDO}}
\titlefoot{Department of Physics, University of Oxford,
     Keble Road, Oxford OX1 3RH, UK
    \label{OXFORD}}
\titlefoot{Dipartimento di Fisica, Universit\`a di Padova and
     INFN, Via Marzolo 8, IT-35131 Padua, Italy
    \label{PADOVA}}
\titlefoot{Rutherford Appleton Laboratory, Chilton, Didcot
     OX11 OQX, UK
    \label{RAL}}
\titlefoot{Dipartimento di Fisica, Universit\`a di Roma II and
     INFN, Tor Vergata, IT-00173 Rome, Italy
    \label{ROMA2}}
\titlefoot{Dipartimento di Fisica, Universit\`a di Roma III and
     INFN, Via della Vasca Navale 84, IT-00146 Rome, Italy
    \label{ROMA3}}
\titlefoot{DAPNIA/Service de Physique des Particules,
     CEA-Saclay, FR-91191 Gif-sur-Yvette Cedex, France
    \label{SACLAY}}
\titlefoot{Instituto de Fisica de Cantabria (CSIC-UC), Avda.
     los Castros s/n, ES-39006 Santander, Spain
    \label{SANTANDER}}
\titlefoot{Inst. for High Energy Physics, Serpukov
     P.O. Box 35, Protvino, (Moscow Region), Russian Federation
    \label{SERPUKHOV}}
\titlefoot{J. Stefan Institute, Jamova 39, SI-1000 Ljubljana, Slovenia
     and Laboratory for Astroparticle Physics,\\
     \indent~~Nova Gorica Polytechnic, Kostanjeviska 16a, SI-5000 Nova Gorica, Slovenia, \\
     \indent~~and Department of Physics, University of Ljubljana,
     SI-1000 Ljubljana, Slovenia
    \label{SLOVENIJA}}
\titlefoot{Fysikum, Stockholm University,
     Box 6730, SE-113 85 Stockholm, Sweden
    \label{STOCKHOLM}}
\titlefoot{Dipartimento di Fisica Sperimentale, Universit\`a di
     Torino and INFN, Via P. Giuria 1, IT-10125 Turin, Italy
    \label{TORINO}}
\titlefoot{INFN,Sezione di Torino, and Dipartimento di Fisica Teorica,
     Universit\`a di Torino, Via P. Giuria 1,\\
     \indent~~IT-10125 Turin, Italy
    \label{TORINOTH}}
\titlefoot{Dipartimento di Fisica, Universit\`a di Trieste and
     INFN, Via A. Valerio 2, IT-34127 Trieste, Italy \\
     \indent~~and Istituto di Fisica, Universit\`a di Udine,
     IT-33100 Udine, Italy
    \label{TU}}
\titlefoot{Univ. Federal do Rio de Janeiro, C.P. 68528
     Cidade Univ., Ilha do Fund\~ao
     BR-21945-970 Rio de Janeiro, Brazil
    \label{UFRJ}}
\titlefoot{Department of Radiation Sciences, University of
     Uppsala, P.O. Box 535, SE-751 21 Uppsala, Sweden
    \label{UPPSALA}}
\titlefoot{IFIC, Valencia-CSIC, and D.F.A.M.N., U. de Valencia,
     Avda. Dr. Moliner 50, ES-46100 Burjassot (Valencia), Spain
    \label{VALENCIA}}
\titlefoot{Institut f\"ur Hochenergiephysik, \"Osterr. Akad.
     d. Wissensch., Nikolsdorfergasse 18, AT-1050 Vienna, Austria
    \label{VIENNA}}
\titlefoot{Inst. Nuclear Studies and University of Warsaw, Ul.
     Hoza 69, PL-00681 Warsaw, Poland
    \label{WARSZAWA}}
\titlefoot{Fachbereich Physik, University of Wuppertal, Postfach
     100 127, DE-42097 Wuppertal, Germany
    \label{WUPPERTAL}}
\addtolength{\textheight}{-10mm}
\addtolength{\footskip}{5mm}
\clearpage
\headsep 30.0pt
\end{titlepage}
%
\pagenumbering{arabic} 
\setcounter{footnote}{0} %
\large
\section{Introduction}
Measurements of the strong coupling, \as, of quantum chromodynamics 
\cite{QCD} (QCD), the theory of strong  interaction, using different
observables and different analysis methods serve as an important 
consistency test of QCD. 
Once $\alpha_s$ is measured at a given scale, QCD predicts its energy 
dependence as described by the renormalisation group 
equation. A measurement of the strong  coupling at different scales allows 
therefore a test of this important prediction, which is related to the 
property of asymptotic freedom \cite{asfreedom}.

At LEP, hadronic final states of the $e^+e^-$ annihilation are used to
study QCD. While the process $e^+e^- \rightarrow q\bar{q}$ is described by 
the electroweak theory alone, the radiation of
gluons carries sensitivity to properties of the strong interaction.
Our analysis uses event shape observables to measure the strong
coupling. These dimensionless quantities characterize the topology of
the events, e.g. whether the radiation of hard gluons gave rise to
further jets.

The strong coupling is measured by comparing experimental cross-sections, 
$\frac{1}{\sigma}\frac{d\sigma}{dy}$, for an observable $y$
with the  theoretical predictions in which $\alpha_s$ enters as a  
free parameter. But since QCD is the theory of (asymptotically) free quarks 
and gluons, hadronisation effects need to be accounted for. This may be done 
either with phenomenological models \cite{CompPhysComm39-347,ariadne,herwig}, 
or with the help of  QCD-inspired power corrections \cite{PhysLettB352_451}.

The QCD calculation  can be performed in different ways as well. The earliest 
results were based on fixed-order perturbation 
theory \cite{NuclPhysB178_412}.  For the observables studied here these 
predictions are limited to the three-jet region. An extension to the 
four-jet region would need  next-to-next-to-leading-order
(NNLO) corrections to be calculated.
On the other hand the applicability of these calculations close to the 
two-jet region is limited as well, since in this kinematic domain 
enhanced logarithms occur \cite{nlla_old}.
To extend the applicability into the two-jet region  the summation of these 
logarithms was  developed, the so called next-to-leading-log 
approximation (NLLA) \cite{nlla_old,nlla}. 
Finally the fixed-order results can be combined with  
NLLA calculations leading to the \oas+NLLA matched 
theory for the cross-section, 
$R=\int~\frac{1}{\sigma}\frac{d\sigma}{dy}dy$, \cite{nlla_old,nlla}. 
According to 
different ``matching schemes'', these calculations are referred to as e.g. 
$R$ or $\log R$ matching \cite{nlla_old}. 

As a consequence of renormalisation, all perturbative QCD 
calculations to finite order depend upon the renormalisation scale $\mu$, 
which is an unphysical parameter. The choice of $\mu$ is conventional and
the effect of its variation is usually used to estimate the theoretical 
uncertainty. In the NLLA and matched theory even more arbitrary parameters 
enter, related to the phase-space boundary. We will discuss this point and 
our definition of the 
theoretical uncertainties in section \ref{errdef}. 

This paper presents the measurements of event shape distributions in $e^+e^-$
collisions between 
183 and 207\gev.  The data have been reprocessed in 2001 and our final results
supersede some earlier DELPHI measurements at the corresponding energies
\cite{daniel_final_draft}. Another change with respect to the previous LEP2 
analysis is the use of improved event generators for both acceptance 
correction and background subtraction (see section \ref{sec_select}).

From the event shapes Thrust, C parameter, heavy jet mass, wide and total jet
broadening, \as\ is extracted with four different methods: the differential 
distributions are compared to predictions in \oas, pure NLLA and \oas+NLLA 
(logR), folded with fragmentation models, and in the fourth method the strong 
coupling is 
extracted from the mean values using an analytical power correction ansatz. 
An extension of this analysis with respect to the previous one 
\cite{daniel_final_draft} is the use of five observables instead of thrust 
and heavy jet mass only. Additionally the matching procedure for the 
\oas+NLLA (logR) prediction has been modified.
For consistency the distributions at $M_Z$ from 
\cite{daniel_final_draft} have been refitted for these five observables 
(and with the same fit ranges). 
The combination of five $\alpha_s$ values for each 
method and energy makes the treatment of correlations more crucial. Section 
\ref{combi} is devoted to this topic.

From here the analysis 
proceeds in two steps: first the $\alpha_s$ values  
(together with the results from previous measurements at 
other LEP2 energies and LEP1 data)  are used to  test the QCD predicted 
scale dependence, i.e. to measure the $\beta$ function of the strong
interaction. 
Second, assuming the QCD $\beta$ function, all \as\ values at LEP1 and LEP2 
are evolved to a reference energy and combined to a single \as\ value for each 
method. It turns out that the weight of the LEP2 data in the combined 
$\alpha_s$ results is comparable to the weight of the LEP1 data alone. This
unexpected result is due to the fact that at LEP2 the bigger 
statistical uncertainties are compensated for by smaller hadronisation and 
scale uncertainties.  

The paper is organized as follows: 
in section \ref{sec_select}
the selection of hadronic events,
the determination of the centre-of-mass energy,
the correction procedures applied to the data, and
the suppression of
WW and ZZ events are briefly discussed.
Section \ref{results}
presents event shapes and the comparison of the data with
predictions from different generators.
The measurements of \as\ from differential distributions are discussed in 
section \ref{alphas}, while section \ref{asfrommeans} describes the \as\ 
determination from mean values with power corrections. In section
\ref{roas} the running of the strong coupling is discussed and 
section \ref{finalas} contains the combination of all \as\ measurements.
Section \ref{dasletzte} gives a summary of the results.

\section{Selection and correction of hadronic data\label{sec_select}}
The analysis is based on data taken with the DELPHI detector
in the years from 1997 to 2000 at centre-of-mass energies between 183 and 
207\gev. Detailed information about the design and performance of
DELPHI can be found in~\cite{NuclInstrMethA303_233,NuclInstrMethA378_57}.

\newcommand{\bmincut}{\leq 0.08}
\newcommand{\sprimecut}{\geq 90 \% \sqrt{s}}
\begin{table}[bt]
\begin{center}
\renewcommand{\arraystretch}{1.2}
\begin{tabular}{|l|l|l|}\hline
 Track       & $0.2\gev/\mbox{c}\leq p \leq 100\gev/\mbox{c}$ \\
 selection   & $ \Delta p / p \leq 1.0$     \\
             & measured track length $\geq 30$ cm \\
             & distance to I.P in $r\phi$ plane $\leq 4$ cm \\
             & distance to I.P. in $z \leq 10$ cm  \\

\hline
\hline

 Event       & $ N_{\mathrm charged}\ge 7 $ \\
 selection   & $ 25^{\circ} \le \theta_{\mathrm Thrust} \le 155^{\circ} $\\
             & $ E_{\mathrm tot} \geq  0.50 \sqrt{s}$  \\

              & $ \sqrt{s'} \sprimecut $ \\

        & $N_{\mathrm{charged}} > 500 B_{\mathrm{min}}+1.5 $\\
                    & $N_{\mathrm charged} \le 42$\\
\hline
\end{tabular}
\caption[gaga]{\label{cuts}%
Selection of tracks and events.
$p$ is the momentum, $\Delta p$ its error,
$r$  the
radial distance to the beam-axis, $z$  the distance to the beam interaction
point (I.P.) along the beam-axis, $\phi$  the azimuthal angle,
$N_{\mathrm  charged}$  the number of charged particles,
$\theta_{\mathrm Thrust}$ the polar angle of the thrust axis with
respect to the beam,
$E_{\mathrm tot}$ the total energy carried by charged and neutral
particles, $\sqrt{s'}$ the reconstructed centre-of-mass energy,
$\sqrt{s}$ the nominal centre-of-mass energy, and
\bmin\ is the minimal jet broadening. The first two cuts apply to
charged and neutral particles, while the other track selection cuts apply
only to charged particles.}
\end{center}
\end{table}

In order to select well-measured charged particle tracks,
the cuts given in the upper part of \tab{cuts} have been applied.
The cuts in the lower part of the table are used to
select \qcd\ events and to
suppress background processes such as two-photon interactions, beam-gas and
beam-wall interactions, leptonic final states, events with hard
initial-state radiation (ISR), WW and  ZZ pair production.

\begin{table}[t]
\begin{center}
\begin{small}
\begin{tabular}{|r|c|c|c|c|c|c|c|c|}\hline
$\sqrt{s}$ [\gev]&183&189 &192  & 196  & 200  & 202  & 205  & 207  \\ \hline
\hline
${\cal L}[pb]^{-1}$ &55.73&157.97& 25.34 & 67.29  & 78.07 & 39.31 & 76.33 &
130.12 \\ \hline
$\sigma^{QCD}_{tot}$[pb]&108.78&100.05& 96.06    & 91.31   & 86.73 & 84.56 &
 81.18 & 79.78 \\
\hline
$\sigma^{QCD}_{s'>90\%}$[pb]   &23.09&21.24& 20.42  & 19.36 & 18.35 & 18.18 &
16.89 &
16.59 \\ \hline
$\sigma^{4F, CC}$[pb]   &17.54&18.74& 19.10    & 19.57  & 19.85 & 19.97 &
 20.10 & 20.14 \\
\hline
$\sigma^{4F, NC}$[pb]   &8.16&8.15& 8.14  & 8.08  & 8.03  & 8.01  & 7.93  &
7.90  \\ \hline
\hline
$\epsilon_{HE}$   &0.721&0.720& 0.736 & 0.740 & 0.735 & 0.734 & 0.736 & 0.749
  \\ \hline
$\epsilon_{CC}$     &0.090&0.100& 0.104 & 0.112 & 0.122 & 0.120 & 0.127 &
 0.124 \\ \hline
$\epsilon_{NC}$     &0.017&0.017& 0.017    & 0.017   & 0.015  & 0.016 &0.016 &
 0.016 \\
\hline
$p_{HE,QCD}$      &0.867&0.848& 0.837 & 0.828 & 0.808 & 0.801 & 0.790 & 0.795
  \\ \hline
$\epsilon_{HE}\cdot p_{HE,QCD}$ &0.625&0.610& 0.617 & 0.612 & 0.594 & 0.588 &
0.581 &
0.593 \\ \hline\hline
\# selected events     &1070&2848&  455  & 1164   & 1303  & 653  & 1203  & 2036  \\
\hline
\# CC background &87.8&296.0& 50.1  & 147.7  & 189.8 & 94.1 & 195.6 & 315.4 \\
 \hline
\# NC background &7.7&21.4& 3.54  & 9.36   & 9.65  & 4.88 & 9.67  & 16.60 \\
 \hline
\end{tabular}
\end{small}
 \end{center}
\caption{\label{lumies} { Luminosities, cross-sections of QCD signal  
and background from four-fermion events (split into  neutral current, NC, 
and charged current, CC), selection efficiencies, $\epsilon$, and  purities, 
$p$. The subscript $HE$ denotes QCD high energy events, i.e. with 
$\sqrt{s^{\prime}}>0.9\cdot \sqrt{s}$. Also given is the total number of 
selected events and the expected number of remaining four-fermion events.}}
\end{table}

At energies well above $M_Z$ the high cross-section of the Z
resonance  raises the probability of events with hard ISR. These 
``radiative return events'' 
constitute a large fraction of all hadronic events. The initial-state photons 
are typically  aligned along the beam direction and are 
identified inside the detector only at a rate of about 10\% .
In order to evaluate the effective hadronic centre-of-mass energy of an event,
considering ISR, an algorithm called \sprime\ is used~\cite{CERN-OPEN98-026}.
\sprime\ is based on a 3C fit imposing transverse momentum and energy 
conservation. Several assumptions
about the event topology are tested. The decision is taken according to the
$\chi^2$ obtained from the constrained fits with different topologies.

\begin{figure}[tb]
\unitlength1cm
 \unitlength1cm
 \begin{center}
 \begin{minipage}[t]{7.5cm}
               \mbox{\epsfig{file=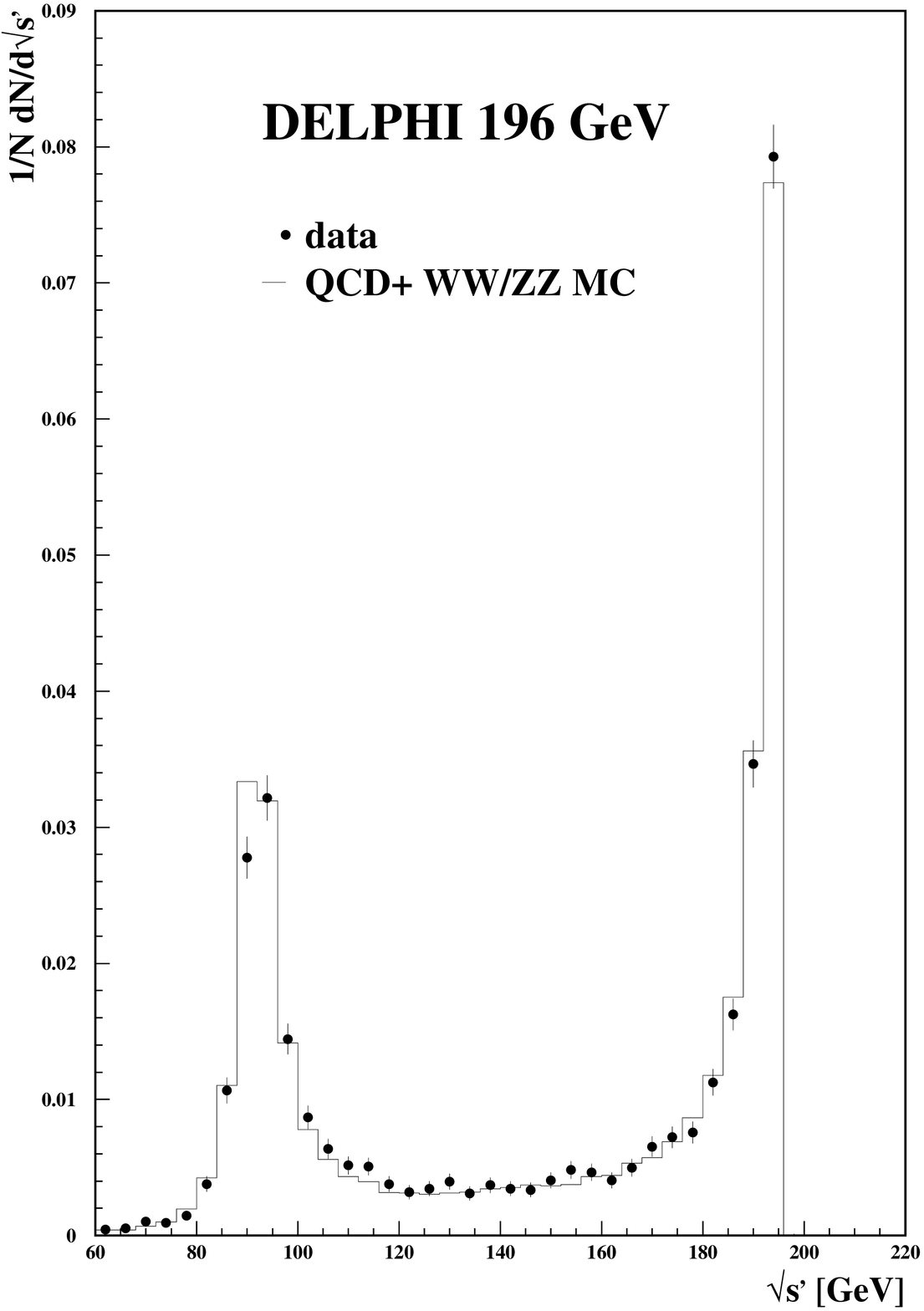,width=7.cm}}
 \end{minipage}
 \begin{minipage}[t]{7.5cm}
           \mbox{\epsfig{file=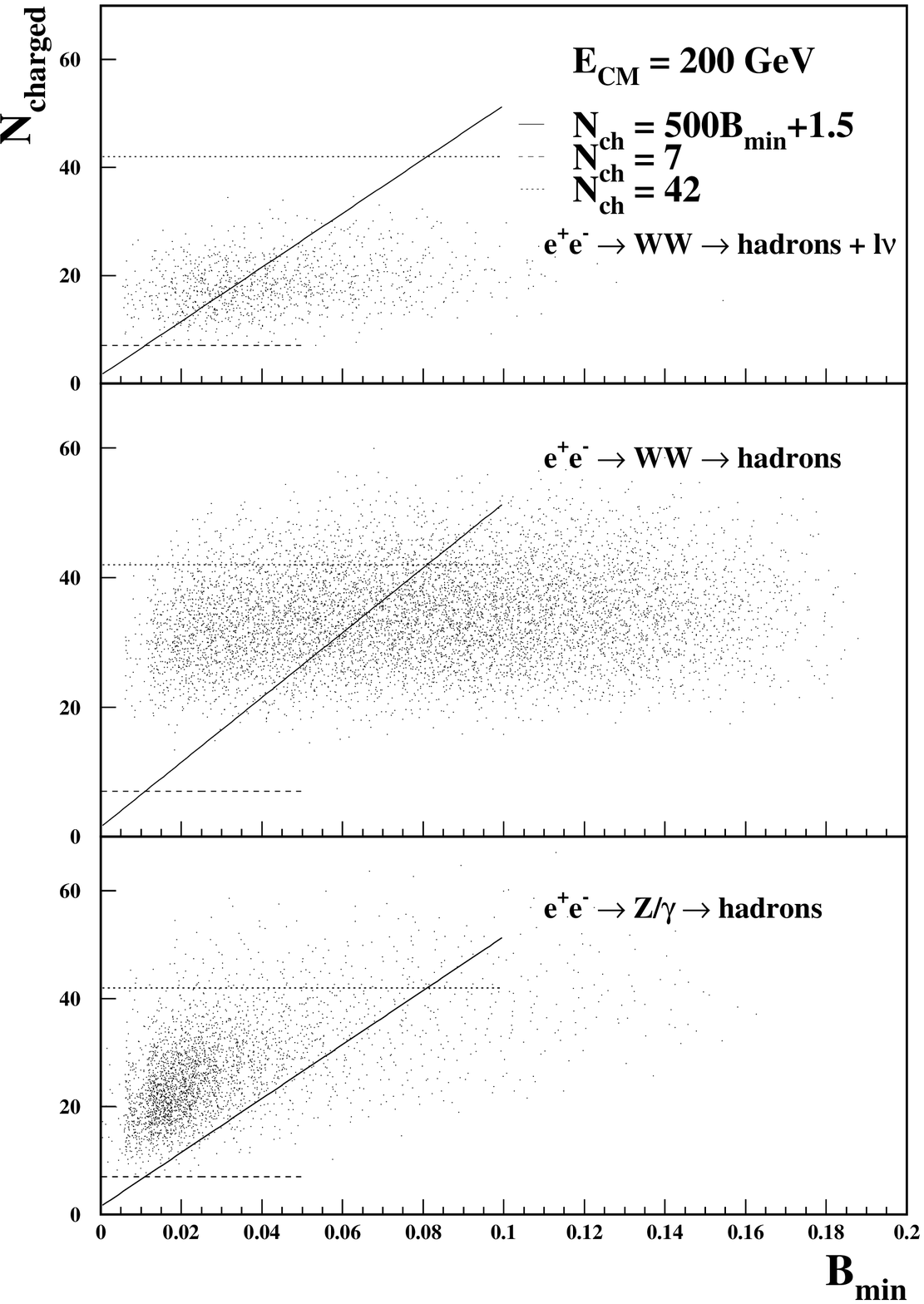,width=7.0cm}}
 \end{minipage}
\end{center}
\caption{\label{plot_isr}
Left: reconstructed centre-of-mass energy $\sqrt{s^{\prime}}$ .
\label{plot_ww} Right: simulation of four-fermion background and QCD events
in the $N_{\mathrm{charged}}$-$B_{\mathrm{min}}$ plane. The lines 
delineate the accepted region.}
\end{figure}

\fig{plot_isr}(left) shows the spectra of the calculated energies for
simulated and measured  events after all but the $\sqrt{s^{\prime}}$ cut.
A cut on the reconstructed centre-of-mass energy
$\sqrt{s'}\sprimecut$
is applied to discard radiative return events.

Two-photon events are strongly suppressed by the cuts. Leptonic
background was found to be negligible in this analysis as well.

Since the topological signatures of QCD four-jet events and hadronic
WW, ZZ and other events with four-fermions (4F) in the final state
are similar, no highly efficient separation of QCD events and backgrounds
is possible. Furthermore any 4F rejection implies a severe bias to the QCD
event shape distributions, which needs to be corrected by simulation.

Our suppression of these backgrounds uses a two-dimensional cut in the plane 
spanned by the charged particle
multiplicity ($N_{\mathrm{charged}}$) and the narrow jet broadening 
$B_{\mathrm{min}}  = \min(B_+,B_-)$. $B_{\pm}$ is defined as the normalized 
sum over  the transverse momentum of charged and neutral particles in the 
two event hemispheres separated 
by the plane perpendicular to the thrust axis $n_T$:  

\begin{eqnarray}
\label{bdef}
B_{\pm} = \left (\sum_{\pm \vec{p_i}\cdot\vec{n_T}>0} 
|\vec{p_i}\times\vec{n_T}|\right) \left /
\left( 2\sum_i|\vec{p_i}| \right)~~~.\right.
\end{eqnarray}

\noindent
By applying a cut on an observable calculated from the narrow event
hemisphere only, the bias to event shape observables mainly sensitive
to the wide event hemisphere is reduced. The charged particle multiplicity
is used to reduce the 4F contribution further.
The two-dimensional cut in the 
$N_{\mathrm{charged}}$-$B_{\mathrm{min}}$ plane exploits the different 
correlation between these observables for QCD and four-fermion
events, as shown 
in Figure \ref{plot_isr} (right). Especially some reduction for
semi-leptonic decaying 4F events is gained. The lines indicate 
the cut values chosen. This cut suppress almost 90\% of the four-fermion 
background.
The remaining 4F contribution is estimated by the WPHACT \cite{wphact} 
generator and subtracted from the measurement.

Table \ref{lumies} contains the integrated luminosities at different 
energies, the cross-sections for signal and background and summarizes the  
selection statistics. The cross-sections were taken from 
the  simulation which was used to correct the data and to subtract the 
background. The cross-sections for the 4F background are quoted for charged 
current (CC) and neutral current (NC) contributions separately. Details on the 
four-fermion simulation in DELPHI can be found in \cite{4fatdelphi}.

The influence of detector effects  was studied by passing
events (generated with ${\cal{KK}}$ \cite{KK}) and fragmented with
\jetset/\pythia~\cite{CompPhysComm39-347} using the
DELPHI tuning described in~\cite{ZPhysC73_11} through a full
detector simulation (\delsim~\cite{NuclInstrMethA303_233}). 
This simulation is improved with respect to the previous LEP2 analysis 
\cite{daniel_final_draft} by including electroweak corrections 
(multiple photon emission, treatment of ISR and FSR etc.).
These simulated events are processed with the same reconstruction program 
and selection cuts as are the real data.
In order to correct for cuts, detector, and ISR effects a
bin-by-bin acceptance correction $C$, obtained from
\qcd\ simulation, is applied to the data:
\begin{equation}
 C_i  = \frac{h(f_i)_{\mathrm gen, no ISR}}
                             {h(f_i)_{\mathrm acc}}
\label{acc}
\end{equation}
where $h(f_i)_{\mathrm gen, no ISR}$ represents
 bin $i$ of the shape distribution $f$
generated with the tuned generator. The subscript $\mathrm no ISR$
indicates that only events without relevant ISR ($\sqrt{s}-\sqrt{s'}<0.1\gev$)
enter the distribution.
$h(f_i)_{\mathrm acc}$ represents the accepted distribution $f$ as 
obtained with the full detector simulation. The more detailed matrix 
correction used for the data measured at the Z peak \cite{siggi} is not  
applied here, because of the smaller statistics at LEP2.

\begin{figure}[hp]
 \begin{center}
  \vspace{-2.5cm}
  \unitlength1cm
  \begin{minipage}[t]{7.0cm}
   \mbox{\epsfig{file=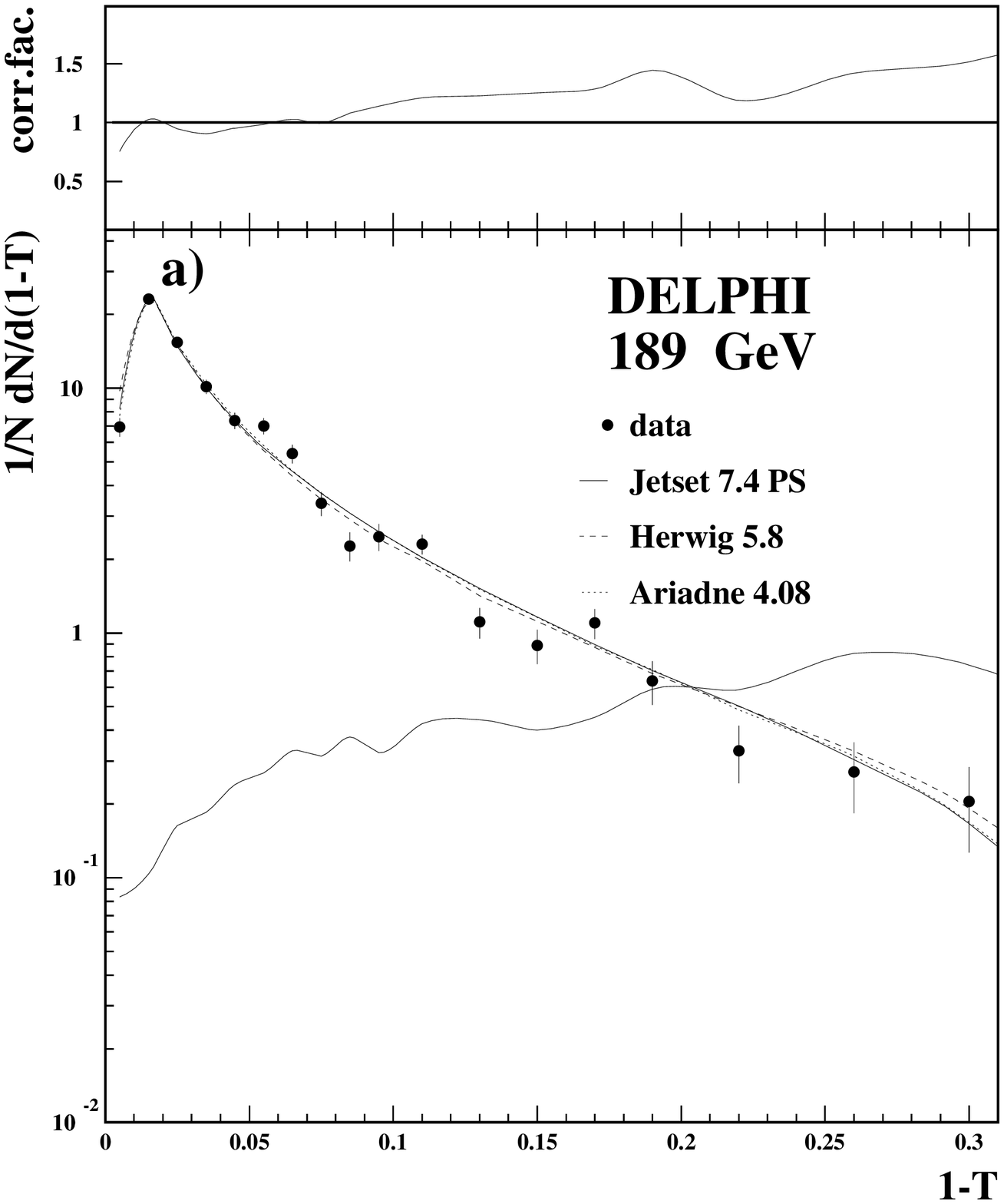,width=7.8cm}}
  \end{minipage}
  \begin{minipage}[t]{7.0cm}
   \mbox{\epsfig{file=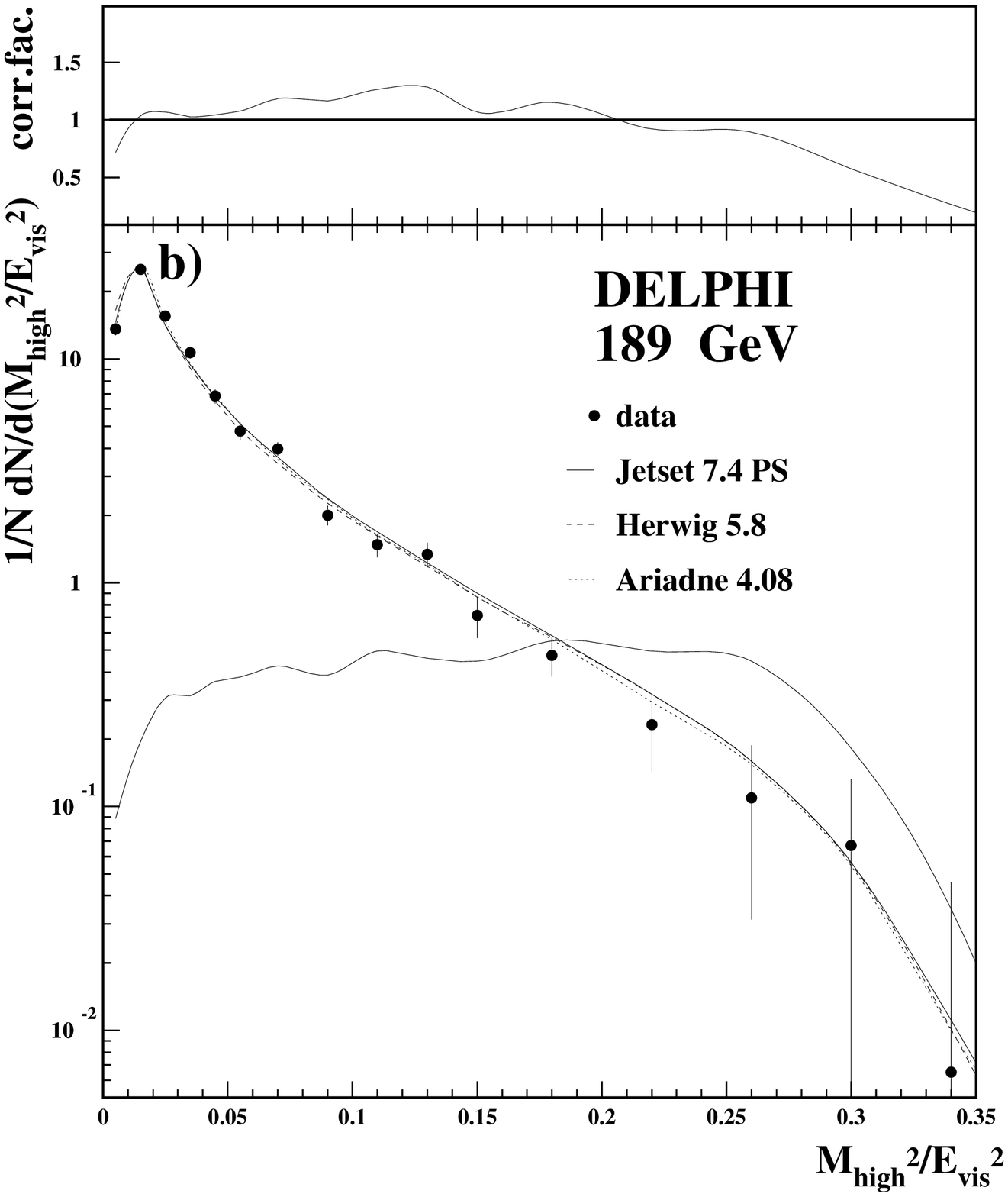,width=7.8cm}}
  \end{minipage}
 \end{center}

\vspace*{-2.5cm}
 \begin{center}
  \unitlength1cm
  \begin{minipage}[t]{7.0cm}
   \mbox{\epsfig{file=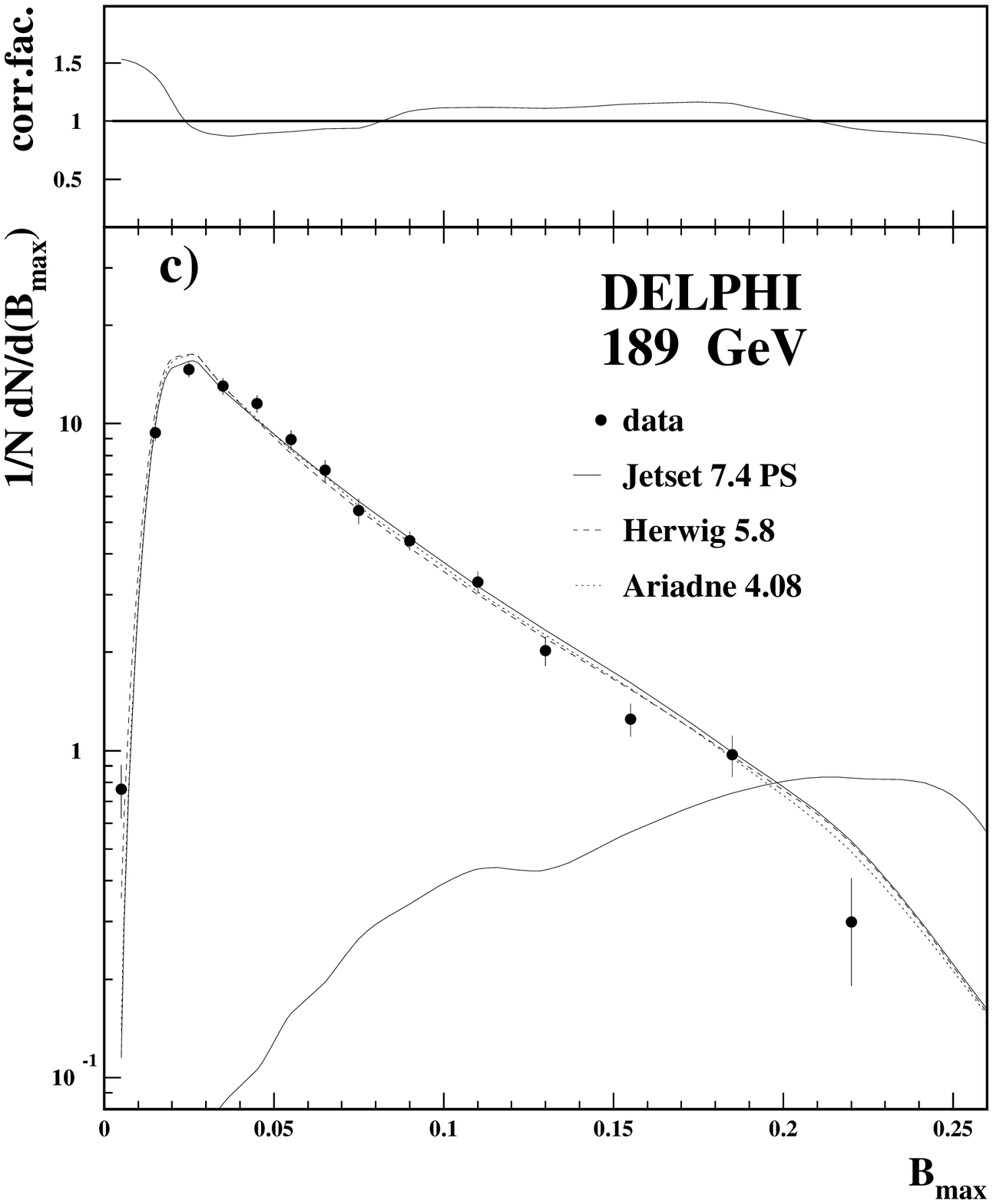,width=7.8cm}}
  \end{minipage}
  \begin{minipage}[t]{7.0cm}
   \mbox{\epsfig{file=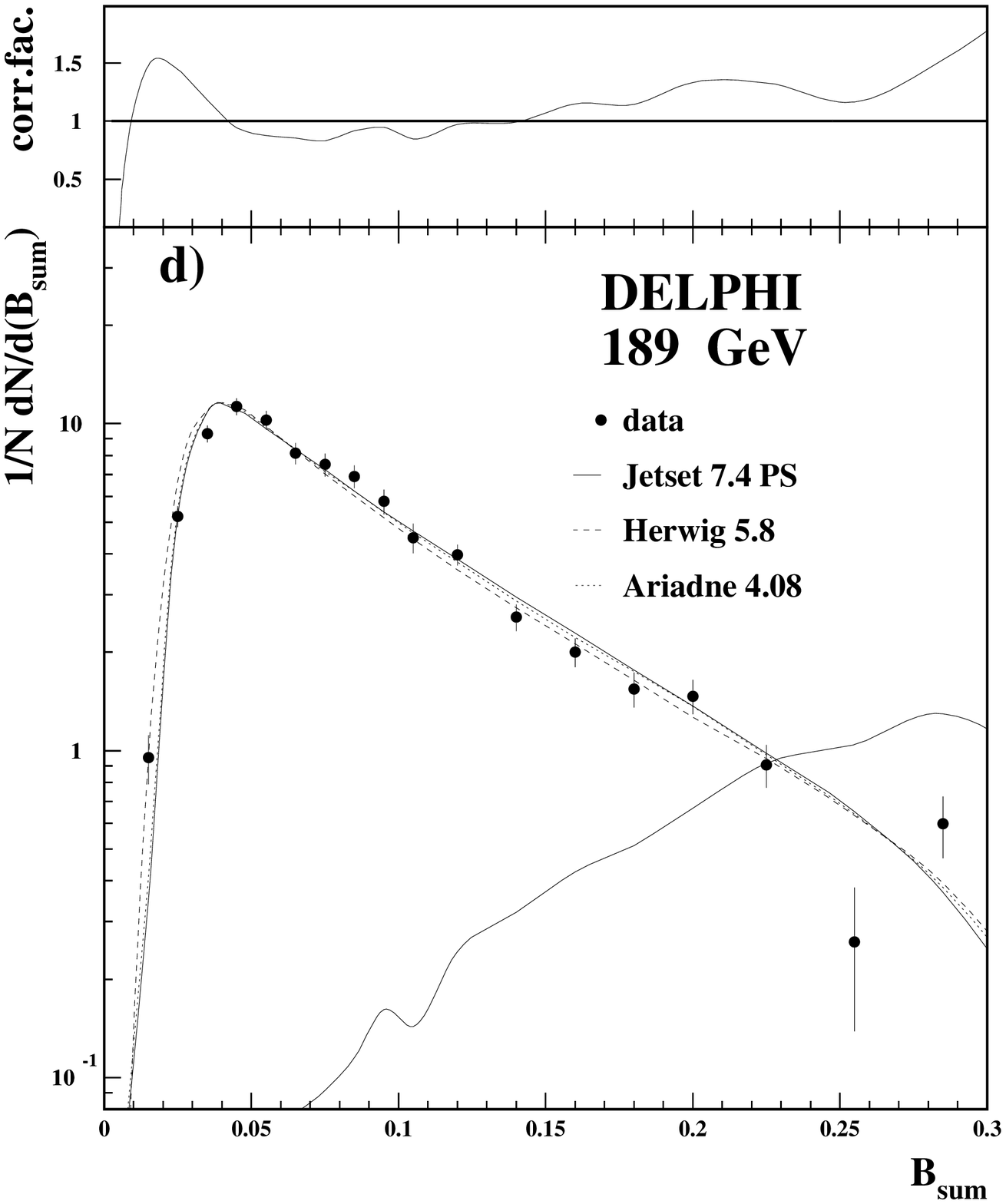,width=7.8cm}}
  \end{minipage}
\vspace*{-1.5cm}
 \caption{\label{shapes183}
Event shape distributions of 1-Thrust ($1-T$), heavy jet mass
(${M_{\mathrm{h}}^2}/{E_{\mathrm{vis}}^2}$),
wide jet broadening ($B_{\mathrm{max}}$) and
total jet broadening ($B_{\mathrm{sum}}$) at 189\gev.
The upper inset shows the acceptance corrections. 
The central part shows data with statistical uncertainties, simulation and 
the four-fermion background  which was subtracted from the data. 
}
\end{center}
\end{figure}


\begin{figure}[hp]
 \begin{center}
  \vspace{-2.5cm}
  \unitlength1cm
  \begin{minipage}[t]{7.0cm}
   \mbox{\epsfig{file=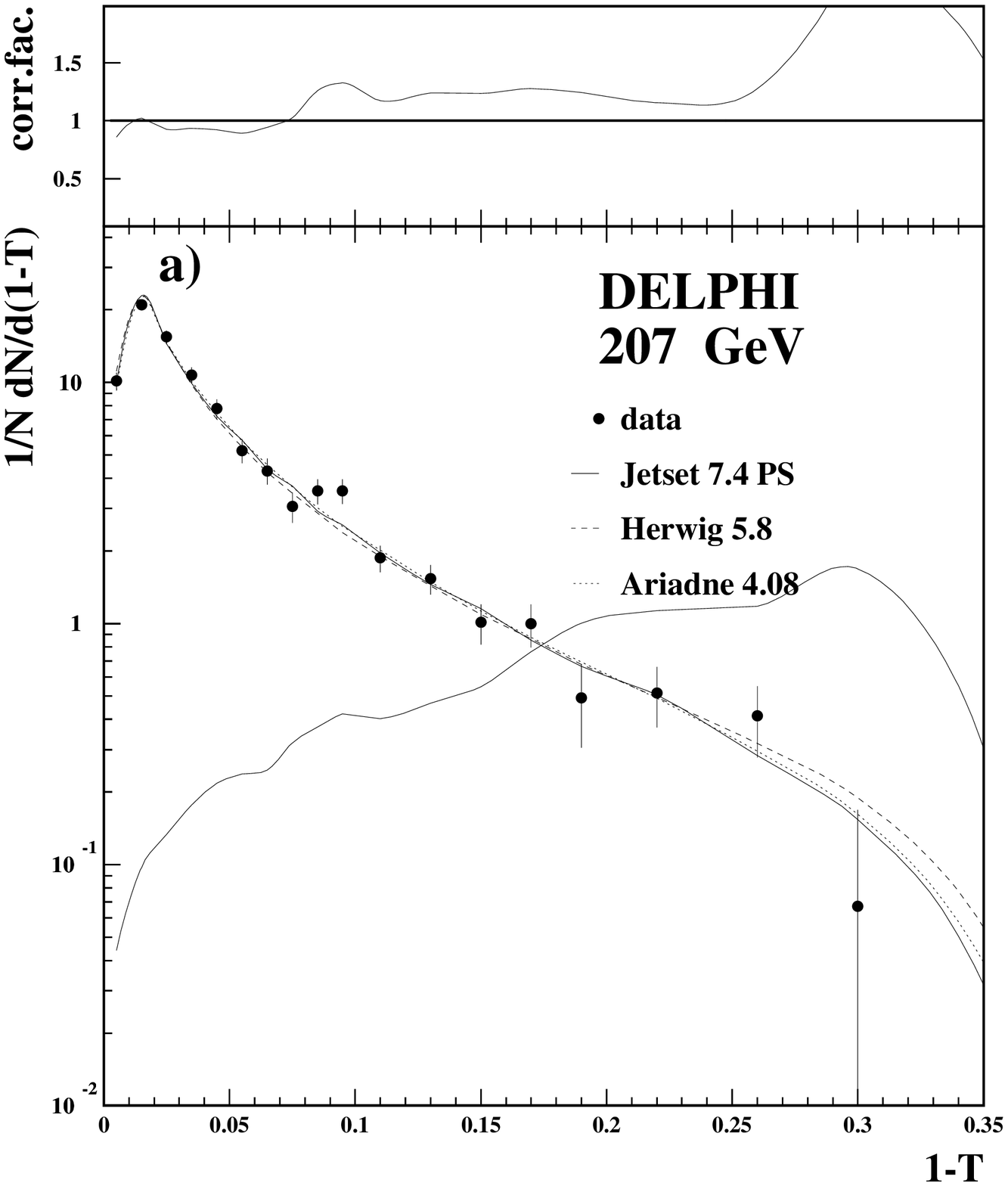,width=7.8cm}}
  \end{minipage}
  \begin{minipage}[t]{7.0cm}
   \mbox{\epsfig{file=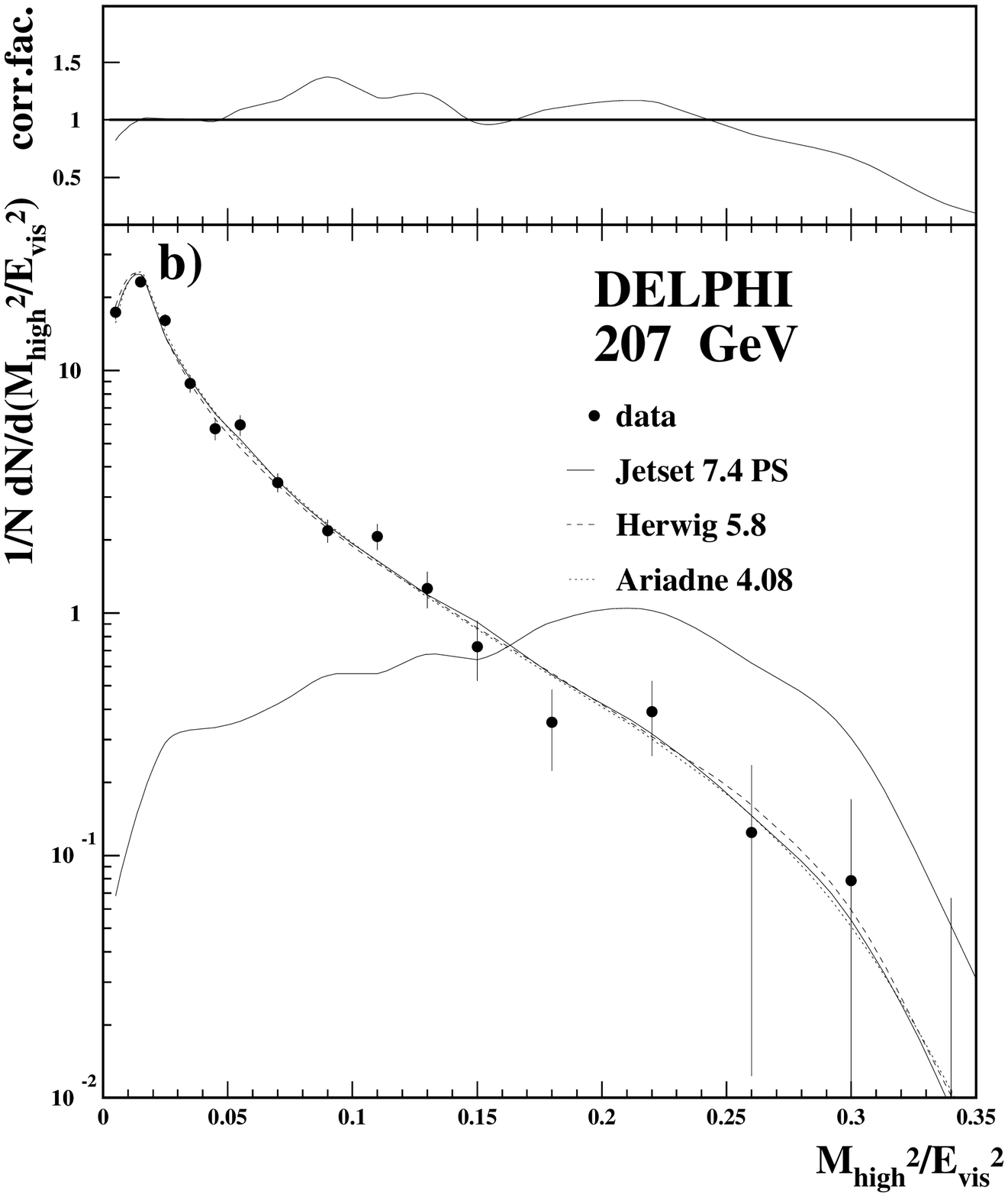,width=7.8cm}}
  \end{minipage}
 \end{center}

\vspace*{-2.5cm}
 \begin{center}
  \unitlength1cm
  \begin{minipage}[t]{7.0cm}
   \mbox{\epsfig{file=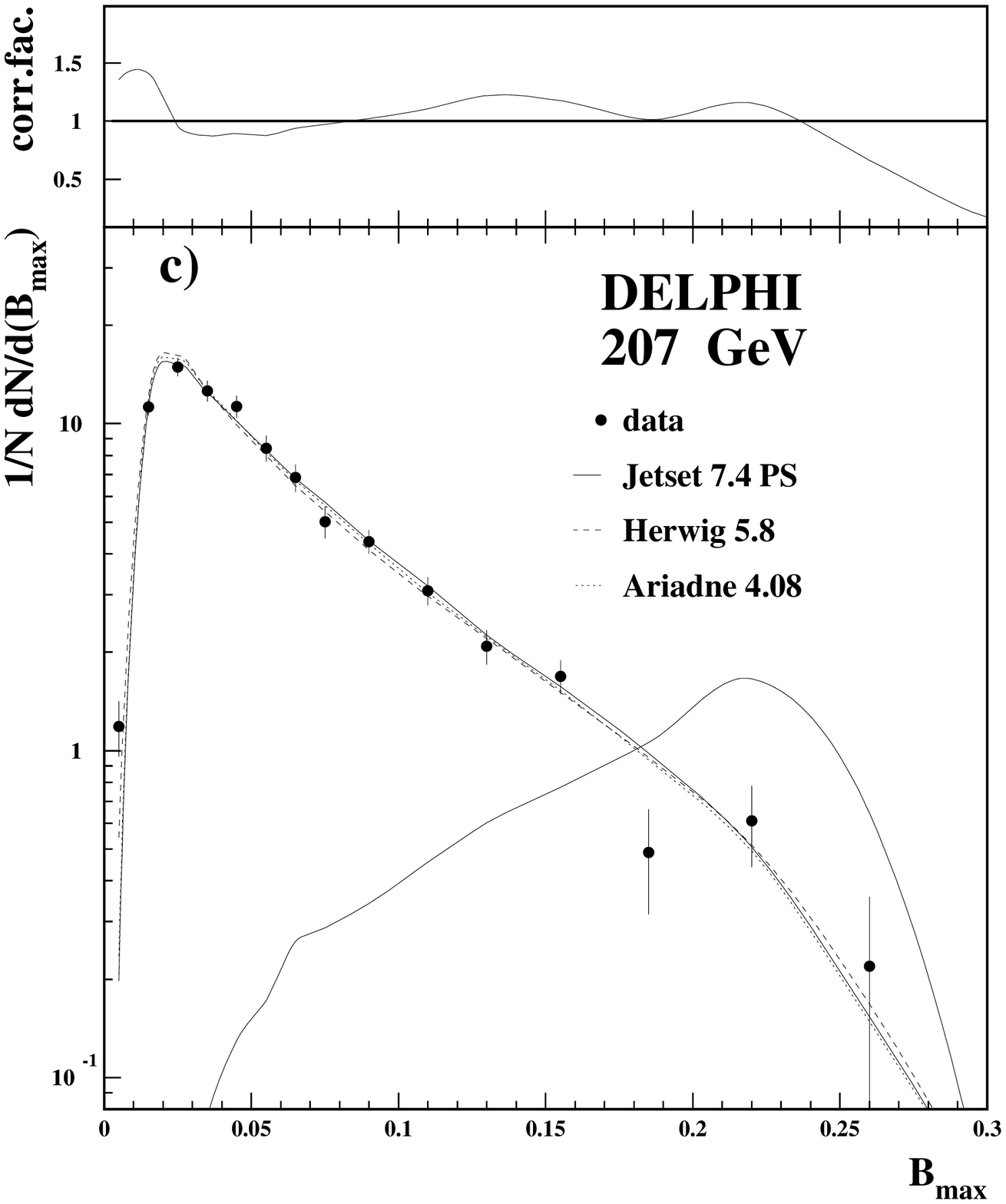,width=7.8cm}}
  \end{minipage}
  \begin{minipage}[t]{7.0cm}
   \mbox{\epsfig{file=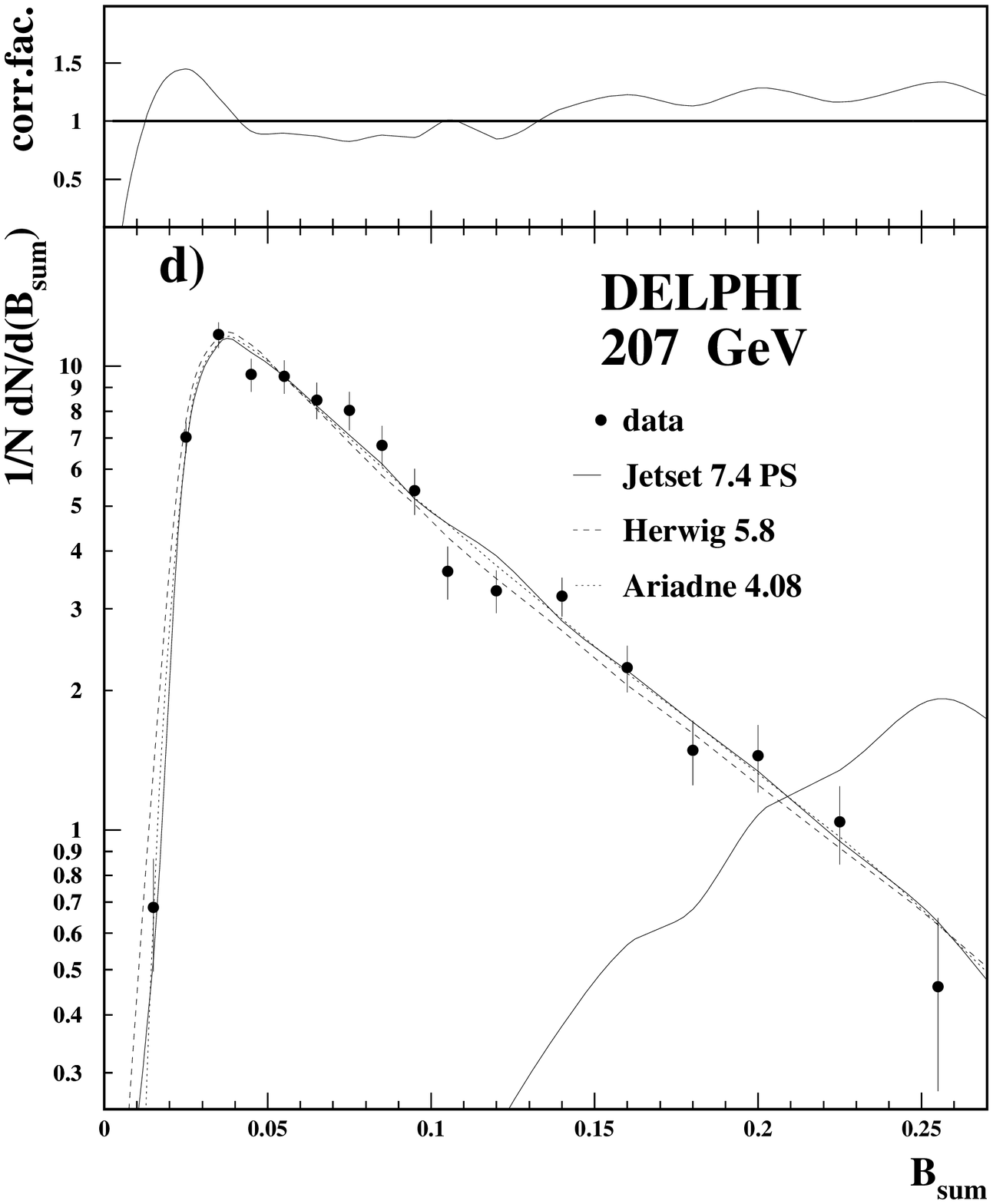,width=7.8cm}}
  \end{minipage}
\vspace*{-1.5cm}
 \caption{\label{shapes207}
Event shape distributions of 1-Thrust ($1-T$), heavy jet mass
(${M_{\mathrm{h}}^2}/{E_{\mathrm{vis}}^2}$),
wide jet broadening ($B_{\mathrm{max}}$) and
total jet broadening ($B_{\mathrm{sum}}$) at 207\gev.
The upper inset shows the acceptance corrections. 
The central part shows data with statistical uncertainties, simulation and the
four-fermion background  
which was subtracted from the data. 
}
\end{center}
\end{figure}

\section{\label{results}Event shape distributions and mean values}
Selected event shape distributions at 189 and 207\gev\ are shown in
Figures \ref{shapes183} and \ref{shapes207}. 
The definitions of these observables are given is Section \ref{alphas}.

The data in Figures \ref{shapes183} and \ref{shapes207} are
corrected to be comparable with \qcd\ simulation of charged
and neutral hadron production. In the data all charged particles are assumed to
have pion mass while neutral particles are considered massless. The monte
carlo correction of the data includes the effects of the simulated particle
masses. The Figures \ref{shapes183} and \ref{shapes207}
compare these data with the \jetset\ 
\cite{CompPhysComm39-347}, \ariadne\ \cite{ariadne} and \herwig\ \cite{herwig}
generators as tuned by DELPHI~\cite{ZPhysC73_11} with LEP1 data. The amount of
4F-background which was subtracted to obtain the 
final data points is also shown. The acceptance corrections 
are plotted in the upper inset. 


The Tables \ref{mean1} and \ref{mean2} at the end of the paper 
contain mean values and higher moments 
for the  event shapes 1-T, C parameter, $M_{\mathrm{h}}^2/E_{\mathrm{vis}}^2$, 
$ B_{\mathrm{max}}$ and $B_{\mathrm{sum}}$. Also included are the results 
for alternative definitions of the heavy jet mass as proposed in 
\cite{gavinunddaniel}. They are obtained if in the definition of the heavy 
jet mass the invariant mass is calculated with the following replacements:
\begin{eqnarray*}
(E_i,\vec{p_i}) &\rightarrow& (|\vec{p_i}|,\vec{p_i}) \\
\mbox{or }\,\,(E_i,\vec{p_i}) &\rightarrow& (E_i,E_i\cdot\vec{p_i}/
|\vec{p_i}|).
\end{eqnarray*}
In what follows we will refer to these observables as p-scheme and E-scheme
definitions of the heavy jet mass. 

In order to estimate the systematic uncertainty from the selection and
correction procedure, the effects of the following  changes with respect to
the standard values have been considered: $N_{\mathrm{ch}}\pm 1$,
$\Theta_{\mathrm thrust}\pm 5^{\circ} $ and $\sqrt{s'}/\sqrt{s}\pm 0.025$.  
For the  4F cross-section a change of $\pm$5\% has been considered and the
uncertainty due to the acceptance correction was estimated by a change of
$\pm$0.02. The last uncertainty agrees with that of~\cite{ZPhysC73_11} where
the systematic uncertainty was verified using independent data.
Half of the difference between up- and downward variation is regarded as one
component of the systematic uncertainty. These five contributions 
are added in quadrature to estimate the experimental systematic uncertainty.
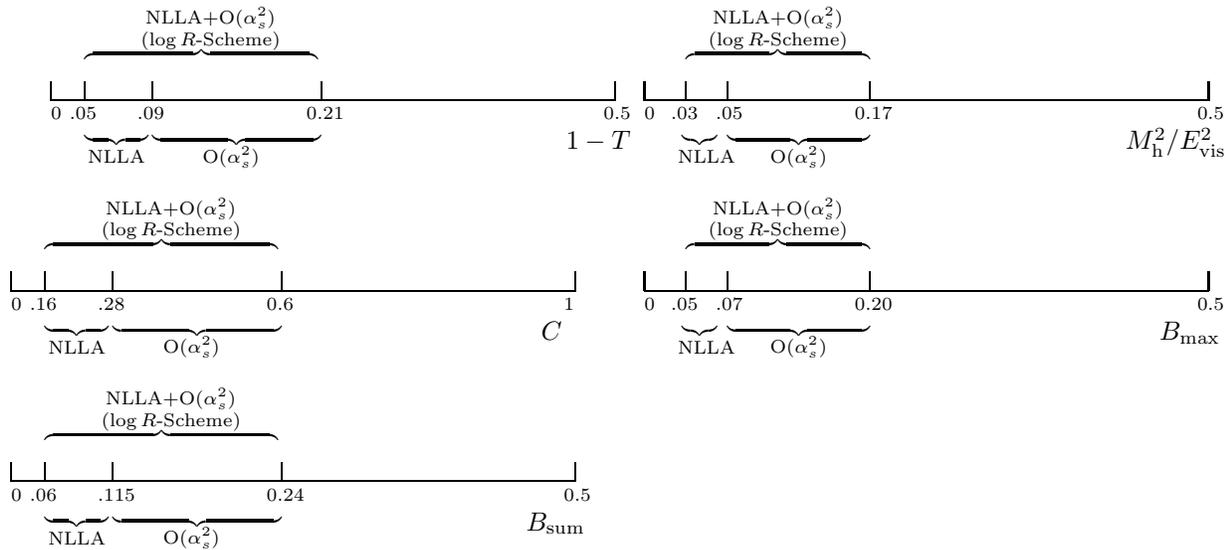
\begin{figure}[bth]
\scriptsize
\unitlength 0.5mm
\begin{picture}(150,50)
\put(  0,20){\line(1,0){150}}
\put(  0,20){\line(0,1){5}}
\put(  9,20){\line(0,1){5}}
\put( 27,20){\line(0,1){5}}
\put( 72,20){\line(0,1){5}}
\put(150,20){\line(0,1){5}}
\put(0,15){0}
\put(5,15){.05}
\put(23,15){.09}
\put( 68,15){0.21}
\put(147,15){0.5}
\put(137,7){\normalsize$1-T$}
\put(9,30){$\overbrace{\makebox(62,0){}}$}
\put(9,12){$\underbrace{\makebox(17,0){}}$}
\put(27,12){$\underbrace{\makebox(45,0){}}$}
\put(16,42){\makebox(53,0){NLLA+$\mbox{O}(\alpha_s^2)$}}
\put(16,36){\makebox(53,0){($\log R$-Scheme)}}
\put(9,5){\makebox(17,0){NLLA}}
\put(27,5){\makebox(42,0){$\mbox{O}(\alpha_s^2)$}}
\end{picture}\hfill
\begin{picture}(150,50)
\put(  0,20){\line(1,0){150}}
\put(  0,20){\line(0,1){5}}
\put( 11,20){\line(0,1){5}}
\put( 22,20){\line(0,1){5}}
\put( 60,20){\line(0,1){5}}
\put(150,20){\line(0,1){5}}
\put(0,15){0}
\put( 7,15){.03}
\put(19,15){.05}
\put(56,15){0.17}
\put(147,15){0.5}
\put(128,7){\normalsize$M_{\mathrm{h}}^2/E_{\mathrm{vis}}^2$}
\put(11,30){$\overbrace{\makebox(49,0){}}$}
\put(10,12){$\underbrace{\makebox(9,0){}}$}
\put(22,12){$\underbrace{\makebox(38,0){}}$}
\put(11,42){\makebox(49,0){NLLA+$\mbox{O}(\alpha_s^2)$}}
\put(11,36){\makebox(49,0){($\log R$-Scheme)}}
\put(11,5){\makebox(11,0){NLLA}}
\put(22,5){\makebox(38,0){$\mbox{O}(\alpha_s^2)$}}
\end{picture}
\begin{picture}(150,50)
\put(  0,20){\line(1,0){150}}
\put(  0,20){\line(0,1){5}}
\put(  9,20){\line(0,1){5}}
\put( 27,20){\line(0,1){5}}
\put( 72,20){\line(0,1){5}}
\put(150,20){\line(0,1){5}}
\put(0,15){0}
\put(5,15){.16}
\put(23,15){.28}
\put( 68,15){0.6}
\put(147,15){1}

\put(141,7){\normalsize$C$}
\put(9,30){$\overbrace{\makebox(62,0){}}$}
\put(9,12){$\underbrace{\makebox(17,0){}}$}
\put(27,12){$\underbrace{\makebox(45,0){}}$}
\put(16,42){\makebox(53,0){NLLA+$\mbox{O}(\alpha_s^2)$}}
\put(16,36){\makebox(53,0){($\log R$-Scheme)}}
\put(9,5){\makebox(17,0){NLLA}}
\put(27,5){\makebox(42,0){$\mbox{O}(\alpha_s^2)$}}
\end{picture}\hfill
\begin{picture}(150,50)
\put(  0,20){\line(1,0){150}}
\put(  0,20){\line(0,1){5}}
\put( 11,20){\line(0,1){5}}
\put( 22,20){\line(0,1){5}}
\put( 60,20){\line(0,1){5}}
\put(150,20){\line(0,1){5}}
\put(0,15){0}
\put( 7,15){.05}
\put(19,15){.07}
\put(56,15){0.20}
\put(147,15){0.5}
\put(137,7){\normalsize$B_{\mathrm{max}}$}
\put(11,30){$\overbrace{\makebox(49,0){}}$}
\put(10,12){$\underbrace{\makebox(9,0){}}$}
\put(22,12){$\underbrace{\makebox(38,0){}}$}
\put(11,42){\makebox(49,0){NLLA+$\mbox{O}(\alpha_s^2)$}}
\put(11,36){\makebox(49,0){($\log R$-Scheme)}}
\put(11,5){\makebox(11,0){NLLA}}
\put(22,5){\makebox(38,0){$\mbox{O}(\alpha_s^2)$}}
\end{picture}
\begin{picture}(150,50)
\put(  0,20){\line(1,0){150}}
\put(  0,20){\line(0,1){5}}
\put(  9,20){\line(0,1){5}}
\put( 27,20){\line(0,1){5}}
\put( 72,20){\line(0,1){5}}
\put(150,20){\line(0,1){5}}
\put(0,15){0}

\put(5,15){.06}
\put(23,15){.115}
\put( 68,15){0.24}
\put(147,15){0.5}
\put(137,7){\normalsize$B_{\mathrm{sum}}$}
\put(9,30){$\overbrace{\makebox(62,0){}}$}
\put(9,12){$\underbrace{\makebox(17,0){}}$}
\put(27,12){$\underbrace{\makebox(45,0){}}$}
\put(16,42){\makebox(53,0){NLLA+$\mbox{O}(\alpha_s^2)$}}
\put(16,36){\makebox(53,0){($\log R$-Scheme)}}
\put(9,5){\makebox(17,0){NLLA}}
\put(27,5){\makebox(42,0){$\mbox{O}(\alpha_s^2)$}}
\end{picture}\hfill
\caption[Anpassungsbereiche dieser Analyse]{\label{fig_fit_ranges}
Fit ranges for the different observables and methods to determine \as.}
\end{figure}

\newcommand{\tsppm}{\hspace{\tabcolsep}$\pm$\hspace{\tabcolsep}}

\section{Determination of \boldmath\as\  from event shape distributions 
\label{alphas}}
Our determination of \as\ is based on the five variables  1-Thrust 
($1-T$), C-parameter, heavy jet mass
(${M_{\mathrm{h}}^2}/{E_{\mathrm{vis}}^2}$), wide jet broadening 
($B_{\mathrm{max}}$) and total jet broadening ($B_{\mathrm{sum}}$).
The thrust, $T$, is defined as:
\begin{eqnarray*}
   T=\max_{\vec{n}} \left\{   \frac{\sum_i |\vec{p}_i
 \cdot \vec{n}|}{\sum_i |\vec{p}_i|}  \right\}
   =  \frac{\sum_i |\vec{p}_i
 \cdot \vec{n}_{T}|}{\sum_i |\vec{p}_i|} ~~~.
\end{eqnarray*}
The vector which maximizes the above expression defines the thrust axis, 
$\vec{n}_{T}$.
The plane perpendicular to the thrust axis divides the event into two 
hemispheres. Based on this separation several other event shapes can be 
defined. One defines the heavy jet mass by the following 
expression:
\begin{eqnarray*}
M^2_{\mathrm{h}}/E_{\mathrm{vis}}^2 &=& 
\max(M_+^2,M_-^2)/E_{\mathrm{vis}}^2 
\end{eqnarray*}
$M_{\pm}^2$ denotes the invariant mass of the two hemispheres: 
\begin{eqnarray*}
M_{\pm}^2=\left( \sum_{\pm \vec{p_i}\cdot\vec{n_T}>0} p_i \right)^2
~~~.
\end{eqnarray*}
Here $p_i$ is the four-momentum of the $i$th particle.
Using the expression $B_{\pm}$ as defined in Equ.\ref{bdef}
the wide and total jet broadening are defined as:
\begin{eqnarray*}
B_{\mathrm{max}}  &=& \max(B_+,B_-)\\
B_{\mathrm{sum}}  &=& B_+ + B_-
\end{eqnarray*}

The linear momentum tensor offers a possibility to define event shapes without
distinguishing an event axis. It is defined as:
\begin{eqnarray*}
\Theta^{ab}=\sum_{i=1}^{n_{\mathrm{track}}}\frac{p_i^ap_i^b}{|\vec{p_i}|}
\left /
\sum_{i=1}^{n_{\mathrm{track}}}|\vec{p_i}| \right. \qquad 
\mbox{with:}\quad \,a,b=x,y,z
\end{eqnarray*}
From its eigenvalues  $\lambda_i$ the C-parameter is defined:
\begin{eqnarray*}
C &=& 3(\lambda_1\lambda_2+\lambda_1\lambda_3+\lambda_2\lambda_3 ) 
\end{eqnarray*}

From these differential distributions \as\ is determined by fitting an 
\as-dependent QCD prediction folded with a hadronisation correction to the
data. 
The following  QCD predictions are used: \oas, pure NLLA, and the modified 
\oas+NLLA in the $\log R$-scheme
\cite{NuclPhysB178_412,nlla_old,nlla,PhysLettB263_491,PhysLettB272_368}.
Hadronisation corrections are calculated using the \mbox{\jetset\ PS} model
(Version 7.4 as tuned by DELPHI~\cite{ZPhysC73_11}).
In each bin the QCD prediction is multiplied by the hadronisation
correction
\begin{equation}
C_{\mathrm had} =
        \frac{f_{\mathrm had}^{\mathrm Sim.}}
             {f_{\mathrm part}^{\mathrm Sim.}}
   \quad\mbox{,}
\label{eq_hadc}
\end{equation}
where $f_{\mathrm had}^{\mathrm Sim.}$
($f_{\mathrm part}^{\mathrm Sim.}$)
is the model prediction on hadron (parton) level.
The parton level is defined as the final state of the parton shower
created by the simulation.

The fit ranges used for the different QCD predictions are shown in
Figure~\ref{fig_fit_ranges}. The lower edges are chosen in such a way, that 
the  hadronisation corrections in the 2-jet region remain small ($\le$ 10\%)
for LEP2 energies.
The upper  limit of the fit ranges ensures that the signal-to-background 
ratio is above 1. The ranges for pure NLLA and \oas\ fits are chosen to
be distinct, so that the results are statistically uncorrelated. 

In~\cite{siggi} it has been shown that fixing the
renormalisation scale to $\mu=\sqrt{s}$ results in a
poor description of the data.
\begin{table}[hbt]
\begin{center}
\begin{tabular}{|c|c|}\hline
observable & experimentally optimized             \\ 
           &   scales ($x_{\mu}=\mu/\sqrt{s}$)              \\\hline 
1-T                                       & 0.057 \\
C                                         & 0.082 \\
${M_{\mathrm{h}}^2}/{E_{\mathrm{vis}}^2}$ & 0.060 \\
$B_{\mathrm{max}}$                        & 0.143 \\
$B_{\mathrm{sum}}$                        & 0.096 \\ \hline
\end{tabular}
  \end{center}
\caption{\label{eos_skalen}
{The values for the experimentally optimized scales from \cite{siggi}.}}
\end{table}
Therefore, the experimentally optimized scales ($\mu_{EOS}$) from \cite{siggi}
(see Table \ref{eos_skalen}) are used for the \oas\ fits. For the NLLA and the
combined  NLLA+\oas\ fits, $\mu$ is still set equal to $\sqrt{s}$. This is the
conventional choice of scale 
for resummed and matched calculations and allows a direct comparison with the
results from other experiments \cite{lepalphas}. Furthermore the
meaning of the renormalisation scale $\mu$  in resummed
calculations is different from its interpretation in the framework of
fixed-order perturbation theory. While in \oas\ $\mu$ paramatrizes the choice 
of the renormalisation scheme this interpretation is lost in resummed
calculations. This difference makes the application of experimentally
optimized scales especially for NLLA+\oas\ predictions meaningless.
Tables \ref{daniel1}-\ref{as_logr2} at the end 
of this paper contains all \as\ values derived from event shapes.

\subsection{Definition of uncertainties \label{errdef}}
Experimental systematic uncertainties are obtained from fits to 
distributions
evaluated with different cuts and corrections. These variations are 
described in section \ref{results}.
The hadronisation uncertainty is taken to be the bigger of the two differences 
when the hadronisation correction is determined from the \ariadne\ 
\cite{ariadne}
and \herwig\ models \cite{herwig} alternatively. The  
\jetset\ result is used as the central value. In all cases the dominant
systematics come from  
the theoretical uncertainty. The  conventional method for estimating this
uncertainty is to 
consider the effect of a renormalisation scale variation 
\cite{daniel_final_draft} when fitting the experimental 
distributions.  This method, however, has at least
two  drawbacks: since the resulting scale uncertainty is positively correlated
with  the measured $\alpha_s$, this definition produces a bias towards small
$\alpha_s$ values when combining the results of e.g. different observables.  
Secondly there are indications that observables calculated only in one 
hemisphere  (like the heavy jet mass or $B_{\mathrm{max}}$)  yield less 
reliable results in  the resummation of leading logarithms \cite{einanpc}. 
This should be reflected in their theoretical uncertainty. Conversely  
the scale  variation yields the  { smallest}  uncertainty for the 
heavy jet mass and especially 
$B_{\mathrm max}$. For these reasons a new definition of the theoretical 
uncertainty for the logR prediction was developed in cooperation with the 
LEP QCD working group.
By construction, 
the NLLA calculations  do not vanish at the phase-space limit 
$y_{max}$ \cite{nlla_old}. In the so-called modified theory 
(NLLA or matched) they are forced to vanish by the replacement:
\begin{eqnarray*}
L=\ln \frac{1}{y} \rightarrow L=\ln \left[\frac{1}{X\cdot y}-\frac{1}{X\cdot
 y_{max}}+1
\right]~~.
\end{eqnarray*}
In agreement with the LEP QCD working group $y_{max}$ is chosen 
as the  maximum value of the parton shower simulation \cite{qcdworking}.
Usually $X=1$ is chosen for the quantity $X$, as suggested by the authors of 
\cite{nlla_old}, although different values for this 
$X$ scale introduce only subleading contributions \cite{xskala}. The 
theoretical uncertainty of the logR prediction in this analysis is now  
defined as half of the difference when $X$ is varied between $2/3$ and $3/2$.
By this new definition of the uncertainty the  observables 
$M^2_{\mathrm{h}}/E_{\mathrm{vis}}^2$ and $B_{\mathrm{max}}$, which are
calculated in one hemisphere only,  get a bigger uncertainty compared to the
uncertainty estimated by $\mu$ variation. The same definition of the
theoretical uncertainty has been adopted for the pure NLLA prediction.

For the \oas\ calculation we use, as in the previous publication 
\cite{daniel_final_draft},  the effect from the variation 
around the experimentally optimized scales, $\mu_{EOS}$, between 
$0.5\mu_{EOS}$ and $2\mu_{EOS}$ to estimate the theoretical uncertainty.

In order to avoid the effect mentioned above of a positive correlation, all 
scale variations have been calculated for a fixed value of \as\ from the 
theoretical distributions for each method separately. The fixed \as\ value 
is chosen as the average 
\as\ value of the combination. To obtain this value the procedure has to be 
iterated. 
\subsection{Method for combining the \boldmath \as\ measurements
  \label{combi}} 
For a combination of the $\alpha_s$ results from different observables 
calculated from the same data sets a proper treatment of the correlation 
is mandatory. The average  value $\bar{y}$ for correlated measurements 
$y_i$ is \cite{glen}:
\begin{eqnarray*}
\bar{y}=\sum_{i=1}^N w_i y_i
\qquad \mbox{with:}\quad 
w_i = \frac{\sum_j (V^{-1})_{ij}}{\sum_{k,l}(V^{-1})_{kl}}~~.
\end{eqnarray*}
Note that the weights $w$ can be negative, if the correlation $\rho_{ij}$ 
between two quantities $i$ and $j$ is bigger than $\sigma_i/\sigma_j$. Here
$\sigma$ is the uncertainty of the corresponding quantity with
$\sigma_i \ge \sigma_j$. The covariance matrix $V$ has an additive structure 
for each source of uncertainty:
$$
V = V^{\mathrm{stat}}+V^{\mathrm{sys.exp.}}+V^{\mathrm{had}}+
V^{\mathrm{scale}}~~.
$$
Its statistical component  is estimated with  simulation 
which yields correlations of typically $\ge$ 80\%. The correlation of 
systematic uncertainties is modeled by the minimum overlap assumption:
\begin{eqnarray*}
V_{ij}=\min(\sigma_i^2,\sigma_j^2 ).
\end{eqnarray*}
The \as\ values evaluated from the distributions and their mean 
values taking correlations into account are given in the Tables
\ref{daniel1}-\ref{as_logr2} at the end of the paper. 

\section{Determination of \boldmath\as\ from mean values with power 
corrections \label{asfrommeans}}

The analytical  power ansatz for non-perturbative corrections by 
\DW~\cite{PhysLettB352_451,hep-ph/9510283} including the Milan factor
established by Dokshitzer et al.~\cite{NuclPhysB511_396,hep-ph/9802381}
is used to determine \as\ from mean event shapes.
This ansatz provides an additive term to the perturbative \oas\ QCD
prediction:
\begin{equation}
\left< f \right> = 
\frac{1}{\sigma_{\mathrm tot}}\int f\frac{df}{d\sigma}d\sigma =
\left< f_{\mathrm pert} \right> + \left< f_{\mathrm pow} \right> 
\label{eq_f},
\end{equation}
where the 2nd order perturbative prediction can be written as
\begin{equation*}
\left< f_{\mathrm pert} \right> = A   \frac{\alpha_s(\mu)}{2\pi}+
               \left(A\cdot 2 \pi b_0 \log\frac{\mu^2}{s} + 
(B-2A)\right)
                \left(\frac{\alpha_s(\mu)}{2\pi}\right)^2{\rm ,}
\end{equation*}
with A and B being the perturbative 
coefficients~\cite{NuclPhysB178_412,CERN89-08vol1}, $\mu$ being the 
renormalisation scale and $b_0=(33-2N_f)/12\pi$.
The power correction is given by
\begin{equation*}
\left < f_{\mathrm{pow}}\right >  =  c_f \frac{4C_F}{\pi^2} {\cal M}
\frac{\mu_I}{\sqrt{s}} 
  \left[{\alpha}_0(\mu_I) - \alpha_s(\mu)
        - \left(b_0 \cdot \log{\frac{\mu^2}{\mu_I^2}} + 
\frac{K}{2\pi} + 2b_0 \right) \alpha_s^2(\mu) 
\right],\nonumber
\end{equation*}
where \asb\ is a non-perturbative parameter accounting for the
contributions to the event shape below an infrared matching scale $\mu_I$ and
$K=(67/18-\pi^2/6)C_A-5N_f/9$. The Milan factor ${\cal M}$ is set to
1.49, which corresponds to three active flavours in the
non-perturbative region. The observable-dependent quantities $A$, $B$  
and $c_f$ are listed in Table \ref{theo_koeff}.  For the jet broadenings 
$c_f$ takes a more complicated form\cite{bpow}:
\begin{eqnarray}
 \label{jetbreitencf}
c_f=c_B\left( \frac{\pi \sqrt{c_B}}{2\sqrt{C_F\alpha_s
(1+K\frac{\alpha_s}{2\pi})}} +\frac{3}{4}-\frac{2\pi b_0 c_B}{3C_F}+\eta_0
\right )~~.
\end{eqnarray}
Here  $c_B$ is 0.5 or 1 for $\langle B_{\mathrm{max}}\rangle$ or
$\langle B_{\mathrm{sum}}\rangle$ respectively, $\eta_0=-0.6137$.
\begin{table}[t]
\begin{center}
\begin{tabular}{|l|c|c|c|}\hline
observable & $A_f$ & $B_f$ & $c_f$ \\ \hline
$\langle 1-T \rangle$               & 2.103 & 44.99 & 2       \\
$\langle C \rangle$                 & 8.638 & 146.8 & 3$\pi$  \\
$\langle M^2_h/E^2_{vis} \rangle$     & 2.103 & 23.24 & 1       \\
$\langle B_{\mathrm{max}} \rangle$  & 4.066 & -9.53 & Eq.\ref{jetbreitencf}\\
    $\langle B_{\mathrm{sum}} \rangle$  & 4.066 & 64.24 & 
Eq.\ref{jetbreitencf} \\ \hline   
\end{tabular}
\end{center}
\caption{\label{theo_koeff}
A and B coefficients for the expansion of the mean values in
$\alpha_s/2\pi$, and values for the observable dependent $c_f$. }
\end{table}
The infrared matching scale is set to 2\gev\, as suggested
by the authors of~\cite{PhysLettB352_451}, the renormalisation scale
$\mu$ is set to $\sqrt{s}$ i.e. the $\overline{MS}$ scheme is used, since the
power corrections are provided only in this scheme.

\begin{figure}[t]
\unitlength1cm
 \unitlength1cm
 \begin{center}
 \begin{minipage}[t]{7.5cm}
               \mbox{\epsfig{file=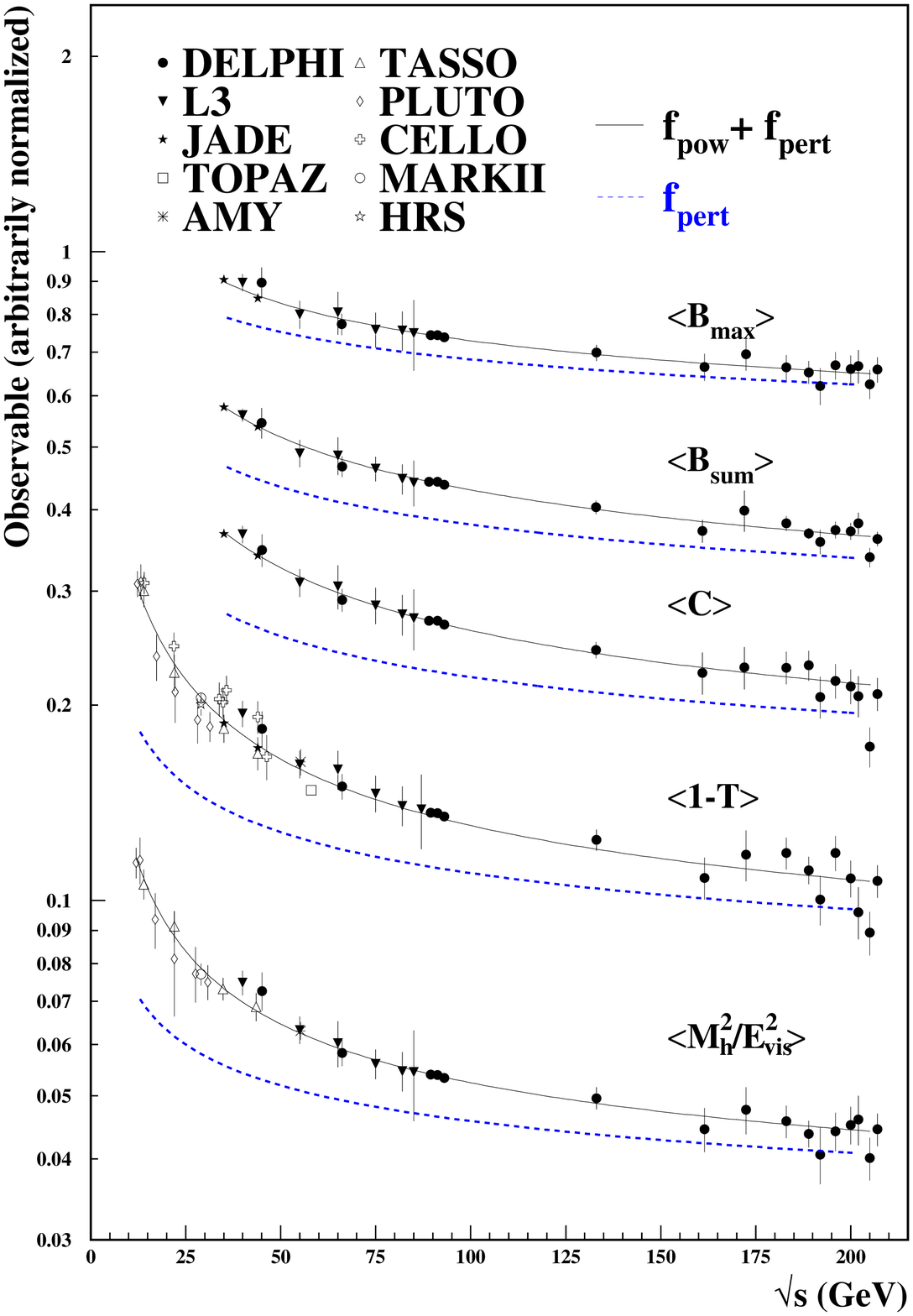,width=7.0cm}}
 \end{minipage}
 \begin{minipage}[t]{7.5cm}
           \mbox{\epsfig{file=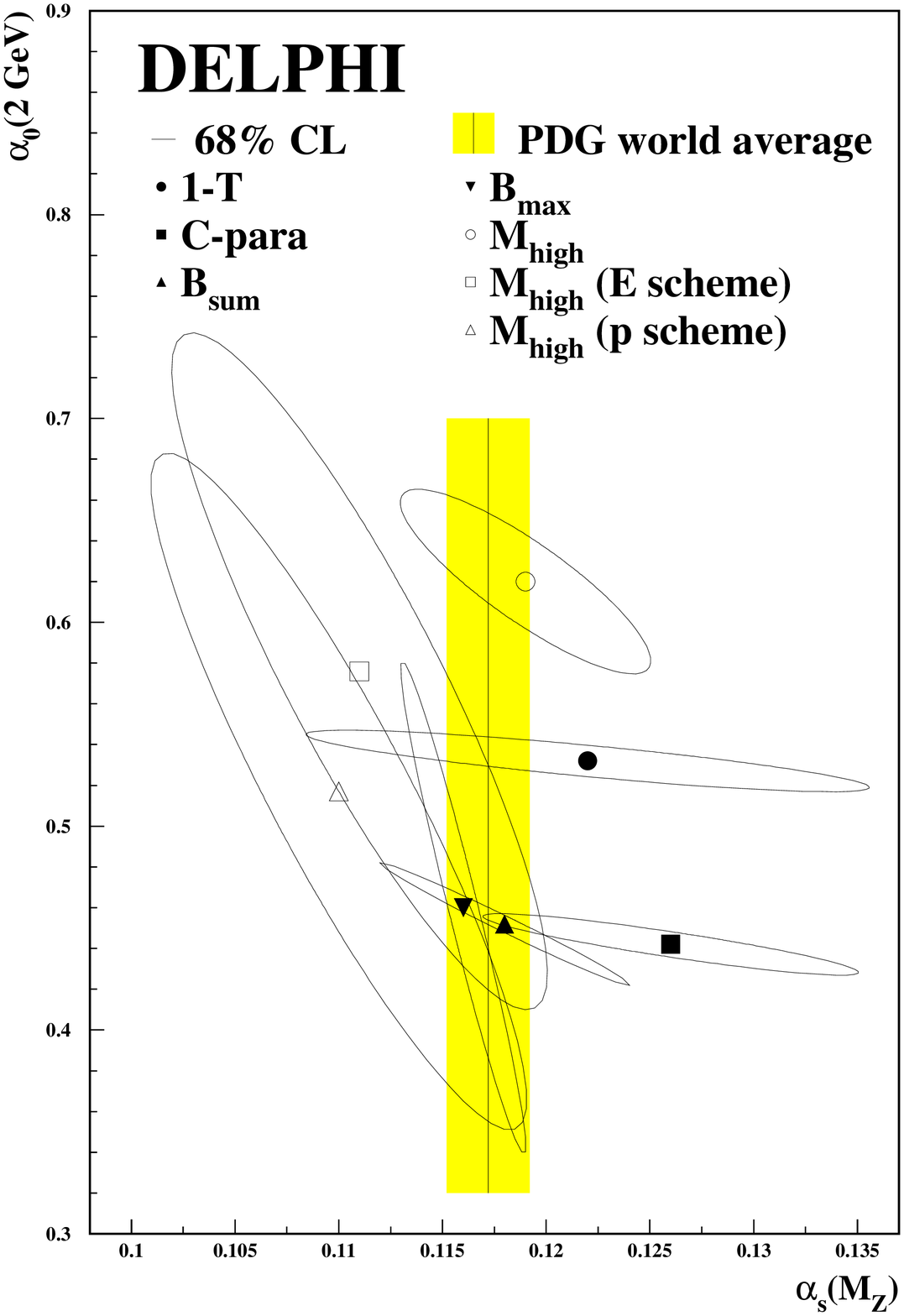,width=7.cm}}
 \end{minipage}
\end{center}
\caption{\label{fig_mess_fit}
Left: Dokshitzer-Webber fit to several mean values. The dotted line
shows the perturbative contribution. Right: the results of the global fits 
in the $\alpha_s$-$\alpha_0$ plane. The vertical line with shading shows the
world average of \as.}
\end{figure}

Besides \as\ these formulae contain \asb\ as the only free parameter. In
order to measure \as\ from the high energy data this quantity 
has to be determined.
\begin{table}[b]
\begin{center}
\begin{tabular}{|l|c|c|c|}\hline
Observable  & $\alpha_0$(2\gev)     & $\alpha_s(M_Z)$  & $\chi^2/ndf$ \\\hline
$\langle 1-T \rangle$               & 0.532 $\pm$ 0.011 $\pm$ 0.002
& 0.122 $\pm$ 0.001 $\pm$ 0.009  &  69/43 \\
$\langle C \rangle$                 & 0.442 $\pm$ 0.010 $\pm$ 0.008
& 0.126 $\pm$ 0.002 $\pm$ 0.006  & 18/22  \\
$\langle M_{h}^2/E_{vis}^2 \rangle$ & 0.620 $\pm$ 0.028 $\pm$ 0.010
& 0.119 $\pm$ 0.002 $\pm$ 0.004  &  10/32 \\
$\langle M_{h}^2/E_{vis}^2 \rangle$ (E def)& 0.576 $\pm$ 0.113 $\pm$ 0.002
& 0.111 $\pm$ 0.005 $\pm$ 0.003  & 5/14  \\
$\langle M_{h}^2/E_{vis}^2 \rangle$ (p def)& 0.517 $\pm$ 0.110 $\pm$ 0.003
& 0.110 $\pm$ 0.005 $\pm$ 0.004  & 3/14   \\
$\langle B_{\mathrm{max}} \rangle$  & 0.460 $\pm$ 0.029 $\pm$ 0.078
& 0.116 $\pm$ 0.001 $\pm$ 0.002  & 7/22  \\
$\langle B_{\mathrm{sum}} \rangle$  & 0.452 $\pm$ 0.014 $\pm$ 0.015
& 0.118 $\pm$ 0.001 $\pm$ 0.004  &  12/22 \\ \hline
\end{tabular}
\end{center}
\caption{\label{tab_mess_fit}
{$\alpha_0$ and $\alpha_s$ values from the global fit of the Dokshitzer-Webber
  ansatz  for mean values to $e^+e^-$ data from several experiments
  ~\cite{collection_eventshapes}. Only the \asb\ values are used further for
  the \as\ determination from single mean values at LEP2.}}
\end{table}
To infer \asb, a combined fit of \as\ and \asb\ to a large set of
measurements at different energies~\cite{collection_eventshapes}
is performed. For $\sqrt{s} \ge M_{\mathrm{Z}}$ only DELPHI measurements are 
included in the fit. Figure~\ref{fig_mess_fit} (left) shows the measured mean 
values of our five observables as a function of
the centre-of-mass energy together with the results of the fit.
The resulting values of \asb\ are summarized in \tab{tab_mess_fit}.
The first uncertainty in \tab{tab_mess_fit} is taken from the fit to 
the data  with full errors, while the second uncertainty reflects the effect 
of a variation $0.5\mu\le\mu\le 2\mu$. Figure \ref{fig_mess_fit} (right) 
shows the  fit  results also  in the $\alpha_s$-$\alpha_0$ plane. 
The extracted \asb\ values are supposed to be observable independent and
around 0.5~\cite{hep-ph/9510283,hep-ph/9802381}. However, higher order effects
are expected to violate this universality. Within the theoretically expected
accuracy of 20\% this universality is fulfilled. 

After fixing  \asb\ for each observable
to the values in Table \ref{tab_mess_fit}, the \as\ values corresponding to 
the high energy data points can be calculated from 
Eq.~(\ref{eq_f}). The effect of an \asb\ variation
within its uncertainty  was found to be well within the systematic
uncertainties of \as. By using the \asb\ value from the global fit, the
determination of \as\ uses the DELPHI data points twice. But 
since the global fit is dominated by the low-energy data the effect
is negligible. \as\ is calculated for all
observables individually and then combined taking correlations into account
as described in section \ref{combi}.
An additional scale uncertainty is calculated by 
varying $\mu$ for a fixed value of \as\ and
the infrared matching  scale $\mu_I$ from $1\gev$ to $3\gev$.
The \as\ results are summarized in the Tables \ref{as_meansm1} to
\ref{as_means2} at the end of the paper. The total error for this method is
smaller than e. g. for NLLA+${\cal{O}}(\alpha_s^2)$ fits.  However, the hadron
level which is experimentally accessible does include the effects of resonance
decays and hadron masses which are not accounted for in the calculation of
power corrections. In order to investigate the influence of different 
hadron level definitions a Monte Carlo study was
performed in \cite{gavinunddaniel}. Three different hadron level definitions
were considered: 
(i) hadrons  which are primary produced, (ii) stable hadrons after resonance 
decays or (iii) particles out of a subsequent decay into two massless
particles. The subsequent determination of the strong coupling from power 
corrections leads to a shift in \as\ of $\pm$0.0035 with respect to the
hadron level (ii). With regard to these extreme assumptions one may therefore
assign 
$0.007/\sqrt{12}=0.002$ as an additional uncertainty which accounts
for the fact that resonance decays and hadron masses are not
considered in the calculation. 

\section{The running of \boldmath\as \label{roas}}
The \as\ values determined at different energies are used to test
the  predicted scale dependence of the coupling. We include also the LEP2 
results at  133, 161 and 172\gev\ from \cite{daniel_final_draft}. For \as\ at 
and around $M_Z$ we have reanalyzed  the distributions from 
\cite{daniel_final_draft} for the five observables and combined the results 
using the same  treatment of correlations as described in section \ref{combi}. 
For \as\ from  mean values the measurements of events with reduced 
centre-of-mass energy  between 44 and 76\gev\ \cite{ralle} and the 
data between 133 and 172\gev\ \cite{daniel_final_draft} 
have been 
included as well. In the Tables \ref{daniel1}-\ref{as_means2} at the end 
of this paper all these  \as\ values are provided.
 
The logarithmic energy slope of the inverse coupling is given by:
\begin{eqnarray} 
\label{einspunkt}
\frac{\mathrm{d}\alpha_s^{-1}}{\mathrm{d}\log \sqrt{s}}=2b_0
+2b_1\alpha_s+\cdots ~~~,
\end{eqnarray}
with $b_0=\frac{33-2N_f}{12\pi}$ and $b_1=\frac{153-19N_f}{24\pi^2 }$
corresponding to the first coefficients of the $\beta$ function. 
The measurement of this quantity allows both a test of QCD and a consistency 
check of the four different methods used to determine \as. 
Equation \ref{einspunkt} shows that in leading order 
d$\alpha_s^{-1}/$d$\log{\sqrt{s}}$ is independent of $\alpha_s$ and  
twice the first coefficient of the $\beta$ function. 
Evaluating this equation in 
{ second order} 
results in a slight dependence on \as. With \as=0.11 (which corresponds to 
$\Lambda_{QCD}=230$\mev\ and $\sqrt{s}=$150\gev, the average energy  
of our measurements) one obtains $\mathrm{d}\alpha_s^{-1}/\mathrm{d}
\log{\sqrt{s}}=1.27$. 

\begin{figure}[hp]
 \begin{center}
  \vspace{-0.1cm}
  \unitlength1cm
  \begin{minipage}[t]{7.4cm}
   \mbox{\epsfig{file=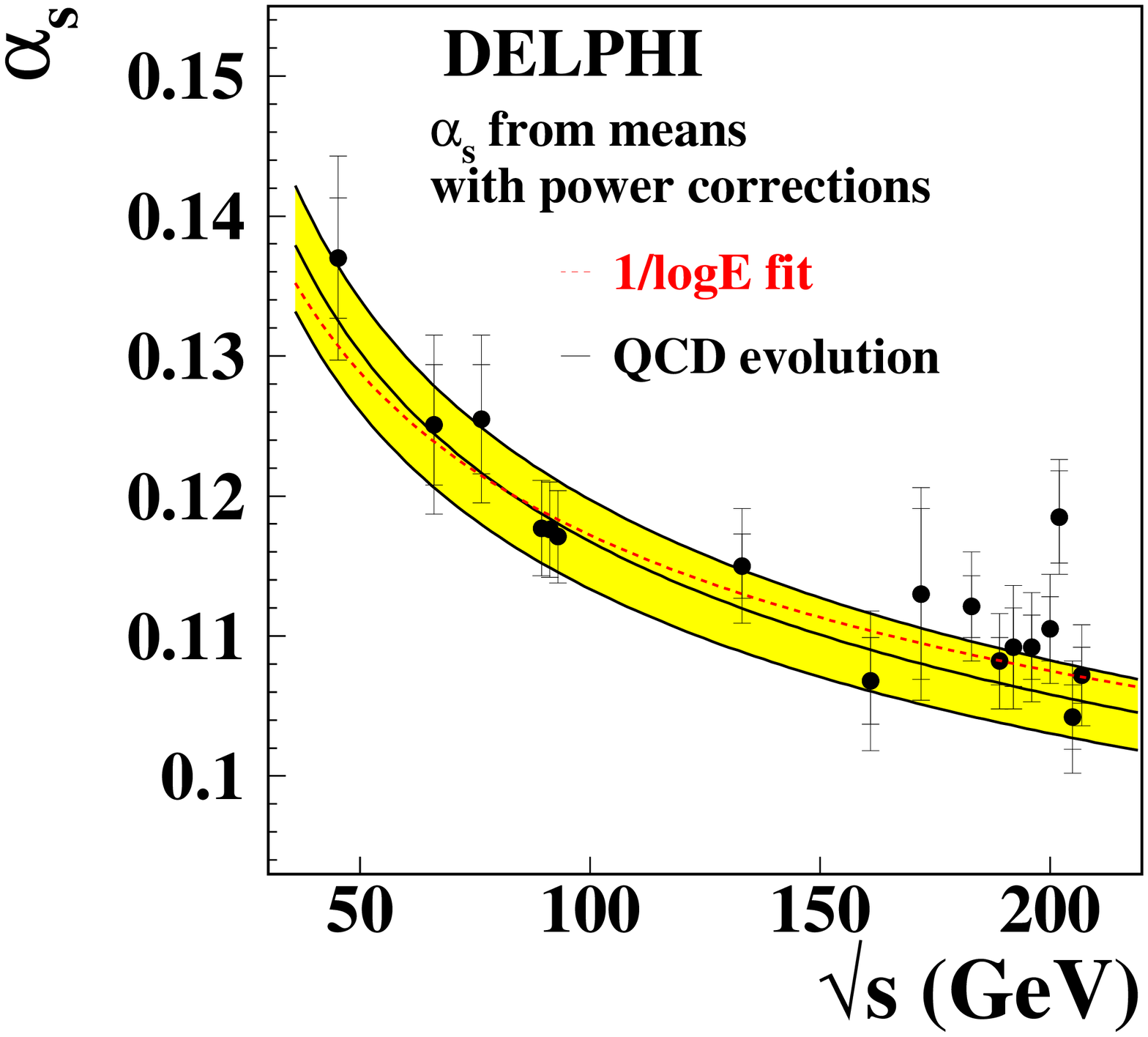,width=8.cm}}
  \end{minipage}
  \begin{minipage}[t]{7.4cm}
   \mbox{\epsfig{file=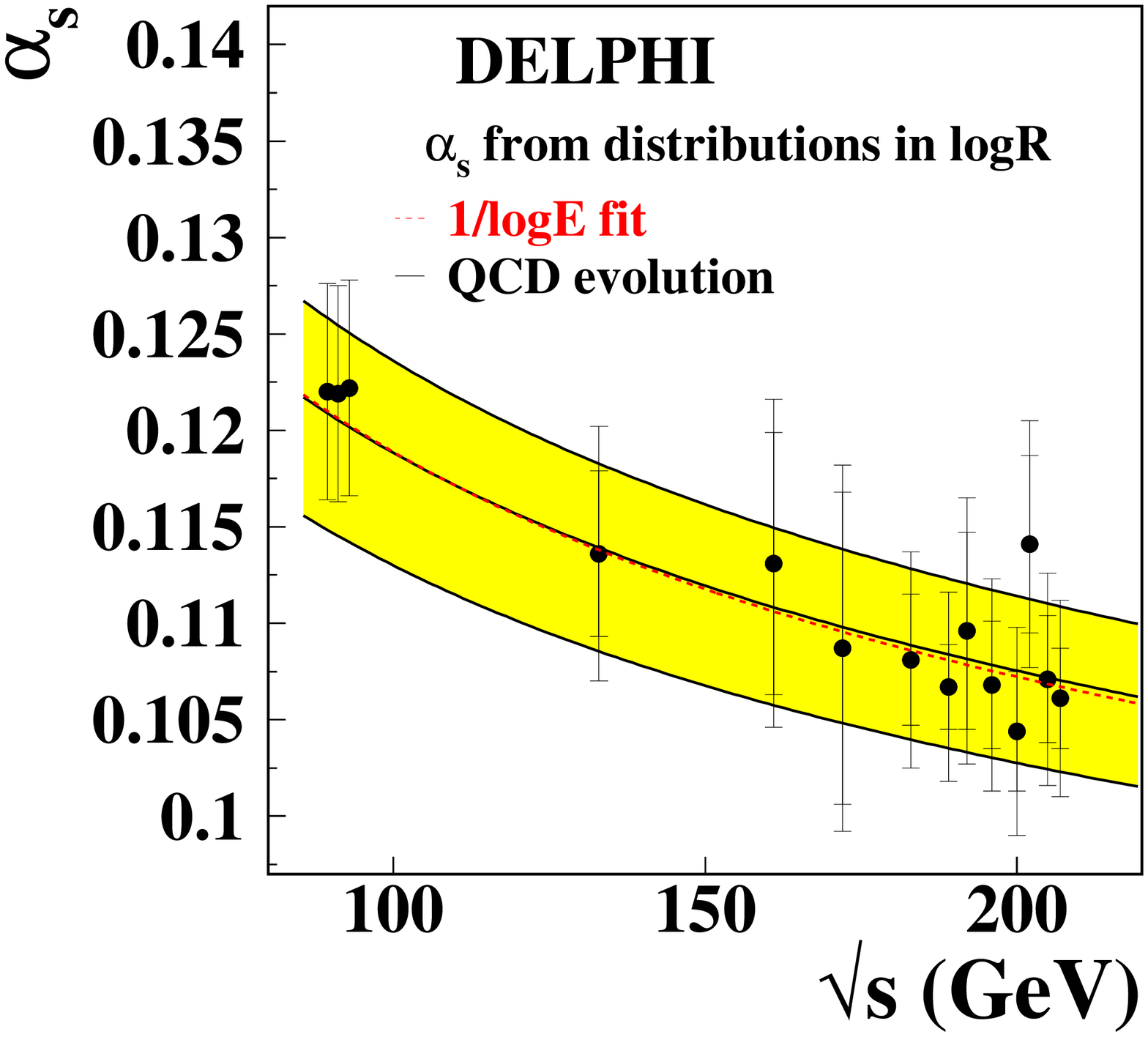,width=8.cm}}
  \end{minipage}
 \end{center}
 \vspace*{-1.2cm}
 \begin{center}
  \unitlength1cm
  \begin{minipage}[t]{7.4cm}
   \mbox{\epsfig{file=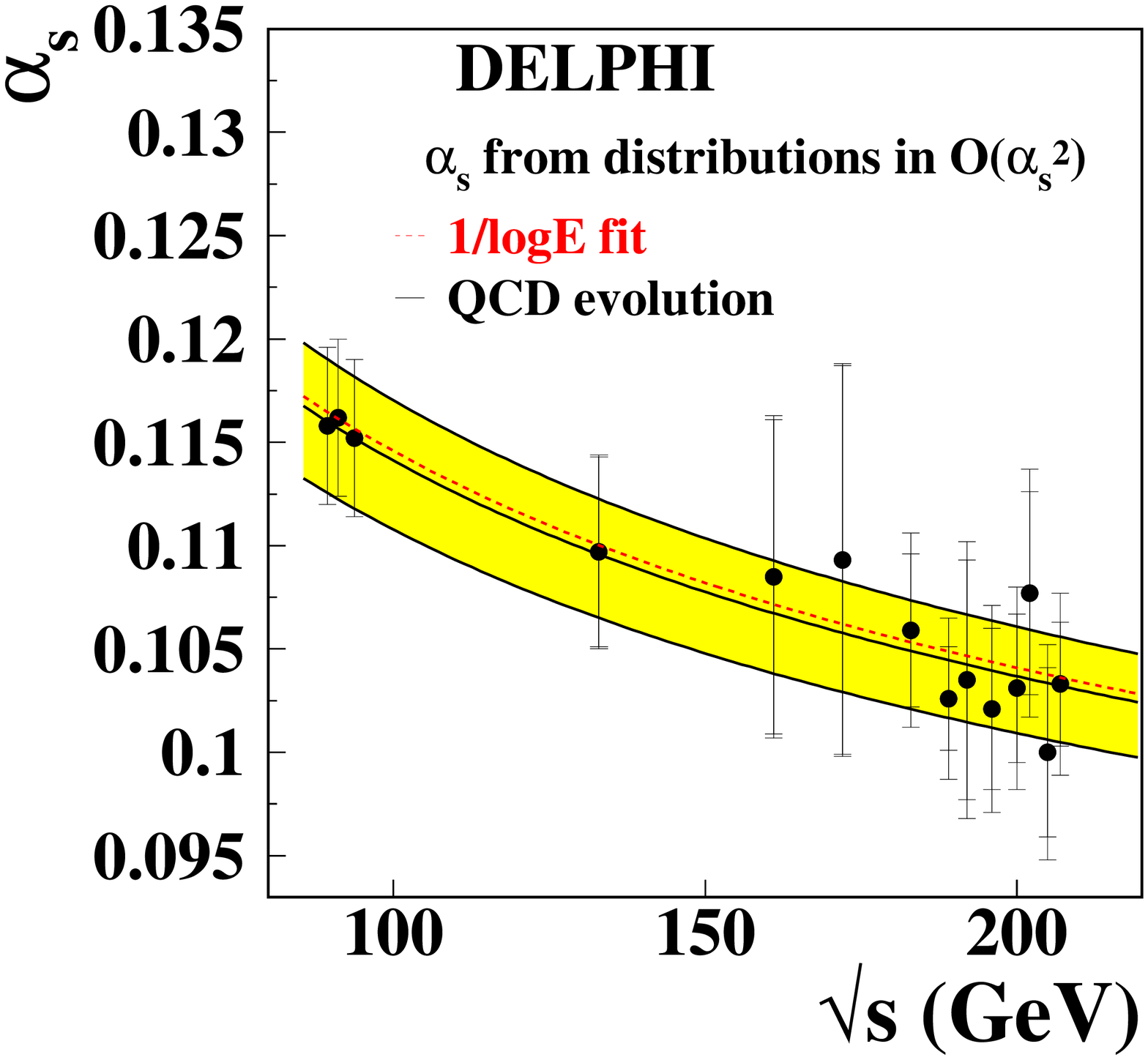,width=8.cm}}
  \end{minipage}
  \begin{minipage}[t]{7.4cm}
   \mbox{\epsfig{file=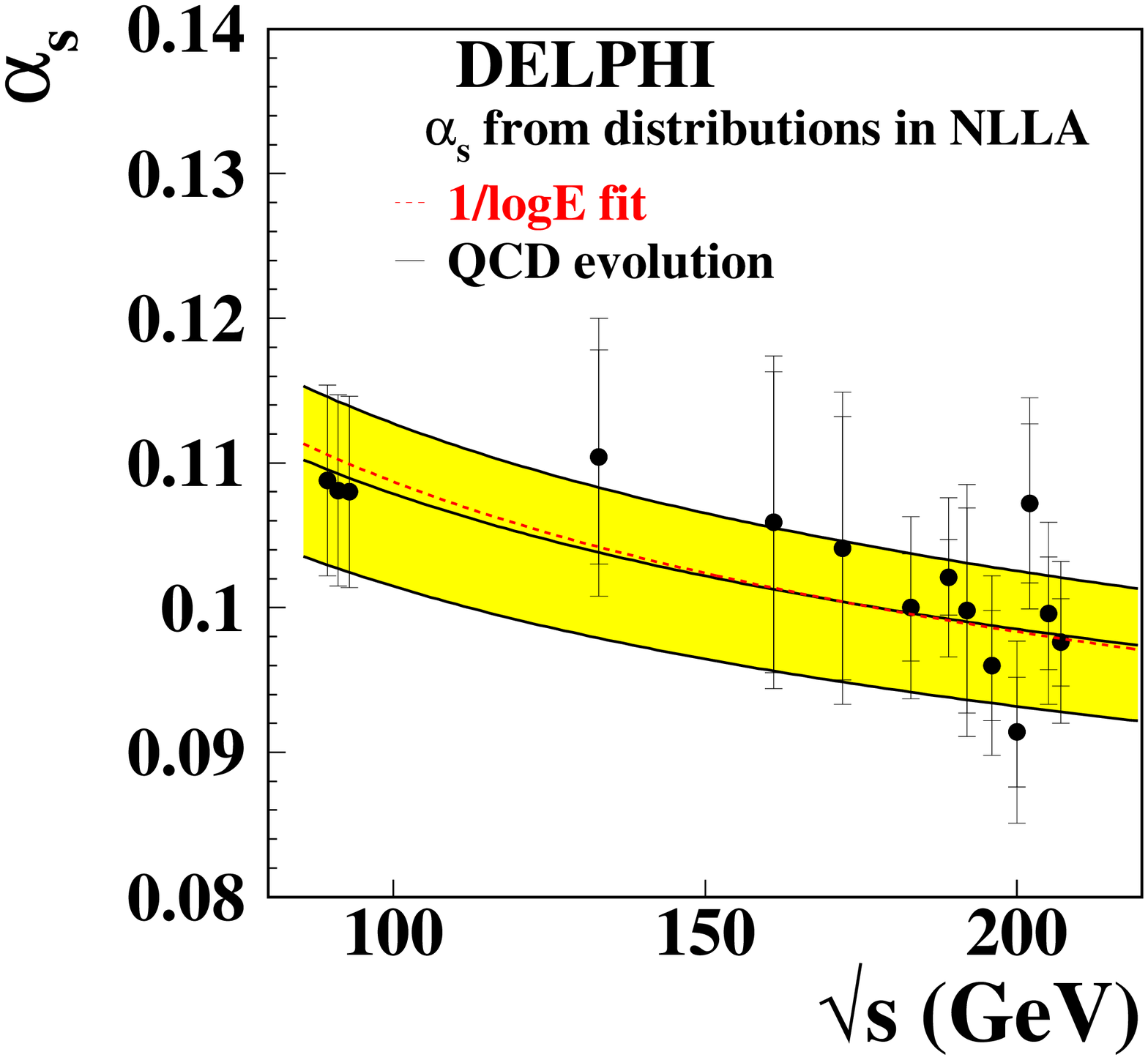,width=8.cm}}
  \end{minipage}
 \caption{\label{run}{ Energy dependence of \as\ as obtained from
event shape distributions using different theoretical calculations.
The  total and statistical (inner error-bars) uncertainties are shown.  
The band displays the average values of these measurements when 
extrapolated according to the QCD prediction. The dashed lines show the result
of the $1/\log \sqrt{s}$ fit.}}
 \end{center}
\end{figure}

Table \ref{betas} gives the slopes when fitting the function
$(b\log{\sqrt{s}}+c)$ to the $\alpha_s^{-1}$ values. The correlation between 
the \as\ measurements is taken  into account by including the full covariance 
matrix in the definition of  the $\chi^2$ function. The correlation is 
modeled  as described in  section \ref{combi}. The only difference
here is that the   statistical  uncertainties are uncorrelated.
The \as\ values and the fit of their energy dependence are also displayed in 
Figure \ref{run}. The results are in good 
agreement with the QCD expectation. Using the definition of the $b_i$ the 
result for the slope can be converted into the number of active flavours, 
$N_f$. These numbers are also included in Table \ref{betas}.

A model-independent way to measure the $\beta$ function is offered 
by applying the renormalisation group invariant (RGI)
perturbation theory to the mean  values of event shapes directly \cite{ralle}.

\begin{figure}[bh]
 \begin{center}
  \vspace{-0.1cm}
  \unitlength1cm
  \begin{minipage}[t]{6.4cm}
   \mbox{\epsfig{file=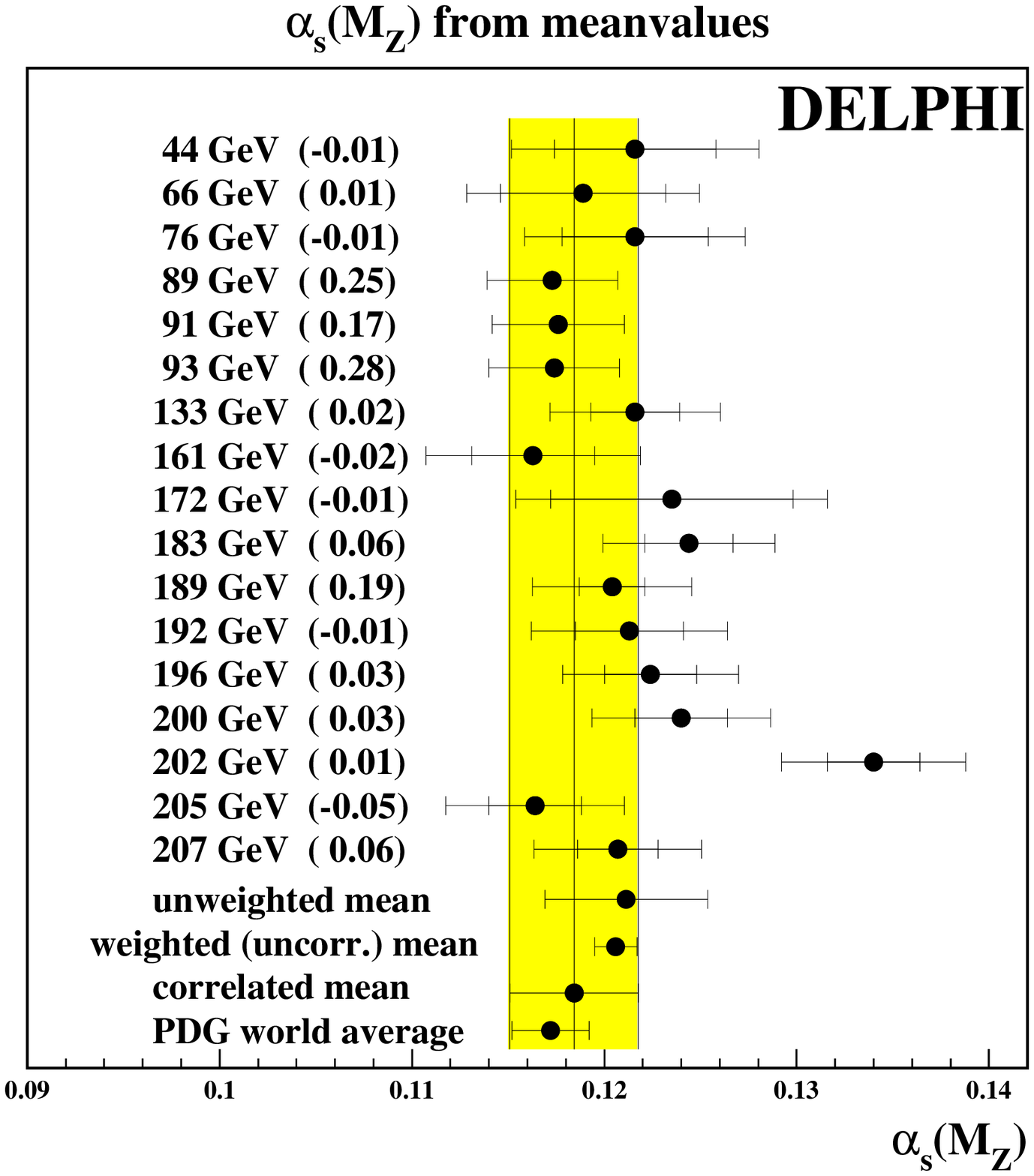,width=7.5cm}}
  \end{minipage}
  \begin{minipage}[t]{6.4cm}
   \mbox{\epsfig{file=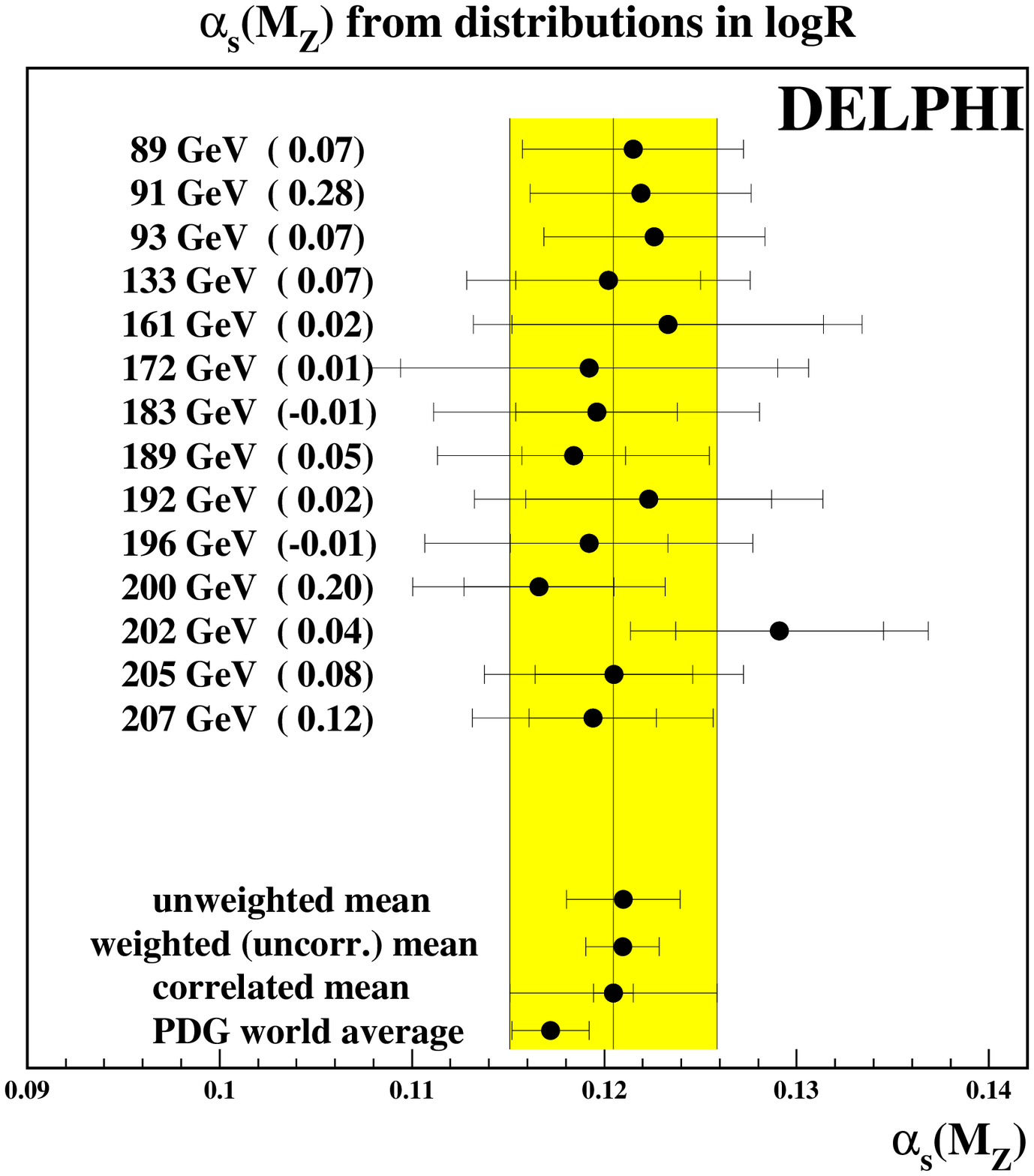,width=7.5cm}}
  \end{minipage}
 \end{center}
 \begin{center}
  \unitlength1cm
  \begin{minipage}[t]{6.4cm}
   \mbox{\epsfig{file=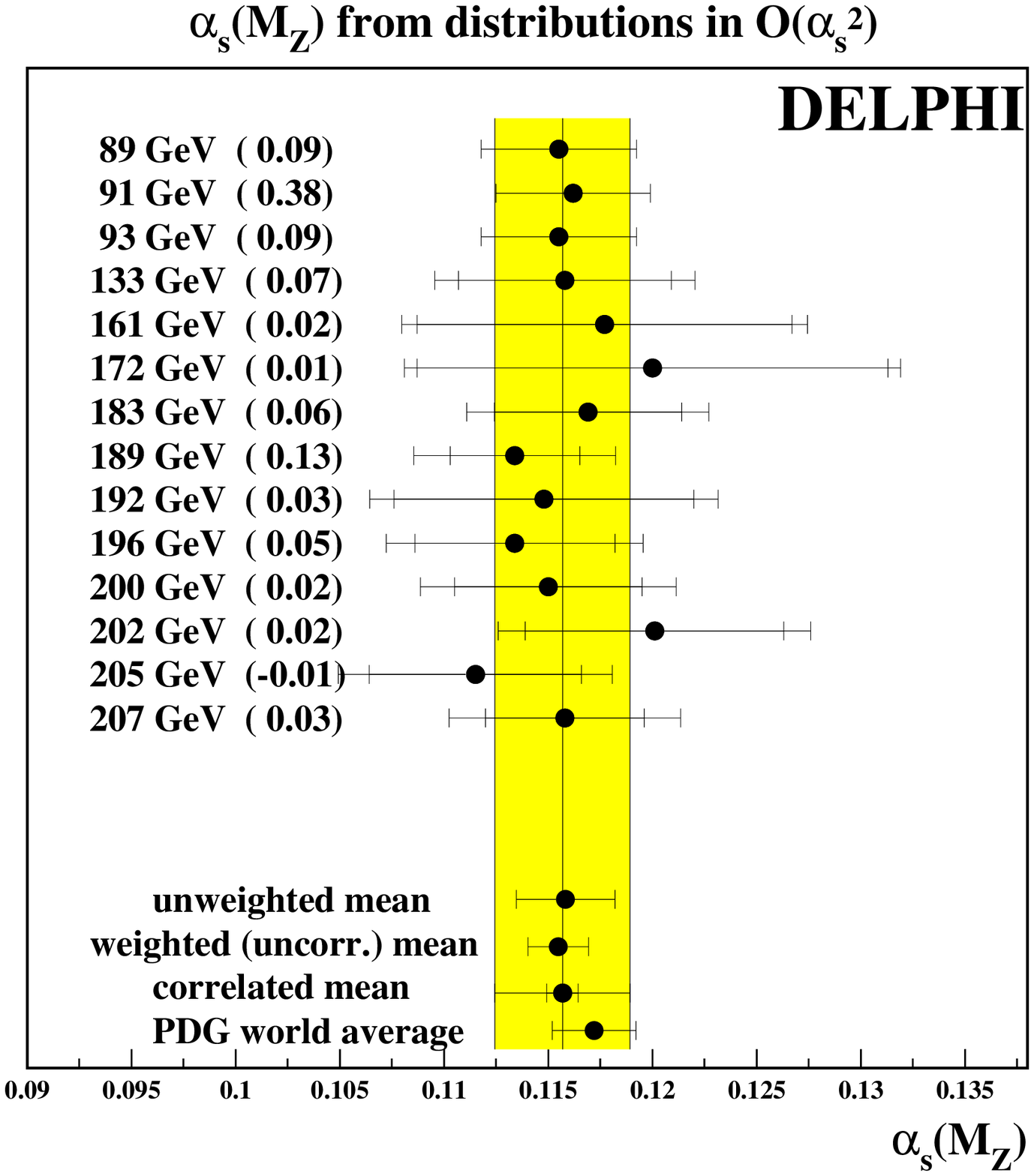,width=7.5cm}}
  \end{minipage}
  \begin{minipage}[t]{6.4cm}
   \mbox{\epsfig{file=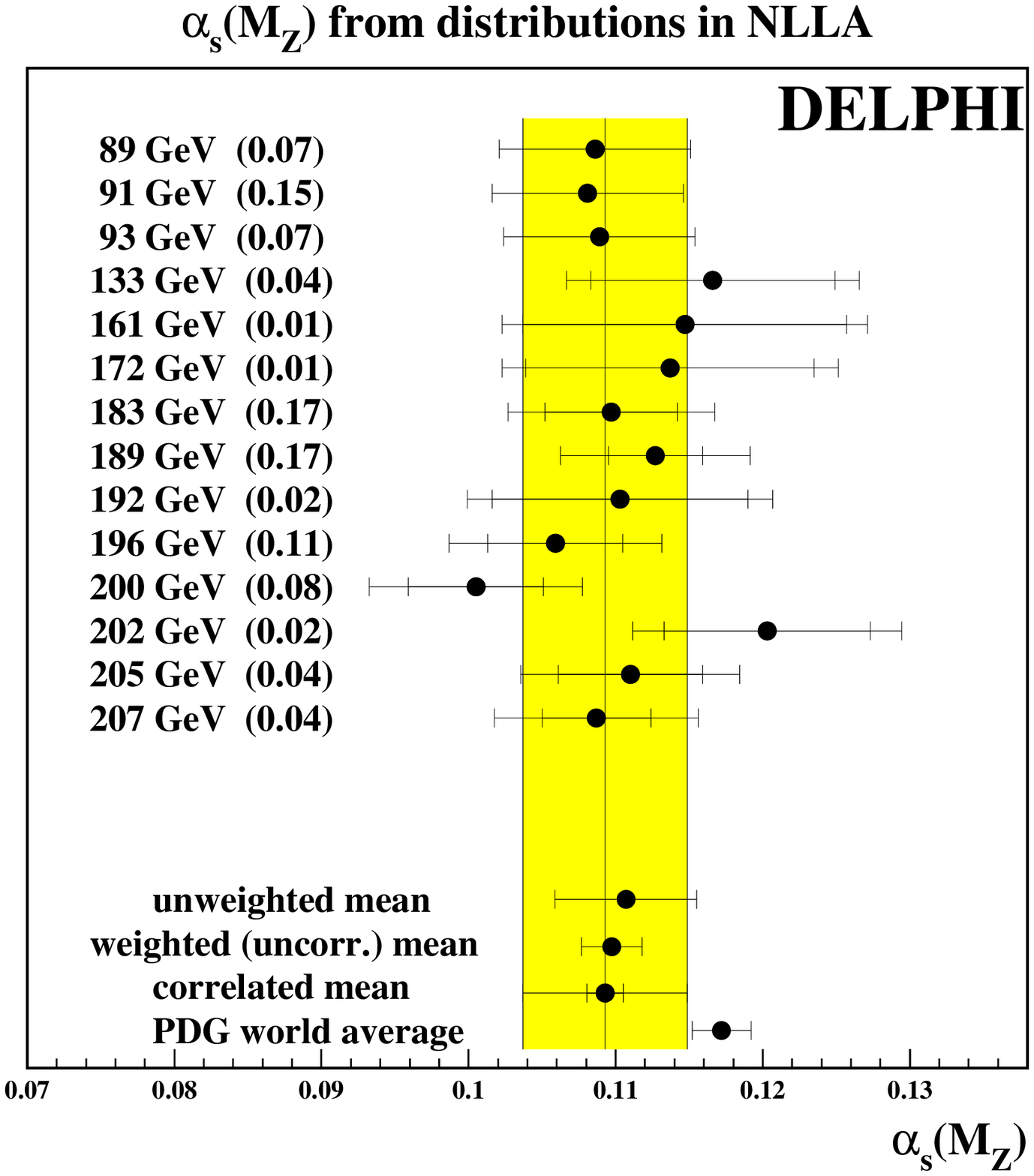,width=7.5cm}}
  \end{minipage}
\caption{\label{final_fig}
{ Results of combining all DELPHI \as\ measurements at LEP1 and LEP2. 
The total and statistical (inner error-bars) uncertainties of the 
individual  measurements are displayed. The central results are the 
 correlated means. For comparison also the unweighted and 
total-error weighted (but uncorrelated) averages are shown. 
For the unweighted mean the size of the error-bars indicate the 
RMS of the measurements. The weights of the 
individual measurements within the correlated average are given in brackets.
Note that these can turn negative in the presence of strong correlation as 
e.g. for the \as\ results from mean values.}}
 \end{center}
\end{figure}
\clearpage
\begin{table}[t]
\begin{center}
\begin{tabular}{|c|c|c||c|}\hline
theory used for measurement & $\frac{d\alpha_s^{-1}}{d\log{\sqrt{s}}}$ 
$\pm$ stat $\pm$
sys & $\chi^2/ndf$ & $N_F$ \\\hline
mean values + power corr.           & 1.11 $\pm$ 0.09 $\pm$ 0.19 & 1.25
&6.3$\pm$1.7 \\
${\cal{O}}(\alpha_s^2)$+NLLA (logR) & 1.32 $\pm$ 0.11 $\pm$ 0.27 & 0.58
& 4.6$\pm$2.3 \\
${\cal{O}}(\alpha_s^2)$             & 1.27 $\pm$ 0.15 $\pm$ 0.33 & 0.29
& 5.0$\pm$2.9\\
NLLA                                & 1.40 $\pm$ 0.17 $\pm$ 0.44 & 0.83
& 4.0$\pm$3.8\\ \hline
QCD expectation                     &  1.27   &        & 5        \\\hline
\end{tabular}
  \end{center}
\caption{\label{betas}
{Results for the slope $b$ when fitting the function $1/(b\log{\sqrt{s}}+c)$
to $\alpha_s$ values obtained for the different energies. Also given is
the corresponding result for the number of active flavours, $N_F$.
}}
\end{table}
\section{Combination of all DELPHI \boldmath\as\ measurements \label{finalas}}
As shown  in the last section, 
the energy dependence of \as\  is shown to be in good  agreement with the QCD 
prediction.  Assuming now the validity of QCD, all \as\ results can be 
evolved to a reference energy, e.g. $M_Z$, and combined 
to a single \as$(M_Z)$ measurement. Again we include results from other LEP2
energies and LEP1 as described in section \ref{roas}. 
Combining the \as\ results is, again, complicated by correlations among these
measurements. 
Although the measurements at different energies are clearly 
statistically independent, the systematic and theoretical uncertainties are 
not.
Again this part of the covariance matrix was modeled assuming minimum 
overlap. 

The results of the combinations are given in Table \ref{final_tab} and 
displayed in Figure \ref{final_fig}. The Figure contains 
in brackets also the weights of the individual measurements within the 
average. As can be seen from these numbers the weight of LEP1 and LEP2
measurements are roughly the same, since smaller theoretical uncertainties 
at LEP2 compensate for the larger statistical error. 
As can be seen from Table \ref{final_tab} the total error is still dominated
by  
the scale uncertainty.  The result with the smallest total 
uncertainty is deduced from the \oas\ prediction from distributions. Here the 
total uncertainty is 0.0033. 
\begin{table}[b]
\begin{center}
\begin{tabular}{|l|c|c|c|c|c||c|}\hline
theory & $\alpha_s(M_Z)$ & stat. & sys.exp. & had. & scale & tot\\ \hline
mean values + power corr.      & 0.1184 & 0.0004 & 0.0008 & 0.0022 & 0.0031
 & 0.0039 \\
${\cal{O}}(\alpha_s^2)$+NLLA (logR)& 0.1205 & 0.0010 & 0.0018 & 0.0013 & 0.0048
 & 0.0054 \\ 
${\cal{O}}(\alpha_s^2)$            & 0.1157 & 0.0008 & 0.0016 & 0.0016 & 0.0022
 & 0.0033 \\ 
NLLA                               & 0.1093 & 0.0012 & 0.0020 & 0.0011 & 0.0050
 & 0.0056 \\ \hline
\end{tabular}
\end{center}
\caption{\label{final_tab}
{Results of combining all DELPHI \as\ measurements at LEP1 and LEP2.
For the \as\ results from mean values with power 
corrections  ``hadronisation uncertainty'' (had.) denotes the combined effect
of the $\mu_I$ variation and the uncertainty related to resonance
decays and particle masses as described at the end of Section 
\ref{asfrommeans}. 
The scale uncertainty is either the effect of a variation of the 
renormalisation scale $\mu$ (${\cal{O}}(\alpha_s^2)$ and power corrections) 
or the effect of changing the $X$ scale (see Section \ref{errdef}).}}
\end{table}
This value can be compared with the central result of the DELPHI analysis
\cite{siggi} from the 
observable jet cone energy fraction (JCEF) alone: \as=0.1180$\pm$0.0018. 
The different precision is mainly  due to the  definition of 
the  scale uncertainty. While we use the variation  $0.5\mu\le\mu\le 2\mu$,
the  
analysis \cite{siggi} changes the corresponding quantity only between 
$\sqrt{0.5}\mu$  and $\sqrt{2}\mu$. The actual choice of the $\mu$ variation 
is not guided by  solid theoretical arguments and was studied in
all possible detail in \cite{siggi}. 
However, our definition of the theoretical 
uncertainty yields good agreement with the average root-mean-square of the 
fits to the 
five different  observables at the same energy (see Tables 
\ref{daniel1}-\ref{as_oas2}). Another reason for the higher precision 
in \cite{siggi} is the use of the observable JCEF, which has particularly 
small uncertainties from hadronisation and scale variation. However, the focus
of this work is the analysis of five observables with several 
different techniques. 
At present the theoretical uncertainties of \as\ measurements from event 
shape distributions are subject of a debate. Further substantial progress can 
only be achieved by the arrival of next-to-next-to-leading-order
(NNLO) calculations.

\section{Conclusion\label{dasletzte}}
A measurement of event shape distributions and mean values is presented 
as obtained from data at centre-of-mass energies from 183 to 207\gev.
The strong coupling constant $\alpha_s$ has been determined from the
event shape variables  Thrust, C parameter, heavy jet mass, wide and total jet
broadening, with four different methods: the differential 
distributions are compared to predictions in \oas, pure NLLA and \oas+NLLA 
(logR), folded with fragmentation models, while from the mean values, \as\ is 
extracted using an analytical power correction ansatz. 
The \as\ values are combined with results obtained at other LEP2 energies and 
at and around $M_Z$. This allows both a combined measurement of $\alpha_s$ 
and a test of the running of $\alpha_s$. In these combinations the full
correlation between energies and observables was taken into account. 

The main aim of this study is the comparison of different methods to extract
$\alpha_s$ and its scale dependence. Within their uncertainties all techniques
yield consistent results.  
The $\alpha_s$ with smallest uncertainty is obtained from \oas\ with
experimentally optimised scales: 
\begin{eqnarray*}
\alpha_s(M_Z) &=& 0.1157 \pm 0.0008 \,\mbox{(stat)} \pm 0.0016 
\,\mbox{(sys.ex.)} \pm 0.0016 \,\mbox{(had)} \pm 0.0022 \,\mbox{(scale)}\\
        &=& 0.1157 \pm 0.0033 \,\mbox{(tot)}
\end{eqnarray*} 
The current world average from the particle data group is
$0.1172\pm0.0020$ \cite{pdg2002}.

For the energy dependence of the strong coupling
the highest precision is obtained for the \as\ values derived from
mean values with power corrections:
\begin{eqnarray*}
\frac{\mathrm{d}\alpha_s^{-1}}{\mathrm{d}\log \sqrt{s}} = 
1.11 \pm 0.09 \,\mbox{(stat)}  \pm 0.19 \,\mbox{(sys)}
\end{eqnarray*}
The last number has to be compared with the QCD expectation of 1.27.

\input{acknow}

\begin{appendix}
\begin{table}[h]
\begin{center}
\begin{small}
\renewcommand{\arraystretch}{1.2}
\begin{tabular}{|c|c|c|c|}\hline
$\sqrt{s}$ & $\langle  1-T    \rangle$  &
          $\langle (1-T)^2 \rangle$  &
          $\langle (1-T)^3 \rangle$   \\ \hline
183 & 0.0592$\pm$ 0.0024$\pm$ 0.0020 & 0.00766$\pm$0.00070$\pm$0.00054 &
0.00154
$\pm$0.00021$\pm$0.00015\\
189 & 0.0557$\pm$ 0.0016$\pm$ 0.0022 & 0.00658$\pm$0.00048$\pm$0.00060 &
0.00121
$\pm$0.00015$\pm$0.00017\\
192 & 0.0502$\pm$ 0.0040$\pm$ 0.0023 & 0.00454$\pm$0.00116$\pm$0.00064 &
0.00055
$\pm$0.00035$\pm$0.00018\\
196 & 0.0592$\pm$ 0.0029$\pm$ 0.0024 & 0.00810$\pm$0.00085$\pm$0.00067 &
0.00171
$\pm$0.00026$\pm$0.00019\\
200 & 0.0541$\pm$ 0.0028$\pm$ 0.0025 & 0.00613$\pm$0.00086$\pm$0.00071 &
0.00101
$\pm$0.00027$\pm$0.00020\\
202 & 0.0480$\pm$ 0.0040$\pm$ 0.0025 & 0.00330$\pm$0.00121$\pm$0.00072 &
0.00003
$\pm$0.00039$\pm$0.00020\\
205 & 0.0446$\pm$ 0.0030$\pm$ 0.0026 & 0.00322$\pm$0.00090$\pm$0.00077 &
0.00019
$\pm$0.00027$\pm$0.00022\\
207 & 0.0536$\pm$ 0.0023$\pm$ 0.0027 & 0.00572$\pm$0.00066$\pm$0.00079 &
0.00086
$\pm$0.00020$\pm$0.00022\\
\hline\hline
          & $\langle  C    \rangle$  &
          $\langle (C)^2 \rangle$  &
          $\langle (C)^3 \rangle$   \\ \hline
183 & 0.2286$\pm$ 0.0070$\pm$ 0.0106 & 0.08846$\pm$0.00544$\pm$0.00952 &
0.04585
$\pm$0.00413$\pm$0.00824\\
189 & 0.2304$\pm$ 0.0046$\pm$ 0.0113 & 0.09252$\pm$0.00369$\pm$0.01030 &
0.05114
$\pm$0.00287$\pm$0.00894\\
192 & 0.2060$\pm$ 0.0115$\pm$ 0.0117 & 0.06634$\pm$0.00915$\pm$0.01073 &
0.02676
$\pm$0.00716$\pm$0.00933\\
196 & 0.2181$\pm$ 0.0080$\pm$ 0.0121 & 0.08091$\pm$0.00657$\pm$0.01118 &
0.03909
$\pm$0.00515$\pm$0.00973\\
200 & 0.2139$\pm$ 0.0079$\pm$ 0.0126 & 0.07882$\pm$0.00658$\pm$0.01170 &
0.03907
$\pm$0.00524$\pm$0.01020\\
202 & 0.2066$\pm$ 0.0111$\pm$ 0.0127 & 0.06730$\pm$0.00919$\pm$0.01184 &
0.02761
$\pm$0.00737$\pm$0.01032\\
205 & 0.1726$\pm$ 0.0088$\pm$ 0.0133 & 0.03792$\pm$0.00743$\pm$0.01251 &
0.00349
$\pm$0.00596$\pm$0.01092\\
207 & 0.2081$\pm$ 0.0065$\pm$ 0.0136 & 0.07123$\pm$0.00540$\pm$0.01278 &
0.03150
$\pm$0.00426$\pm$0.01116\\
\hline\hline
          & $\langle B_{\mathrm{sum}}     \rangle$  &
          $\langle (B_{\mathrm{sum}} )^2 \rangle$  &
          $\langle (B_{\mathrm{sum}} )^3 \rangle$   \\ \hline
183 & 0.0953$\pm$ 0.0023$\pm$ 0.0010 & 0.01334$\pm$0.00070$\pm$0.00028 &
0.00247
$\pm$0.00020$\pm$0.00011\\
189 & 0.0920$\pm$ 0.0015$\pm$ 0.0010 & 0.01192$\pm$0.00047$\pm$0.00030 &
0.00199
$\pm$0.00013$\pm$0.00012\\
192 & 0.0893$\pm$ 0.0038$\pm$ 0.0010 & 0.01113$\pm$0.00117$\pm$0.00031 &
0.00178
$\pm$0.00034$\pm$0.00013\\
196 & 0.0931$\pm$ 0.0026$\pm$ 0.0010 & 0.01266$\pm$0.00082$\pm$0.00032 &
0.00224
$\pm$0.00024$\pm$0.00014\\
200 & 0.0927$\pm$ 0.0026$\pm$ 0.0010 & 0.01254$\pm$0.00081$\pm$0.00034 &
0.00222
$\pm$0.00023$\pm$0.00015\\
202 & 0.0954$\pm$ 0.0035$\pm$ 0.0010 & 0.01344$\pm$0.00111$\pm$0.00034 &
0.00257
$\pm$0.00032$\pm$0.00015\\
205 & 0.0845$\pm$ 0.0028$\pm$ 0.0010 & 0.00952$\pm$0.00086$\pm$0.00036 &
0.00131
$\pm$0.00025$\pm$0.00016\\
207 & 0.0902$\pm$ 0.0021$\pm$ 0.0010 & 0.01151$\pm$0.00066$\pm$0.00036 &
0.00188
$\pm$0.00019$\pm$0.00016\\
\hline\hline
    & $\langle B_{\mathrm{max}}     \rangle$  &
          $\langle (B_{\mathrm{max}} )^2 \rangle$  &
          $\langle (B_{\mathrm{max}} )^3 \rangle$   \\ \hline
183 & 0.0663$\pm$ 0.0021$\pm$ 0.0021 & 0.00688$\pm$0.00053$\pm$0.00034 &
0.00095
$\pm$0.00013$\pm$0.00007\\
189 & 0.0652$\pm$ 0.0014$\pm$ 0.0022 & 0.00656$\pm$0.00037$\pm$0.00036 &
0.00089
$\pm$0.00009$\pm$0.00008\\
192 & 0.0621$\pm$ 0.0035$\pm$ 0.0022 & 0.00557$\pm$0.00096$\pm$0.00038 &
0.00061
$\pm$0.00026$\pm$0.00008\\
196 & 0.0668$\pm$ 0.0024$\pm$ 0.0022 & 0.00719$\pm$0.00064$\pm$0.00039 &
0.00105
$\pm$0.00016$\pm$0.00009\\
200 & 0.0659$\pm$ 0.0024$\pm$ 0.0023 & 0.00699$\pm$0.00063$\pm$0.00041 &
0.00100
$\pm$0.00016$\pm$0.00009\\
202 & 0.0666$\pm$ 0.0033$\pm$ 0.0023 & 0.00671$\pm$0.00089$\pm$0.00041 &
0.00087
$\pm$0.00023$\pm$0.00009\\
205 & 0.0625$\pm$ 0.0025$\pm$ 0.0023 & 0.00585$\pm$0.00068$\pm$0.00044 &
0.00073
$\pm$0.00017$\pm$0.00010\\
207 & 0.0658$\pm$ 0.0020$\pm$ 0.0023 & 0.00695$\pm$0.00053$\pm$0.00044 &
0.00100
$\pm$0.00013$\pm$0.00010\\
\hline
\end{tabular}
\end{small}
 \end{center}
\caption{\label{mean1} Mean values and higher moments of the Thrust, 
C, $B_{\mathrm{max}}$ and $B_{\mathrm{sum}}$ distributions with statistical 
and  systematic errors.}
\end{table}

\begin{table}[h]
\begin{center}
\begin{small}
\renewcommand{\arraystretch}{1.2}
\begin{tabular}{|c|c|c|c|}\hline
$\sqrt{s}$ & $\langle M_{\mathrm{h}}^2/E_{\mathrm{vis}}^2     \rangle$  &
          $\langle (M_{\mathrm{h}}^2/E_{\mathrm{vis}}^2)^2 \rangle$  &
          $\langle (M_{\mathrm{h}}^2/E_{\mathrm{vis}}^2)^3 \rangle$   \\ \hline
183 & 0.0457$\pm$ 0.0023$\pm$ 0.0012 & 0.00451$\pm$0.00066$\pm$0.00027 &
0.00068
$\pm$0.00021$\pm$0.00006\\
189 & 0.0437$\pm$ 0.0016$\pm$ 0.0013 & 0.00408$\pm$0.00045$\pm$0.00030 &
0.00060
$\pm$0.00014$\pm$0.00007\\
192 & 0.0406$\pm$ 0.0039$\pm$ 0.0013 & 0.00285$\pm$0.00117$\pm$0.00032 &
0.00024
$\pm$0.00040$\pm$0.00008\\
196 & 0.0441$\pm$ 0.0027$\pm$ 0.0014 & 0.00421$\pm$0.00079$\pm$0.00034 &
0.00060
$\pm$0.00024$\pm$0.00008\\
200 & 0.0451$\pm$ 0.0027$\pm$ 0.0015 & 0.00458$\pm$0.00078$\pm$0.00036 &
0.00071
$\pm$0.00025$\pm$0.00009\\
202 & 0.0460$\pm$ 0.0038$\pm$ 0.0015 & 0.00470$\pm$0.00112$\pm$0.00036 &
0.00083
$\pm$0.00037$\pm$0.00009\\
205 & 0.0401$\pm$ 0.0028$\pm$ 0.0016 & 0.00338$\pm$0.00080$\pm$0.00039 &
0.00045
$\pm$0.00023$\pm$0.00009\\
207 & 0.0444$\pm$ 0.0022$\pm$ 0.0016 & 0.00439$\pm$0.00066$\pm$0.00040 &
0.00068
$\pm$0.00021$\pm$0.00010\\
\hline\hline
       & $\langle  M_{\mathrm{h}}^2/E_{\mathrm{vis}}^2 p \rangle$   &
          $\langle (M_{\mathrm{h}}^2/E_{\mathrm{vis}}^2 p)^2 \rangle$ &
          $\langle (M_{\mathrm{h}}^2/E_{\mathrm{vis}}^2 p)^3 \rangle$ \\
\hline
183 & 0.0427$\pm$ 0.0023$\pm$ 0.0012 & 0.00421$\pm$0.00066$\pm$0.00027 &
0.00068
$\pm$0.00021$\pm$0.00006\\
189 & 0.0411$\pm$ 0.0016$\pm$ 0.0013 & 0.00383$\pm$0.00045$\pm$0.00030 &
0.00060
$\pm$0.00014$\pm$0.00007\\
192 & 0.0384$\pm$ 0.0039$\pm$ 0.0013 & 0.00274$\pm$0.00117$\pm$0.00032 &
0.00025
$\pm$0.00039$\pm$0.00008\\
196 & 0.0413$\pm$ 0.0027$\pm$ 0.0014 & 0.00396$\pm$0.00079$\pm$0.00034 &
0.00057
$\pm$0.00024$\pm$0.00008\\
200 & 0.0424$\pm$ 0.0027$\pm$ 0.0015 & 0.00426$\pm$0.00078$\pm$0.00036 &
0.00065
$\pm$0.00025$\pm$0.00009\\
202 & 0.0436$\pm$ 0.0038$\pm$ 0.0015 & 0.00451$\pm$0.00112$\pm$0.00036 &
0.00081
$\pm$0.00037$\pm$0.00009\\
205 & 0.0380$\pm$ 0.0029$\pm$ 0.0016 & 0.00320$\pm$0.00080$\pm$0.00039 &
0.00043
$\pm$0.00023$\pm$0.00009\\
207 & 0.0420$\pm$ 0.0023$\pm$ 0.0016 & 0.00419$\pm$0.00067$\pm$0.00040 &
0.00065
$\pm$0.00021$\pm$0.00010\\
\hline\hline
    & $\langle  M_{\mathrm{h}}^2/E_{\mathrm{vis}}^2 E   \rangle$  &
          $\langle (M_{\mathrm{h}}^2/E_{\mathrm{vis}}^2 E)^2 \rangle$  &
          $\langle (M_{\mathrm{h}}^2/E_{\mathrm{vis}}^2 E)^3 \rangle$   \\
\hline
183 & 0.0434$\pm$ 0.0023$\pm$ 0.0012 & 0.00426$\pm$0.00066$\pm$0.00027 &
0.00064
$\pm$0.00021$\pm$0.00006\\
189 & 0.0417$\pm$ 0.0016$\pm$ 0.0013 & 0.00390$\pm$0.00045$\pm$0.00030 &
0.00057
$\pm$0.00014$\pm$0.00007\\
192 & 0.0391$\pm$ 0.0039$\pm$ 0.0013 & 0.00282$\pm$0.00117$\pm$0.00032 &
0.00026
$\pm$0.00039$\pm$0.00008\\
196 & 0.0420$\pm$ 0.0027$\pm$ 0.0014 & 0.00403$\pm$0.00079$\pm$0.00034 &
0.00058
$\pm$0.00024$\pm$0.00008\\
200 & 0.0430$\pm$ 0.0027$\pm$ 0.0015 & 0.00434$\pm$0.00078$\pm$0.00036 &
0.00067
$\pm$0.00025$\pm$0.00009\\
202 & 0.0440$\pm$ 0.0038$\pm$ 0.0015 & 0.00450$\pm$0.00112$\pm$0.00036 &
0.00080
$\pm$0.00037$\pm$0.00009\\
205 & 0.0385$\pm$ 0.0029$\pm$ 0.0016 & 0.00325$\pm$0.00080$\pm$0.00039 &
0.00044
$\pm$0.00023$\pm$0.00009\\
207 & 0.0426$\pm$ 0.0023$\pm$ 0.0016 & 0.00425$\pm$0.00067$\pm$0.00040 &
0.00066
$\pm$0.00021$\pm$0.00010\\
\hline
\end{tabular}
\end{small}
 \end{center}
\caption{\label{mean2} Mean values and higher moments for the 
$M^2_h/E^2_{vis}$ distributions
in the standard, E-scheme and p-scheme definitions
 with statistical and
 systematic errors.}
\end{table}
\begin{table}[t]
\begin{center}
\begin{tabular}{|l|c|c|c|c|c||c|}\hline
 \multicolumn{7}{|c|}{$\alpha_s$ in ${\cal{O}}(\alpha_s^2)$}     \\ \hline
\hline 
observable  & 1-T & C  & $M^2_{\mathrm{h}}/E^2_{\mathrm{vis}}$ 
& $B_{\mathrm{max}}$ & $B_{\mathrm{sum}}$ & average\\ \hline 
fit range           & 0.09-0.21 & 0.28-0.6  & 0.05-0.17 & 0.07-0.20 &
0.115-0.24 &   \\ \hline \hline
$\alpha_s(89.5\gev)$  & 0.1135 & 0.1149 & 0.1222 & 0.1186 & 0.1123 & 0.1158  
\\ \hline
$\pm\Delta$ stat.     & 0.0004 & 0.0004 & 0.0005 & 0.0005 & 0.0003 & 0.0004  \\ 
$\pm\Delta$ sys. exp. & 0.0009 & 0.0021 & 0.0014 & 0.0022 & 0.0030 & 0.0015  
\\ \hline
$\pm\Delta$ had.      & 0.0026 & 0.0023 & 0.0025 & 0.0031 & 0.0032 & 0.0027  \\ 
$\pm\Delta$ $\mu_R$ scale& 0.0040 & 0.0054 & 0.0040 & 0.0018 & 0.0068 & 0.0022
  \\ \hline  
$\pm\Delta_{\mathrm{tot}}$& 0.0049 & 0.0062 & 0.0051 & 0.0042 & 0.0081 & 0.0038
  \\ \hline    
\multicolumn{6}{r||}{RMS} & 0.0040  \\ \hline\hline
$\alpha_s(91.2\gev)$  & 0.1139 & 0.1155 & 0.1230 & 0.1191 & 0.1128 & 0.1162  
\\ \hline
$\pm\Delta$ stat.     & 0.0002 & 0.0002 & 0.0002 & 0.0002 & 0.0002 & 0.0002  \\ 
$\pm\Delta$ sys. exp. & 0.0009 & 0.0021 & 0.0014 & 0.0022 & 0.0030 & 0.0015  
\\ \hline
$\pm\Delta$ had.      & 0.0026 & 0.0023 & 0.0025 & 0.0031 & 0.0032 & 0.0027  \\ 
$\pm\Delta$ $\mu_R$ scale& 0.0040 & 0.0054 & 0.0040 & 0.0018 & 0.0068 & 0.0022
  \\ \hline  
$\pm\Delta_{\mathrm{tot}}$& 0.0049 & 0.0062 & 0.0051 & 0.0042 & 0.0081 & 0.0038
  \\ \hline    
\multicolumn{6}{r||}{RMS} & 0.0042  \\ \hline\hline
$\alpha_s(93.0\gev)$  & 0.1128 & 0.1140 & 0.1208 & 0.1179 & 0.1117 & 0.1152  \\
 \hline
$\pm\Delta$ stat.     & 0.0004 & 0.0004 & 0.0005 & 0.0005 & 0.0003 & 0.0003  \\ 
$\pm\Delta$ sys. exp. & 0.0009 & 0.0021 & 0.0014 & 0.0022 & 0.0030 & 0.0015  \\
 \hline
$\pm\Delta$ had.      & 0.0026 & 0.0023 & 0.0025 & 0.0031 & 0.0032 & 0.0027  \\ 
$\pm\Delta$ $\mu_R$ scale& 0.0040 & 0.0054 & 0.0040 & 0.0018 & 0.0068 & 0.0022
  \\ \hline  
$\pm\Delta_{\mathrm{tot}}$& 0.0049 & 0.0062 & 0.0051 & 0.0042 & 0.0081 & 0.0038
  \\ \hline    
\multicolumn{6}{|r||}{RMS} & 0.0039  \\ \hline
\end{tabular}
\end{center}
\caption{\label{daniel1} 
{Results of \as\ measurements from distributions in \oas. The data of 
\cite{daniel_final_draft} have been reanalyzed for this analysis.}}
\end{table}
\begin{table}[h]
\begin{center}
\begin{tabular}{|l|c|c|c|c|c||c|}\hline
 \multicolumn{7}{|c|}{$\alpha_s$ in ${\cal{O}}(\alpha_s^2)$}     \\ \hline
\hline
observable  & 1-T & C  & $M^2_{\mathrm{h}}/E^2_{\mathrm{vis}}$
& $B_{\mathrm{max}}$ & $B_{\mathrm{sum}}$ & average\\ \hline
fit range           & 0.09-0.21 & 0.28-0.6  & 0.05-0.17 & 0.07-0.20 &
0.115-0.24 &   \\ \hline \hline
$\alpha_s(183\gev)$   & 0.1075 & 0.1053 & 0.1053 & 0.1065 & 0.1024 & 0.1059  \\
 \hline
$\pm\Delta$ stat.     & 0.0042 & 0.0039 & 0.0047 & 0.0044 & 0.0034 & 0.0037  \\
$\pm\Delta$ sys. exp. & 0.0022 & 0.0023 & 0.0024 & 0.0026 & 0.0012 & 0.0020
\\ \hline
$\pm\Delta$ had.      & 0.0015 & 0.0010 & 0.0033 & 0.0003 & 0.0009 & 0.0008  \\
$\pm\Delta$ $\mu_R$ scale & 0.0030 & 0.0033 & 0.0030 & 0.0013 & 0.0050 &
0.0020  \\ \hline
$\pm\Delta_{\mathrm{tot}}$& 0.0058 & 0.0057 & 0.0069 & 0.0053 & 0.0062 &
0.0047  \\ \hline
\multicolumn{6}{ r||}{RMS} & 0.0019  \\ \hline\hline
$\alpha_s(189\gev)$     & 0.1000 & 0.1038 & 0.1056 & 0.1042 & 0.1007 & 0.1026
 \\ \hline
$\pm\Delta$ stat.       & 0.0030 & 0.0025 & 0.0030 & 0.0028 & 0.0022 & 0.0025
  \\
$\pm\Delta$ sys. exp.   & 0.0024 & 0.0023 & 0.0024 & 0.0028 & 0.0012 & 0.0022
 \\ \hline
$\pm\Delta$ had.        & 0.0010 & 0.0014 & 0.0032 & 0.0005 & 0.0011 & 0.0007
  \\
$\pm\Delta$ $\mu_R$ scale & 0.0029 & 0.0032 & 0.0029 & 0.0013 & 0.0050 &
0.0018  \\ \hline
$\pm\Delta_{\mathrm{tot}}$& 0.0049 & 0.0049 & 0.0058 & 0.0042 & 0.0057 &
0.0039  \\ \hline
\multicolumn{6}{r||}{RMS} & 0.0024  \\ \hline\hline
$\alpha_s(192\gev)$   & 0.1047 & 0.1053 & 0.1143 & 0.1079 & 0.1029 & 0.1035
\\ \hline
$\pm\Delta$ stat.     & 0.0072 & 0.0064 & 0.0074 & 0.0073 & 0.0055 & 0.0058
 \\
$\pm\Delta$ sys. exp. & 0.0026 & 0.0024 & 0.0025 & 0.0029 & 0.0012 & 0.0019
\\ \hline
$\pm\Delta$ had.      & 0.0003 & 0.0009 & 0.0030 & 0.0004 & 0.0006 & 0.0006  \\
$\pm\Delta$ $\mu_R$ scale & 0.0029 & 0.0032 & 0.0029 & 0.0013 & 0.0050 &
0.0028  \\ \hline
$\pm\Delta_{\mathrm{tot}}$& 0.0081 & 0.0076 & 0.0088 & 0.0079 & 0.0076 &
0.0067  \\ \hline
\multicolumn{6}{r||}{RMS} & 0.0044  \\ \hline\hline
$\alpha_s(196\gev)$     & 0.0977 & 0.1041 & 0.1086 & 0.1034 & 0.1016 & 0.1021
 \\ \hline
$\pm\Delta$ stat.       & 0.0050 & 0.0041 & 0.0041 & 0.0046 & 0.0036 & 0.0039
  \\
$\pm\Delta$ sys. exp.   & 0.0027 & 0.0024 & 0.0024 & 0.0030 & 0.0012 & 0.0021
  \\ \hline
$\pm\Delta$ had.        & 0.0011 & 0.0009 & 0.0038 & 0.0003 & 0.0005 & 0.0005
 \\
$\pm\Delta$ $\mu_R$ scale & 0.0029 & 0.0032 & 0.0029 & 0.0013 & 0.0049 & 0.0023
  \\ \hline
$\pm\Delta_{\mathrm{tot}}$& 0.0065 & 0.0058 & 0.0072 & 0.0057 & 0.0062 & 0.0050
  \\ \hline
\multicolumn{6}{|r||}{RMS} &  0.0039 \\\hline
\end{tabular}
\caption{\label{as_oas1}{Results of \as\ measurements from distributions in 
\oas.}}
\end{center}
\end{table}
\begin{table}[h]
\begin{center}
\begin{tabular}{|l|c|c|c|c|c||c|}\hline
 \multicolumn{7}{|c|}{$\alpha_s$ in ${\cal{O}}(\alpha_s^2)$}     \\ \hline
\hline
observable  & 1-T & C  & $M^2_{\mathrm{h}}/E^2_{\mathrm{vis}}$
& $B_{\mathrm{max}}$ & $B_{\mathrm{sum}}$ & average\\ \hline
fit range           & 0.09-0.21 & 0.28-0.6  & 0.05-0.17 & 0.07-0.20 &
0.115-0.24 &   \\ \hline \hline
$\alpha_s(200\gev)$   & 0.1074 & 0.1039 & 0.1085 & 0.1026 & 0.1032 & 0.1031\\ \hline
$\pm\Delta$ stat.     & 0.0042 & 0.0039 & 0.0045 & 0.0044 & 0.0034 & 0.0036  \\
$\pm\Delta$ sys. exp. & 0.0028 & 0.0025 & 0.0026 & 0.0031 & 0.0012 & 0.0022
\\ \hline
$\pm\Delta$ had.      & 0.0014 & 0.0015 & 0.0032 & 0.0006 & 0.0011 & 0.0009
 \\
$\pm\Delta$ $\mu_R$ scale & 0.0029 & 0.0032 & 0.0029 & 0.0013 & 0.0049 &
0.0023  \\ \hline
$\pm\Delta_{\mathrm{tot}}$& 0.0060 & 0.0058 & 0.0068 & 0.0055 & 0.0062 &
0.0049  \\ \hline
\multicolumn{6}{r||}{RMS} & 0.0027  \\ \hline\hline
$\alpha_s(202\gev)$     & 0.1054 & 0.1107 & 0.1169 & 0.1114 & 0.1055 & 0.1077
 \\ \hline
$\pm\Delta$ stat.       & 0.0064 & 0.0052 & 0.0062 & 0.0061 & 0.0047 & 0.0049
 \\
$\pm\Delta$ sys. exp.   & 0.0028 & 0.0025 & 0.0026 & 0.0031 & 0.0012 & 0.0020
  \\ \hline
$\pm\Delta$ had.        & 0.0016 & 0.0009 & 0.0034 & 0.0004 & 0.0005 & 0.0007
 \\
$\pm\Delta$ $\mu_R$ scale & 0.0029 & 0.0031 & 0.0029 & 0.0013 & 0.0049 &
0.0026  \\ \hline
$\pm\Delta_{\mathrm{tot}}$& 0.0077 & 0.0066 & 0.0080 & 0.0070 & 0.0069 &
0.0060  \\ \hline
\multicolumn{6}{r||}{RMS} & 0.0048  \\ \hline\hline
$\alpha_s(205\gev)$     & 0.0980 & 0.0978 & 0.1017 & 0.1033 & 0.0976 & 0.1000
 \\ \hline
$\pm\Delta$ stat.       & 0.0053 & 0.0044 & 0.0051 & 0.0047 & 0.0039 & 0.0041
 \\
$\pm\Delta$ sys. exp.   & 0.0030 & 0.0026 & 0.0026 & 0.0033 & 0.0012 & 0.0022
 \\ \hline
$\pm\Delta$ had.        & 0.0006 & 0.0012 & 0.0028 & 0.0005 & 0.0012 & 0.0008
 \\
$\pm\Delta$ $\mu_R$ scale & 0.0028 & 0.0031 & 0.0028 & 0.0013 & 0.0048 & 0.0023
  \\ \hline
$\pm\Delta_{\mathrm{tot}}$& 0.0067 & 0.0061 & 0.0070 & 0.0059 & 0.0064 & 0.0052
  \\ \hline
\multicolumn{6}{r||}{RMS} & 0.0026  \\ \hline\hline
$\alpha_s(207\gev)$     & 0.1057 & 0.1031 & 0.1090 & 0.1040 & 0.1032 & 0.1033
  \\ \hline
$\pm\Delta$ stat.       & 0.0035 & 0.0032 & 0.0036 & 0.0035 & 0.0028 & 0.0030
 \\
$\pm\Delta$ sys. exp.   & 0.0031 & 0.0026 & 0.0027 & 0.0034 & 0.0013 & 0.0023
  \\ \hline
$\pm\Delta$ had.        & 0.0010 & 0.0010 & 0.0027 & 0.0002 & 0.0006 & 0.0004
  \\
$\pm\Delta$ $\mu_R$ scale & 0.0028 & 0.0031 & 0.0028 & 0.0013 & 0.0048 & 0.0022
  \\ \hline
$\pm\Delta_{\mathrm{tot}}$& 0.0056 & 0.0052 & 0.0060 & 0.0051 & 0.0057 & 0.0044
  \\ \hline
\multicolumn{6}{|r||}{RMS} & 0.0025  \\\hline
\end{tabular}
\caption{\label{as_oas2}{Results of \as\ measurements from distributions in 
\oas.}}
\end{center}
\end{table}
\begin{table}[h]
\begin{center}
\begin{tabular}{|l|c|c|c|c|c||c|}\hline
 \multicolumn{7}{|c|}{$\alpha_s$ in NLLA}     \\ \hline
\hline 
observable  & 1-T & C  & $M^2_{\mathrm{h}}/E^2_{\mathrm{vis}}$ 
& $B_{\mathrm{max}}$ & $B_{\mathrm{sum}}$ & average\\ \hline 
fit range           & 0.09-0.21 & 0.16-0.28  & 0.05-0.17 & 0.07-0.20 &
0.115-0.24 &   \\ \hline \hline
$\alpha_s(89.5\gev)$  & 0.1158 & 0.1034 & 0.1190 & 0.1062 & 0.1080 & 0.1088  
\\ \hline
$\pm\Delta$ stat.     & 0.0006 & 0.0003 & 0.0007 & 0.0009 & 0.0002 &
0.0003  \\  
$\pm\Delta$ sys. exp. & 0.0014 & 0.0023 & 0.0041 & 0.0042 & 0.0033 & 0.0018  
\\ \hline
$\pm\Delta$ had.      & 0.0042 & 0.0032 & 0.0062 & 0.0031 & 0.0042 &
0.0032  \\  
$\pm\Delta$ X scale& 0.0059 & 0.0054 & 0.0063 & 0.0067 & 0.0067 & 0.0054  
\\ \hline  
$\pm\Delta_{\mathrm{tot}}$& 0.0074 & 0.0067 & 0.0097 & 0.0085 & 0.0086 & 0.0066
  \\ \hline    
\multicolumn{6}{r||}{RMS} & 0.0066  \\ \hline\hline
$\alpha_s(91.2\gev)$  & 0.1167 & 0.1034 & 0.1197 & 0.10056 & 0.1075 & 0.1081  
\\ \hline
$\pm\Delta$ stat.     & 0.0003 & 0.0002 & 0.0003 & 0.0003 & 0.0002 &
0.0002  \\  
$\pm\Delta$ sys. exp. & 0.0014 & 0.0023 & 0.0041 & 0.0042 & 0.0033 & 0.0018  
\\ \hline
$\pm\Delta$ had.      & 0.0042 & 0.0032 & 0.0062 & 0.0031 & 0.0042 & 0.0032  \\ 
$\pm\Delta$ X scale& 0.0059 & 0.0054 & 0.0063 & 0.0067 & 0.0067 & 0.0054  
\\ \hline  
$\pm\Delta_{\mathrm{tot}}$& 0.0074 & 0.0067 & 0.0097 & 0.0085 & 0.0086 & 0.0066 
 \\ \hline    
\multicolumn{6}{r||}{RMS} & 0.0083  \\ \hline\hline
$\alpha_s(93.0\gev)$  & 0.1141 & 0.1013 & 0.1176 & 0.1038 & 0.1062 & 0.1080  
\\ \hline
$\pm\Delta$ stat.     & 0.0005 & 0.0003 & 0.0006 & 0.0009 & 0.0002 & 0.0003  \\ 
$\pm\Delta$ sys. exp. & 0.0014 & 0.0023 & 0.0041 & 0.0042 & 0.0033 & 0.0018 
 \\ \hline
$\pm\Delta$ had.      & 0.0042 & 0.0032 & 0.0062 & 0.0031 & 0.0042 & 0.0032  \\ 
$\pm\Delta$ X scale& 0.0059 & 0.0054 & 0.0063 & 0.0067 & 0.0067 & 0.0054  
\\ \hline  
$\pm\Delta_{\mathrm{tot}}$& 0.0074 & 0.0067 & 0.0097 & 0.0085 & 0.0086 & 0.0066
  \\ \hline    
\multicolumn{6}{|r||}{RMS} & 0.0070  \\ \hline
\end{tabular}
\end{center}
\caption{\label{daniel2} 
{Results of \as\ measurements from distributions in NLLA. The data of 
\cite{daniel_final_draft} have been reanalyzed for this analysis.}}
\end{table}
\begin{table}[h]
\begin{center}
\begin{tabular}{|l|c|c|c|c|c||c|}\hline
 \multicolumn{7}{|c|}{$\alpha_s$ in NLLA}     \\ \hline
\hline
observable & 1-T & C  & $M^2_{\mathrm{h}}/E^2_{\mathrm{vis}}$
& $B_{\mathrm{max}}$ & $B_{\mathrm{sum}}$ & average\\ \hline
fit range           & 0.05-0.09 & 0.16-0.28  & 0.03-0.05 & 0.05-0.07 &
0.06-0.115 &   \\ \hline \hline
$\alpha_s(183\gev)$     & 0.0942 & 0.0971 & 0.1088 & 0.1029 & 0.1000 & 0.1000 
 \\ \hline
$\pm\Delta$ stat.       & 0.0055 & 0.0052 & 0.0041 & 0.0097 & 0.0055 & 0.0037 
 \\ 
 $\pm\Delta$ sys. exp.  & 0.0022 & 0.0023 & 0.0024 & 0.0026 & 0.0012 & 0.0016
  \\ \hline
$\pm\Delta$ had.        & 0.0015 & 0.0023 & 0.0032 & 0.0005 & 0.0016 & 0.0020
  \\ 
$\pm\Delta$ X scale     & 0.0044 & 0.0042 & 0.0035 & 0.0037 & 0.0054 & 0.0044 
 \\ \hline  
$\pm\Delta_{\mathrm{tot}}$& 0.0075 & 0.0074 & 0.0067 & 0.0107 & 0.0079 & 0.0063
\\ \hline    
\multicolumn{6}{r||}{RMS} & 0.0056  \\ \hline\hline
$\alpha_s(189\gev)$     & 0.1052 & 0.1005 & 0.1066 & 0.1026 & 0.1031 & 0.1021 
 \\ \hline
$\pm\Delta$ stat.       & 0.0035 & 0.0031 & 0.0038 & 0.0057 & 0.0031 & 0.0026
 \\ 
$\pm\Delta$ sys. exp.   & 0.0024 & 0.0023 & 0.0024 & 0.0028 & 0.0012 & 0.0015 
 \\ \hline
$\pm\Delta$ had.        & 0.0025 & 0.0025 & 0.0031 & 0.0005 & 0.0019 & 0.0021
  \\ 
$\pm\Delta$ X scale    & 0.0044 & 0.0041 & 0.0035 & 0.0037 & 0.0053 & 0.0041
  \\ \hline  
$\pm\Delta_{\mathrm{tot}}$& 0.0066 & 0.0062 & 0.0065 & 0.0074 & 0.0065 & 
0.0055\\ \hline    
\multicolumn{6}{r||}{RMS} & 0.0024  \\ \hline\hline
$\alpha_s(192\gev)$     & 0.1058 & 0.0906 & 0.1007 & 0.0890 & 0.0998 & 0.0998
  \\ \hline
$\pm\Delta$ stat.       & 0.0101 & 0.0090 & 0.0110 & 0.0099 & 0.0082 & 0.0071
  \\ 
$\pm\Delta$ sys. exp.   & 0.0026 & 0.0024 & 0.0025 & 0.0029 & 0.0012 & 0.0016
  \\ \hline
$\pm\Delta$ had.        & 0.0028 & 0.0017 & 0.0033 & 0.0007 & 0.0019 & 0.0019
  \\ 
$\pm\Delta$ X scale    & 0.0043 & 0.0041 & 0.0035 & 0.0037 & 0.0053 & 0.0044 
 \\ \hline  
$\pm\Delta_{\mathrm{tot}}$& 0.0116 & 0.0103 & 0.0123 & 0.0110 & 0.0100 & 
0.0087\\ \hline    
\multicolumn{6}{r||}{RMS} & 0.0071  \\ \hline\hline
$\alpha_s(196\gev)$     & 0.1007 & 0.0945 & 0.0974 & 0.1013 & 0.0962 & 0.0960 
 \\ \hline
$\pm\Delta$ stat.       & 0.0057 & 0.0050 & 0.0060 & 0.0090 & 0.0047 & 0.0038
  \\ 
$\pm\Delta$ sys. exp.   & 0.0027 & 0.0024 & 0.0025 & 0.0030 & 0.0012 & 0.0015
  \\ \hline
$\pm\Delta$ had.        & 0.0015 & 0.0024 & 0.0017 & 0.0008 & 0.0017 & 0.0019 
 \\ 
$\pm\Delta$ X scale    & 0.0043 & 0.0041 & 0.0034 & 0.0036 & 0.0052 & 0.0042 
 \\ \hline  
$\pm\Delta_{\mathrm{tot}}$& 0.0078 & 0.0073 & 0.0075 & 0.0101 & 0.0073 & 
0.0062\\ \hline    
\multicolumn{6}{|r||}{RMS} & 0.0029  \\\hline 
\end{tabular}
\caption{\label{as_nlla1}{Results of \as\ measurements from distributions in 
NLLA.}}
\end{center}
\end{table}

\begin{table}[h]
\begin{center}
\begin{tabular}{|l|c|c|c|c|c||c|}\hline
 \multicolumn{7}{|c|}{$\alpha_s$ in NLLA}     \\ \hline
\hline
observable & 1-T & C  & $M^2_{\mathrm{h}}/E^2_{\mathrm{vis}}$
& $B_{\mathrm{max}}$ & $B_{\mathrm{sum}}$ & average\\ \hline
fit range           & 0.05-0.09 & 0.16-0.28  & 0.03-0.05 & 0.05-0.07 &
0.06-0.115 &   \\ \hline \hline
$\alpha_s(200\gev)$     & 0.0981 & 0.0864 & 0.0974 & 0.1012 & 0.0953 & 0.0914
  \\ \hline
$\pm\Delta$ stat.       & 0.0058 & 0.0047 & 0.0057 & 0.0085 & 0.0045 & 0.0038\\ 
$\pm\Delta$ sys. exp.   & 0.0028 & 0.0025 & 0.0026 & 0.0031 & 0.0014 & 0.0017
\\ \hline
$\pm\Delta$ had.        & 0.0024 & 0.0020 & 0.0028 & 0.0006 & 0.0020 & 0.0023\\ 
$\pm\Delta$ X scale   & 0.0043 & 0.0040 & 0.0034 & 0.0036 & 0.0052 & 0.0041  
\\ \hline  
$\pm\Delta_{\mathrm{tot}}$& 0.0081 & 0.0070 & 0.0077 & 0.0098 & 0.0073 & 
0.0063\\ \hline    
\multicolumn{6}{r||}{RMS} & 0.0056  \\ \hline\hline
$\alpha_s(202\gev)$     & 0.1165 & 0.1026 & 0.1005 & 0.1013 & 0.1039 & 0.1072
  \\ \hline
$\pm\Delta$ stat.       & 0.0076 & 0.0071 & 0.0088 & 0.0146 & 0.0072 & 0.0055 
 \\ 
$\pm\Delta$ sys. exp.   & 0.0028 & 0.0025 & 0.0026 & 0.0031 & 0.0013 & 0.0016 
 \\ \hline
$\pm\Delta$ had.        & 0.0016 & 0.0017 & 0.0026 & 0.0005 & 0.0020 & 0.0021 
 \\ 
$\pm\Delta$ X scale   & 0.0043 & 0.0040 & 0.0034 & 0.0036 & 0.0052 & 0.0041  
\\ \hline  
$\pm\Delta_{\mathrm{tot}}$& 0.0093 & 0.00 87& 0.0101 & 0.0154 & 0.0092 & 
0.0073\\ \hline    
\multicolumn{6}{r||}{RMS} & 0.0066  \\ \hline\hline
$\alpha_s(205\gev)$     & 0.0928 & 0.0970 & 0.1032 & 0.1039 & 0.1036 & 0.0996 
 \\ \hline
$\pm\Delta$ stat.       & 0.0056 & 0.0050 & 0.0062 & 0.0098 & 0.0049 & 0.0039
  \\ 
$\pm\Delta$ sys. exp.   & 0.0030 & 0.0026 & 0.0026 & 0.0031 & 0.0013 & 0.0015
  \\ \hline
$\pm\Delta$ had.        & 0.0029 & 0.0023 & 0.0024 & 0.0005 & 0.0017 & 0.0022
  \\ 
$\pm\Delta$ X scale   & 0.0042 & 0.0040 & 0.0034 & 0.0036 & 0.0051 & 0.0041 
 \\ \hline  
$\pm\Delta_{\mathrm{tot}}$& 0.0081 & 0.0073 & 0.0079 & 0.0109 & 0.0074 & 
0.0063\\ \hline    
\multicolumn{6}{r||}{RMS} & 0.0050  \\ \hline\hline
$\alpha_s(207\gev)$     & 0.1054 & 0.0935 & 0.1012 & 0.0972 & 0.0975 & 0.0976 
\\ \hline
$\pm\Delta$ stat.       & 0.0043 & 0.0038 & 0.0047 & 0.0066 & 0.0036 & 0.0030 
 \\ 
$\pm\Delta$ sys. exp.   & 0.0031 & 0.0026 & 0.0027 & 0.0034 & 0.0013 & 0.0016 
 \\ \hline
$\pm\Delta$ had.        & 0.0019 & 0.0018 & 0.0025 & 0.0005 & 0.0016 & 0.0019
  \\ 
$\pm\Delta$ X scale   & 0.0042 & 0.0040 & 0.0034 & 0.0036 & 0.0051 & 0.0040  
\\ \hline  
$\pm\Delta_{\mathrm{tot}}$& 0.0070 & 0.0064 & 0.0069 & 0.0083 & 0.0066 & 
0.0056\\ \hline    
\multicolumn{6}{|r||}{RMS} & 0.0045  \\\hline 
\end{tabular}
\caption{\label{as_nlla2}{Results of \as\ measurements from distributions in 
NLLA.}}
\end{center}
\end{table}
\begin{table}[h]
\begin{center}
\begin{tabular}{|l|c|c|c|c|c||c|}\hline
 \multicolumn{7}{|c|}{$\alpha_s$ in ${\cal{O}}(\alpha_s^2)$+NLLA (logR)}     
\\ \hline
\hline 
observable  & 1-T & C  & $M^2_{\mathrm{h}}/E^2_{\mathrm{vis}}$ 
& $B_{\mathrm{max}}$ & $B_{\mathrm{sum}}$ & average\\ \hline 
fit range           & 0.09-0.21 & 0.16-0.6  & 0.05-0.17 & 0.07-0.20 &
0.115-0.24 &   \\ \hline \hline
$\alpha_s(89.5\gev)$  & 0.1257 & 0.1211 & 0.1226 & 0.1155 & 0.1249 & 0.1220  
\\ \hline
$\pm\Delta$ stat.     & 0.0002 & 0.0002 & 0.0004 & 0.0004 & 0.0001 & 0.0004  \\ 
$\pm\Delta$ sys. exp. & 0.0021 & 0.0022 & 0.0023 & 0.0021 & 0.0019 & 0.0020  
\\ \hline
$\pm\Delta$ had.      & 0.0031 & 0.0021 & 0.0019 & 0.0023 & 0.0029 & 0.0020  \\ 
$\pm\Delta$ X scale  & 0.0060 & 0.0057 & 0.0048 & 0.0051 & 0.0073 & 0.0048  
\\ \hline  
$\pm\Delta_{\mathrm{tot}}$& 0.0071 & 0.0065 & 0.0057 & 0.0060 & 0.0081 & 0.0056
 \\ \hline    
\multicolumn{6}{r||}{RMS} & 0.0040  \\ \hline\hline
$\alpha_s(91.2\gev)$  & 0.1256 & 0.1211 & 0.1230 & 0.1156 & 0.1250 & 0.1219  \\
 \hline
$\pm\Delta$ stat.     & 0.0002 & 0.0002 & 0.0002 & 0.0001 & 0.0001 & 0.0002  \\ 
$\pm\Delta$ sys. exp. & 0.0021 & 0.0022 & 0.0023 & 0.0021 & 0.0019 & 0.0020  
\\ \hline
$\pm\Delta$ had.      & 0.0031 & 0.0021 & 0.0019 & 0.0023 & 0.0029 & 0.0020  \\ 
$\pm\Delta$ X scale  & 0.0060 & 0.0057 & 0.0048 & 0.0051 & 0.0073 & 0.0048  
\\ \hline  
$\pm\Delta_{\mathrm{tot}}$& 0.0071 & 0.0065 & 0.0057 & 0.0060 & 0.0081 & 0.0056
  \\ \hline    
\multicolumn{6}{r||}{RMS} & 0.0040  \\ \hline\hline
$\alpha_s(93.0\gev)$  & 0.1257 & 0.1211 & 0.1208 & 0.1144 & 0.1235 & 0.1222  
\\ \hline
$\pm\Delta$ stat.     & 0.0002 & 0.0002 & 0.0004 & 0.0004 & 0.0001 & 0.0004  \\ 
$\pm\Delta$ sys. exp. & 0.0021 & 0.0022 & 0.0023 & 0.0021 & 0.0019 & 0.0020  
\\ \hline
$\pm\Delta$ had.      & 0.0031 & 0.0021 & 0.0019 & 0.0023 & 0.0029 & 0.0020  \\ 
$\pm\Delta$ X   scale& 0.0060 & 0.0057 & 0.0048 & 0.0051 & 0.0073 & 0.0048  
\\ \hline  
$\pm\Delta_{\mathrm{tot}}$& 0.0071 & 0.0065 & 0.0057 & 0.0060 & 0.0081 & 0.0056
  \\ \hline    
\multicolumn{6}{|r||}{RMS} & 0.0042  \\ \hline
\end{tabular}
\end{center}
\caption{\label{daniel3} 
{Results of \as\ measurements from distributions in \oas+NLLA. The data of 
\cite{daniel_final_draft} have been reanalyzed for this analysis.}}
\end{table}
\begin{table}[h]
\begin{center}
\begin{tabular}{|l|c|c|c|c|c||c|}\hline
 \multicolumn{7}{|c|}{$\alpha_s$ in ${\mathbf{\cal{O}}(\alpha_s^2)}$+NLLA
(logR)}\\
\hline  \hline
observable & 1-T & C  & $M^2_{\mathrm{h}}/E^2_{\mathrm{vis}}$
& $B_{\mathrm{max}}$ & $B_{\mathrm{sum}}$ & average\\ \hline
fit range           & 0.05-0.21 & 0.16-0.6  & 0.03-0.17 & 0.05-0.20 &
0.06-0.24 &   \\ \hline \hline
$\alpha_s(183\gev)$    & 0.1072 & 0.1111 & 0.1094 & 0.1049 & 0.1119 & 0.1081  
\\ \hline
$\pm\Delta$ stat.      & 0.0043 & 0.0036 & 0.0037 & 0.0037 & 0.0039 & 0.0034 
\\ 
$\pm\Delta$ sys. exp.  & 0.0022 & 0.0023 & 0.0024 & 0.0026 & 0.0012 & 0.0021 
 \\ \hline
$\pm\Delta$ had.       & 0.0017 & 0.0017 & 0.0031 & 0.0005 & 0.0016 & 0.0011
  \\ 
$\pm\Delta$ X scale    & 0.0044 & 0.0042 & 0.0035 & 0.0037 & 0.0054 & 0.0037 
 \\ \hline  
$\pm\Delta_{\mathrm{tot}}$& 0.0068 & 0.0062 & 0.0064 & 0.0059 & 0.0070 & 
0.0056 \\ \hline    
\multicolumn{6}{r||}{RMS}  & 0.0029  \\ \hline\hline
$\alpha_s(189\gev)$    & 0.1105 & 0.1101 & 0.1079 & 0.1032 & 0.1122 & 0.1067 
 \\ \hline
$\pm\Delta$ stat.      & 0.0027 & 0.0023 & 0.0027 & 0.0024 & 0.0021 & 0.0022 
 \\ 
$\pm\Delta$ sys. exp.  & 0.0024 & 0.0023 & 0.0024 & 0.0028 & 0.0012 & 0.0020 
 \\ \hline
$\pm\Delta$ had.       & 0.0020 & 0.0020 & 0.0031 & 0.0005 & 0.0016 & 0.0010 
 \\ 
$\pm\Delta$ X scale    & 0.0044 & 0.0041 & 0.0035 & 0.0037 & 0.0053 & 0.0038
 \\ \hline  
$\pm\Delta_{\mathrm{tot}}$& 0.0060 & 0.0056 & 0.0059 & 0.0052 & 0.0060 & 
0.0049 \\ \hline    
\multicolumn{6}{r||}{RMS}  & 0.0035  \\ \hline\hline
$\alpha_s(192\gev)$    & 0.1124 & 0.1083 & 0.1082 & 0.1051 & 0.1139 & 0.1096
  \\ \hline
$\pm\Delta$ stat.      & 0.0068 & 0.0060 & 0.0063 & 0.0060 & 0.0052 & 0.0051
  \\ 
$\pm\Delta$ sys. exp.  & 0.0026 & 0.0024 & 0.0025 & 0.0029 & 0.0012 & 0.0018  
 \\ \hline
$\pm\Delta$ had.       & 0.0017 & 0.0014 & 0.0032 & 0.0007 & 0.0019 & 0.0015 
  \\ 
$\pm\Delta$ X scale    & 0.0043 & 0.0041 & 0.0035 & 0.0037 & 0.0053 & 0.0040
  \\ \hline  
$\pm\Delta_{\mathrm{tot}}$& 0.0086 & 0.0078 & 0.0083 & 0.0077 & 0.0078 & 
0.0069 \\ \hline    
\multicolumn{6}{r||}{RMS} & 0.0035  \\ \hline\hline
$\alpha_s(196\gev)$   & 0.1045 & 0.1079 & 0.1060 & 0.1024 & 0.1107 & 0.1068
   \\ \hline
$\pm\Delta$ stat.     & 0.0042 & 0.0038 & 0.0040 & 0.0038 & 0.0032 & 0.0033 
  \\ 
$\pm\Delta$ sys. exp. & 0.0027 & 0.0024 & 0.0025 & 0.0030 & 0.0012 & 0.0019 
  \\ \hline
$\pm\Delta$ had.      & 0.0013 & 0.0013 & 0.0029 & 0.0008 & 0.0017 & 0.0014 
  \\ 
$\pm\Delta$ X scale    & 0.0043 & 0.0041 & 0.0034 & 0.0036 & 0.0052 & 0.0038 
 \\ \hline  
$\pm\Delta_{\mathrm{tot}}$& 0.0067 & 0.0062 & 0.0065 & 0.0061 & 0.0065 & 
0.0055\\ \hline    
\multicolumn{6}{|r||}{RMS} &  0.0036 \\\hline 
\end{tabular}
\caption{\label{as_logr1}{Results of \as\ measurements from distributions 
in \oas +NLLA.}}
\end{center}
\end{table}
\begin{table}[h]
\begin{center}
\begin{tabular}{|l|c|c|c|c|c||c|}\hline
 \multicolumn{7}{|c|}{$\alpha_s$ in ${\mathbf{\cal{O}}(\alpha_s^2)}$+NLLA
(logR)}\\
\hline  \hline
observable & 1-T & C  & $M^2_{\mathrm{h}}/E^2_{\mathrm{vis}}$
& $B_{\mathrm{max}}$ & $B_{\mathrm{sum}}$ & average\\ \hline
fit range           & 0.05-0.21 & 0.16-0.6  & 0.03-0.17 & 0.05-0.20 &
0.06-0.24 &   \\ \hline \hline
$\alpha_s(200\gev)$   & 0.1095 & 0.1044 & 0.1045 & 0.1021 & 0.1101 & 0.1044  
\\ \hline
$\pm\Delta$ stat.     & 0.0041 & 0.0036 & 0.0034 & 0.0036 & 0.0032 & 0.0031 
 \\ 
$\pm\Delta$ sys. exp. & 0.0028 & 0.0025 & 0.0026 & 0.0031 & 0.0012 & 0.0020  
\\ \hline
$\pm\Delta$ had.      & 0.0023 & 0.0022 & 0.0030 & 0.0006 & 0.0020 & 0.0015 
 \\ 
$\pm\Delta$ X scale   & 0.0043 & 0.0040 & 0.0034 & 0.0036 & 0.0052 & 0.0037  
\\ \hline  
$\pm\Delta_{\mathrm{tot}}$& 0.0071 & 0.0063 & 0.0062 & 0.0060 & 0.0066 & 
0.0054\\ \hline    
\multicolumn{6}{r||}{RMS} &  0.0036 \\ \hline\hline
$\alpha_s(202\gev)$   & 0.1188 & 0.1160 & 0.1105 & 0.1103 & 0.1194 & 0.1141  
\\ \hline
$\pm\Delta$ stat.     & 0.0058 & 0.0050 & 0.0055 & 0.0053 & 0.0047 & 0.0046  
\\ 
$\pm\Delta$ sys. exp. & 0.0028 & 0.0025 & 0.0026 & 0.0031 & 0.0012 & 0.0019  
 \\ \hline
$\pm\Delta$ had.      & 0.0018 & 0.0015 & 0.0031 & 0.0005 & 0.0016 & 0.0014
  \\ 
$\pm\Delta$ X scale   & 0.0043 & 0.0040 & 0.0034 & 0.0036 & 0.0052 & 0.0038 
 \\ \hline  
$\pm\Delta_{\mathrm{tot}}$& 0.0081 & 0.0070 & 0.0074 & 0.0071 & 0.0074 & 
0.0064\\ \hline    
\multicolumn{6}{r||}{RMS} &  0.0044 \\ \hline\hline
$\alpha_s(205\gev)$   & 0.1023 & 0.1064 & 0.1041 & 0.1031 & 0.1109 & 0.1071 
  \\ \hline
$\pm\Delta$ stat.     & 0.0045 & 0.0037 & 0.0041 & 0.0038 & 0.0033 & 0.0033 
 \\ 
$\pm\Delta$ sys. exp. & 0.0030 & 0.0026 & 0.0026 & 0.0033 & 0.0012 & 0.0019 
 \\ \hline
$\pm\Delta$ had.      & 0.0021 & 0.0017 & 0.0027 & 0.0005 & 0.0017 & 0.0012 \\ 
$\pm\Delta$ X scale   & 0.0042 & 0.0040 & 0.0034 & 0.0036 & 0.0051 & 0.0037 
 \\ \hline  
$\pm\Delta_{\mathrm{tot}}$& 0.0072 & 0.0063 & 0.0065 & 0.0061 & 0.0064 & 
0.0055\\ \hline    
\multicolumn{6}{r||}{RMS} &  0.0035 \\ \hline\hline
$\alpha_s(207\gev)$   & 0.1118 & 0.1074 & 0.1058 & 0.1021 & 0.1134 & 0.1061  
 \\ \hline
$\pm\Delta$ stat.     & 0.0036 & 0.0026 & 0.0031 & 0.0029 & 0.0025 & 0.0026  \\ 
$\pm\Delta$ sys. exp. & 0.0031 & 0.0026 & 0.0027 & 0.0034 & 0.0013 & 0.0020 
 \\ \hline
$\pm\Delta$ had.      & 0.0016 & 0.0014 & 0.0027 & 0.0005 & 0.0016 & 0.0012  \\ 
$\pm\Delta$ X scale   & 0.0042 & 0.0040 & 0.0034 & 0.0036 & 0.0051 & 0.0037 
 \\ \hline  
$\pm\Delta_{\mathrm{tot}}$& 0.0064 & 0.0056 & 0.0060 & 0.0058 & 0.0060 & 
0.0051\\ \hline    
\multicolumn{6}{|r||}{RMS} & 0.0050  \\\hline
\end{tabular}
\caption{\label{as_logr2}{Results of \as\ measurements from distributions in 
\oas +NLLA.}}
\end{center}
\end{table}
\begin{table}[h]
\begin{center}
\begin{tabular}{|l|c|c|c|c|c||c|}\hline
 \multicolumn{7}{|c|}{$\alpha_s$ from mean values with power corrections}\\ 
\hline  \hline
Observable  & $\langle$1-T$\rangle$ & $\langle$C$\rangle$  & 
$\langle M^2_{\mathrm{h}}/E^2_{\mathrm{vis}}\rangle$ 
& $\langle B_{\mathrm{max}}\rangle$ & $\langle B_{\mathrm{sum}}\rangle$ & 
average\\ \hline 
$\alpha_s(45\gev)$        & 0.1341 & 0.1418 & 0.1268 & 0.1358 & 0.1421 & 0.1370
  \\ \hline
$\pm\Delta$ stat.         & 0.0154 & 0.0165 & 0.0140 & 0.0127 & 0.0091 & 0.0043
  \\ 
$\pm\Delta$ sys. exp.     & 0.0036 & 0.0037 & 0.0026 & 0.0005 & 0.0005 & 0.0023
  \\ \hline 
$\pm\Delta$ $\mu_R$ scale& 0.0072 & 0.0055 & 0.0041 & 0.0062 & 0.0052 & 0.0051
  \\ 
$\pm\Delta$ $\mu_I$ scale& 0.0052 & 0.0038 & 0.0035 & 0.0005 & 0.0001 & 0.0016
  \\ \hline  
$\pm\Delta_{\mathrm{tot}}$& 0.0181 & 0.0181 & 0.0152 & 0.0142 & 0.0105 & 0.0073
\\ \hline    
\multicolumn{6}{r||}{RMS}  & 0.0063  \\ \hline\hline
$\alpha_s(66\gev)$        & 0.1159 & 0.1252 & 0.1140 & 0.1189 & 0.1265 & 0.1251
  \\ \hline
$\pm\Delta$ stat.         & 0.0080 & 0.0081 & 0.0076 & 0.0070 & 0.0055 & 0.0043
  \\ 
$\pm\Delta$ sys. exp.     & 0.0023 & 0.0028 & 0.0032 & 0.0009 & 0.0006 & 0.0010
  \\ \hline 
$\pm\Delta$ $\mu_R$ scale& 0.0059 & 0.0045 & 0.0034 & 0.0051 & 0.0043 & 0.0041
  \\ 
$\pm\Delta$ $\mu_I$ scale& 0.0054 & 0.0049 & 0.0030 & 0.0012 & 0.0009 & 0.0019
  \\ \hline  
$\pm\Delta_{\mathrm{tot}}$& 0.0116 & 0.0109 & 0.0094 & 0.0088 & 0.0070 & 0.0064
\\ \hline    
\multicolumn{6}{r||}{RMS}  & 0.0055  \\ \hline\hline
$\alpha_s(76\gev)$        & 0.1302 & 0.1388 & 0.1235 & 0.1244 & 0.1313 & 0.1255
  \\ \hline
$\pm\Delta$ stat.         & 0.0074 & 0.0077 & 0.0072 & 0.0070 & 0.0053 & 0.0039
  \\ 
$\pm\Delta$ sys. exp.     & 0.0032 & 0.0040 & 0.0034 & 0.0020 & 0.0016 & 0.0023
  \\ \hline 
$\pm\Delta$ $\mu_R$ scale& 0.0055 & 0.0042 & 0.0032 & 0.0048 & 0.0040 & 0.0038
  \\ 
$\pm\Delta$ $\mu_I$ scale& 0.0024 & 0.0013 & 0.0018 & 0.0006 & 0.0004 & 0.0008
  \\ \hline  
$\pm\Delta_{\mathrm{tot}}$& 0.0101 & 0.0098 & 0.0087 & 0.0088 & 0.0068 & 0.0060
\\ \hline    
\multicolumn{6}{r||}{RMS}  &  0.0062 \\ \hline\hline
$\alpha_s(89\gev)$        & 0.1189 & 0.1263 & 0.1152 & 0.1163 & 0.1235 & 0.1177
  \\ \hline
$\pm\Delta$ stat.         & 0.0004 & 0.0004 & 0.0004 & 0.0003 & 0.0002 & 0.0003
  \\ 
$\pm\Delta$ sys. exp.     & 0.0015 & 0.0013 & 0.0019 & 0.0003 & 0.0004 & 0.0009
  \\ \hline 
$\pm\Delta$ $\mu_R$ scale& 0.0052 & 0.0039 & 0.0029 & 0.0044 & 0.0037 & 0.0031
  \\ 
$\pm\Delta$ $\mu_I$ scale& 0.0031 & 0.0027 & 0.0019 & 0.0009 & 0.0007 & 0.0010
  \\ \hline  
$\pm\Delta_{\mathrm{tot}}$& 0.0063 & 0.0050 & 0.0040 & 0.0046 & 0.0038 & 0.0034
\\ \hline    
\multicolumn{6}{r||}{RMS}  & 0.0048 \\ \hline\hline
$\alpha_s(91.2\gev)$      & 0.1193 & 0.1270 & 0.1153 & 0.1167 & 0.1238 & 0.1176
  \\ \hline
$\pm\Delta$ stat.         & 0.0002 & 0.0001 & 0.0001 & 0.0001 & 0.0001 & 0.0001
  \\ 
$\pm\Delta$ sys. exp.     & 0.0015 & 0.0013 & 0.0019 & 0.0008 & 0.0007 & 0.0011
  \\ \hline 
$\pm\Delta$ $\mu_R$ scale& 0.0051 & 0.0039 & 0.0029 & 0.0044 & 0.0037 & 0.0031
  \\ 
$\pm\Delta$ $\mu_I$ scale& 0.0030 & 0.0025 & 0.0019 & 0.0008 & 0.0007 & 0.0010
  \\ \hline  
$\pm\Delta_{\mathrm{tot}}$& 0.0061 & 0.0048 & 0.0039 & 0.0046 & 0.0038 & 0.0034
\\ \hline    
\multicolumn{6}{|r||}{RMS}  & 0.0049  \\ \hline
\end{tabular}
\caption{\label{as_meansm1}{Results of \as\ measurements 
from mean values with power corrections. The data of  
\cite{daniel_final_draft} and \cite{ralle} have been reanalyzed for this 
analysis}}
\end{center}
\end{table}
\begin{table}[h]
\begin{center}
\begin{tabular}{|l|c|c|c|c|c||c|}\hline
 \multicolumn{7}{|c|}{$\alpha_s$  from mean values with power corrections}\\ 
\hline  \hline
Observable  & $\langle$1-T$\rangle$ & $\langle$C$\rangle$  & 
$\langle M^2_{\mathrm{h}}/E^2_{\mathrm{vis}}\rangle$ 
& $\langle B_{\mathrm{max}}\rangle$ & $\langle B_{\mathrm{sum}}\rangle$ & 
average\\ \hline 
$\alpha_s(93\gev)$        & 0.1182 & 0.1255 & 0.1145 & 0.1157 & 0.1226 & 0.1171
 \\ \hline
$\pm\Delta$ stat.         & 0.0004 & 0.0003 & 0.0004 & 0.0003 & 0.0002 & 0.0003
  \\ 
$\pm\Delta$ sys. exp.     & 0.0013 & 0.0011 & 0.0017 & 0.0006 & 0.0004 & 0.0009
  \\ \hline 
$\pm\Delta$ $\mu_R$ scale& 0.0051 & 0.0039 & 0.0029 & 0.0044 & 0.0037 & 0.0031
  \\ 
$\pm\Delta$ $\mu_I$ scale& 0.0030 & 0.0026 & 0.0019 & 0.0009 & 0.0007 & 0.0010
  \\ \hline  
$\pm\Delta_{\mathrm{tot}}$& 0.0061 & 0.0048 & 0.0039 & 0.0045 & 0.0038 & 0.0033
\\ \hline    
\multicolumn{6}{r||}{RMS}  & 0.0047  \\ \hline\hline
$\alpha_s(133\gev)$       & 0.1158 & 0.1203 & 0.1120 & 0.1109 & 0.1163 & 0.1150
  \\ \hline
$\pm\Delta$ stat.         & 0.0047 & 0.0039 & 0.0048 & 0.0038 & 0.0026 & 0.0023
  \\ 
$\pm\Delta$ sys. exp.     & 0.0010 & 0.0011 & 0.0009 & 0.0016 & 0.0011 & 0.0011
  \\ \hline 
$\pm\Delta$ $\mu_R$ scale& 0.0043 & 0.0033 & 0.0025 & 0.0037 & 0.0031 & 0.0030
  \\ 
$\pm\Delta$ $\mu_I$ scale& 0.0020 & 0.0019 & 0.0013 & 0.0007 & 0.0006 & 0.0010
  \\ \hline  
$\pm\Delta_{\mathrm{tot}}$& 0.0067 & 0.0056 & 0.0056 & 0.0056 & 0.0043 & 0.0041
\\ \hline    
\multicolumn{6}{r||}{RMS}  & 0.0037  \\ \hline\hline
$\alpha_s(161\gev)$       & 0.1037 & 0.1131 & 0.1020 & 0.1053 & 0.1083 & 0.1068
  \\ \hline
$\pm\Delta$ stat.         & 0.0069 & 0.0067 & 0.0068 & 0.0053 & 0.0038 & 0.0031
  \\ 
$\pm\Delta$ sys. exp.     & 0.0052 & 0.0066 & 0.0060 & 0.0038 & 0.0023 & 0.0026
  \\ \hline 
$\pm\Delta$ $\mu_R$ scale& 0.0040 & 0.0031 & 0.0023 & 0.0034 & 0.0029 & 0.0028
  \\ 
$\pm\Delta$ $\mu_I$ scale& 0.0023 & 0.0020 & 0.0014 & 0.0007 & 0.0008 & 0.0010
  \\ \hline  
$\pm\Delta_{\mathrm{tot}}$& 0.0098 & 0.0101 & 0.0094 & 0.0074 & 0.0053 & 0.0050
\\ \hline    
\multicolumn{6}{r||}{RMS}  & 0.0044  \\ \hline\hline
$\alpha_s(172\gev)$       & 0.1140 & 0.1164 & 0.1107 & 0.1119 & 0.1167 & 0.1130
  \\ \hline
$\pm\Delta$ stat.         & 0.0112 & 0.0090 & 0.0106 & 0.0080 & 0.0069 & 0.0061
  \\ 
$\pm\Delta$ sys. exp.     & 0.0021 & 0.0036 & 0.0011 & 0.0020 & 0.0048 & 0.0031
  \\ \hline 
$\pm\Delta$ $\mu_R$ scale& 0.0039 & 0.0030 & 0.0022 & 0.0034 & 0.0028 & 0.0031
  \\ 
$\pm\Delta$ $\mu_I$ scale& 0.0015 & 0.0016 & 0.0010 & 0.0004 & 0.0004 & 0.0011
  \\ \hline  
$\pm\Delta_{\mathrm{tot}}$& 0.0121 & 0.0103 & 0.0109 & 0.0089 & 0.0089 & 0.0076
\\ \hline    
\multicolumn{6}{|r||}{RMS}  & 0.0027  \\ \hline
\end{tabular}
\caption{\label{as_means0}{Results of \as\ measurements 
from mean values with power corrections. The data of \cite{daniel_final_draft}
have been reanalyzed for this analysis}}
\end{center}
\end{table}

\begin{table}[h]
\begin{center}
\begin{tabular}{|l|c|c|c|c|c||c|}\hline
 \multicolumn{7}{|c|}{$\alpha_s$ from mean values with power corrections}\\
\hline  \hline
observable & $\langle$1-T$\rangle$ & $\langle$C$\rangle$  &
$\langle M^2_{\mathrm{h}}/E^2_{\mathrm{vis}}\rangle$
& $\langle B_{\mathrm{max}}\rangle$ & $\langle B_{\mathrm{sum}}\rangle$ &
average\\ \hline
$\alpha_s(183\gev)$       & 0.1156 & 0.1172 & 0.1067 & 0.1059 & 0.1121 &
0.1121  \\ \hline
$\pm\Delta$ stat.         & 0.0047 & 0.0038 & 0.0058 & 0.0042 & 0.0026 &
0.0022  \\
$\pm\Delta$ sys. exp.     & 0.0045 & 0.0061 & 0.0032 & 0.0044 & 0.0011 &
0.0013  \\ \hline
$\pm\Delta$ $\mu_R$ scale & 0.0038 & 0.0029 & 0.0022 & 0.0033 & 0.0027 &
0.0029  \\
$\pm\Delta$ $\mu_I$ scale & 0.0013 & 0.0014 & 0.0010 & 0.0005 & 0.0005 &
0.0005  \\ \hline
$\pm\Delta_{\mathrm{tot}}$& 0.0077 & 0.0079 & 0.0071 & 0.0069 & 0.0040 &
0.0039\\ \hline
\multicolumn{6}{r||}{RMS}  & 0.0051  \\ \hline\hline
$\alpha_s(189\gev)$       & 0.1092 & 0.1185 & 0.1022 & 0.1039 & 0.1085 &
0.1082  \\ \hline
$\pm\Delta$ stat.         & 0.0032 & 0.0024 & 0.0040 & 0.0028 & 0.0017 &
0.0017  \\
$\pm\Delta$ sys. exp.     & 0.0044 & 0.0061 & 0.0032 & 0.0044 & 0.0011 &
0.0012  \\ \hline
$\pm\Delta$ $\mu_R$ scale & 0.0037 & 0.0029 & 0.0021 & 0.0032 & 0.0027 &
0.0027  \\
$\pm\Delta$ $\mu_I$ scale & 0.0016 & 0.0013 & 0.0011 & 0.0006 & 0.0006 &
0.0006  \\ \hline
$\pm\Delta_{\mathrm{tot}}$& 0.0068 & 0.0073 & 0.0057 & 0.0061 & 0.0035 &
0.0034\\ \hline
\multicolumn{6}{r||}{RMS}  & 0.0064  \\ \hline\hline
$\alpha_s(192\gev)$       & 0.0984 & 0.1056 & 0.0947 & 0.0980 & 0.1055 &
0.1092  \\ \hline
$\pm\Delta$ stat.         & 0.0093 & 0.0066 & 0.0127 & 0.0074 & 0.0045 &
0.0028  \\
$\pm\Delta$ sys. exp.     & 0.0040 & 0.0057 & 0.0030 & 0.0042 & 0.0010 &
0.0015  \\ \hline
$\pm\Delta$ $\mu_R$ scale & 0.0037 & 0.0028 & 0.0021 & 0.0032 & 0.0027 &
0.0030  \\
$\pm\Delta$ $\mu_I$ scale & 0.0021 & 0.0021 & 0.0013 & 0.0007 & 0.0007 &
0.0009  \\ \hline
$\pm\Delta_{\mathrm{tot}}$& 0.0110 & 0.0094 & 0.0133 & 0.0091 & 0.0054 &
0.0044\\ \hline
\multicolumn{6}{r||}{RMS}  & 0.0049  \\ \hline\hline
$\alpha_s(196\gev)$       & 0.1165 & 0.1124 & 0.1036 & 0.1072 & 0.1100 &
0.1092  \\ \hline
$\pm\Delta$ stat.         & 0.0057 & 0.0044 & 0.0070 & 0.0048 & 0.0030 &
0.0023  \\
$\pm\Delta$ sys. exp.     & 0.0045 & 0.0059 & 0.0032 & 0.0044 & 0.0011 &
0.0014  \\ \hline
$\pm\Delta$ $\mu_R$ scale & 0.0037 & 0.0028 & 0.0021 & 0.0032 & 0.0027 &
0.0029  \\
$\pm\Delta$ $\mu_I$ scale & 0.0012 & 0.0016 & 0.0010 & 0.0005 & 0.0005 &
0.0006  \\ \hline
$\pm\Delta_{\mathrm{tot}}$& 0.0082 & 0.0080 & 0.0081 & 0.0072 & 0.0042 &
0.0039\\ \hline
\multicolumn{6}{|r||}{RMS}  & 0.0049  \\ \hline
\end{tabular}
\caption{\label{as_means1}{Results of \as\ measurements 
from mean values with power corrections.}}
\end{center}
\end{table}
\begin{table}[h]
\begin{center}
\begin{tabular}{|l|c|c|c|c|c||c|}\hline
\multicolumn{7}{|c|}{$\alpha_s$ from mean values with power corrections}\\
\hline  \hline
observable & $\langle$1-T$\rangle$ & $\langle$C$\rangle$  &
$\langle M^2_{\mathrm{h}}/E^2_{\mathrm{vis}}\rangle$
& $\langle B_{\mathrm{max}}\rangle$ & $\langle B_{\mathrm{sum}}\rangle$ &
average\\ \hline
$\alpha_s(200\gev)$       & 0.1068 & 0.1104 & 0.1062 & 0.1056 & 0.1096 &
0.1105  \\ \hline
$\pm\Delta$ stat.         & 0.0057 & 0.0043 & 0.0069 & 0.0048 & 0.0030 &
0.0023  \\
$\pm\Delta$ sys. exp.     & 0.0042 & 0.0058 & 0.0032 & 0.0044 & 0.0011 &
0.0014  \\ \hline
$\pm\Delta$ $\mu_R$ scale & 0.0037 & 0.0028 & 0.0021 & 0.0032 & 0.0026 &
0.0028  \\
$\pm\Delta$ $\mu_I$ scale & 0.0016 & 0.0017 & 0.0009 & 0.0005 & 0.0005 &
0.0006  \\ \hline
$\pm\Delta_{\mathrm{tot}}$& 0.0082 & 0.0079 & 0.0079 & 0.0072 & 0.0042 &
0.0039\\ \hline
\multicolumn{6}{r||}{RMS}  & 0.0022  \\ \hline\hline
$\alpha_s(202\gev)$       & 0.0946 & 0.1066 & 0.1085 & 0.1070 & 0.1127 &
0.1185  \\ \hline
$\pm\Delta$ stat.         & 0.0097 & 0.0063 & 0.0101 & 0.0067 & 0.0040 &
0.0023  \\
$\pm\Delta$ sys. exp.     & 0.0040 & 0.0057 & 0.0032 & 0.0044 & 0.0011 &
0.0015  \\ \hline
$\pm\Delta$ $\mu_R$ scale & 0.0036 & 0.0028 & 0.0021 & 0.0031 & 0.0026 &
0.0029  \\
$\pm\Delta$ $\mu_I$ scale & 0.0022 & 0.0019 & 0.0009 & 0.0005 & 0.0004 &
0.0008  \\ \hline
$\pm\Delta_{\mathrm{tot}}$& 0.0114 & 0.0091 & 0.0108 & 0.0086 & 0.0049 &
0.0041\\ \hline
\multicolumn{6}{r||}{RMS}  & 0.0068  \\ \hline\hline
$\alpha_s(205\gev)$       & 0.0877 & 0.0879 & 0.0942 & 0.0991 & 0.1003 &
0.1042  \\ \hline
$\pm\Delta$ stat.         & 0.0072 & 0.0054 & 0.0078 & 0.0050 & 0.0033 &
0.0023  \\
$\pm\Delta$ sys. exp.     & 0.0037 & 0.0052 & 0.0030 & 0.0042 & 0.0011 &
0.0013  \\ \hline
$\pm\Delta$ $\mu_R$ scale & 0.0036 & 0.0028 & 0.0021 & 0.0031 & 0.0026 &
0.0028  \\
$\pm\Delta$ $\mu_I$ scale & 0.0025 & 0.0030 & 0.0012 & 0.0006 & 0.0008 &
0.0010  \\ \hline
$\pm\Delta_{\mathrm{tot}}$& 0.0092 & 0.0085 & 0.0087 & 0.0072 & 0.0044 &
0.0040\\ \hline
\multicolumn{6}{r||}{RMS}  & 0.0060  \\ \hline\hline
$\alpha_s(207\gev)$       & 0.1063 & 0.1077 & 0.1049 & 0.1055 & 0.1070 &
0.1072  \\ \hline
$\pm\Delta$ stat.         & 0.0047 & 0.0036 & 0.0055 & 0.0040 & 0.0024 &
 0.0020  \\
$\pm\Delta$ sys. exp.     & 0.0042 & 0.0057 & 0.0032 & 0.0044 & 0.0010 &
0.0012  \\ \hline
$\pm\Delta$ $\mu_R$ scale & 0.0036 & 0.0028 & 0.0020 & 0.0031 & 0.0026 &
0.0027  \\
$\pm\Delta$ $\mu_I$ scale & 0.0015 & 0.0018 & 0.0009 & 0.0005 & 0.0006 &
0.0006  \\ \hline
$\pm\Delta_{\mathrm{tot}}$& 0.0074 & 0.0075 & 0.0068 & 0.0067 & 0.0037 &
0.0036\\ \hline
\multicolumn{6}{|r||}{RMS}  & 0.0011  \\ \hline
\end{tabular}
\caption{\label{as_means2}{Results of \as\ measurements 
from mean values with power corrections.}}
\end{center}
\end{table}
\end{appendix}
\clearpage
\newpage

\end{document}

%% file: acknow.tex
\subsection*{Acknowledgements}
\vskip 3 mm
 We are greatly indebted to our technical 
collaborators, to the members of the CERN-SL Division for the excellent 
performance of the LEP collider, and to the funding agencies for their
support in building and operating the DELPHI detector.\\
We acknowledge in particular the support of \\
Austrian Federal Ministry of Education, Science and Culture,
GZ 616.364/2-III/2a/98, \\
FNRS--FWO, Flanders Institute to encourage scientific and technological 
research in the industry (IWT), Belgium,  \\
FINEP, CNPq, CAPES, FUJB and FAPERJ, Brazil, \\
Czech Ministry of Industry and Trade, GA CR 202/99/1362,\\
Commission of the European Communities (DG XII), \\
Direction des Sciences de la Mati$\grave{\mbox{\rm e}}$re, CEA, France, \\
Bundesministerium f$\ddot{\mbox{\rm u}}$r Bildung, Wissenschaft, Forschung 
und Technologie, Germany,\\
General Secretariat for Research and Technology, Greece, \\
National Science Foundation (NWO) and Foundation for Research on Matter (FOM),
The Netherlands, \\
Norwegian Research Council,  \\
State Committee for Scientific Research, Poland, SPUB-M/CERN/PO3/DZ296/2000,
SPUB-M/CERN/PO3/DZ297/2000, 2P03B 104 19 and 2P03B 69 23(2002-2004)\\
FCT - Funda\c{c}\~ao para a Ci\^encia e Tecnologia, Portugal, \\
Vedecka grantova agentura MS SR, Slovakia, Nr. 95/5195/134, \\
Ministry of Science and Technology of the Republic of Slovenia, \\
CICYT, Spain, AEN99-0950 and AEN99-0761,  \\
The Swedish Natural Science Research Council,      \\
Particle Physics and Astronomy Research Council, UK, \\
Department of Energy, USA, DE-FG02-01ER41155. \\
EEC RTN contract HPRN-CT-00292-2002. \\
